\documentclass[fleqn,usenatbib]{mnras}

\usepackage{newtxtext,newtxmath}

\usepackage[T1]{fontenc}

\DeclareRobustCommand{\VAN}[3]{#2}
\let\VANthebibliography\thebibliography
\def\thebibliography{\DeclareRobustCommand{\VAN}[3]{##3}\VANthebibliography}

\usepackage{graphicx}	
\usepackage{amsmath}	
\usepackage{subcaption}
\usepackage{orcidlink}
\usepackage{gensymb}
\usepackage{multicol}
\usepackage{booktabs,caption}
\usepackage[flushleft]{threeparttable}
\usepackage{pdflscape}
\usepackage{float}

\newcommand{\kepler}{{\it Kepler}}

\newcommand{\TESS}{{\it TESS}}
\newcommand{\tess}{{\it TESS}}

\newcommand{\plato}{{\it PLATO}}
\newcommand{\gaia}{{\it Gaia}}

\newcommand{\jwst}{{\it JWST}}
\newcommand{\cheops}{{\it CHEOPS}}

\newcommand{\ngts}{NGTS}

\newcommand{\harps}{HARPS}

\newcommand{\feros}{FEROS}

\newcommand{\coralie}{CORALIE}

\newcommand{\lco}{LCOGT}

\newcommand{\LCOGT}{LCOGT}
\newcommand{\waspsouth}{WASP-South}
\newcommand{\wasp}{WASP}

\newcommand{\elt}{ELT}

\newcommand{\elsauce}{El Sauce}

\newcommand{\espresso}{ESPRESSO}
\newcommand{\tycho}{Tycho}
\newcommand{\assasn}{ASAS-SN}

\newcommand{\alles}{\textsc{allesfitter}}
\newcommand{\paws}{\textsc{paws}}

\newcommand{\isochrones}{\textsc{isochrones}}
\newcommand{\tiara}{\textsc{TIaRA}}
\newcommand{\bsproc}{\textsc{bsproc}}

\newcommand{\completo}{\textsc{completo}}
\newcommand{\exorings}{\textsc{exorings}}
\newcommand{\aij}{\textsc{AstroImageJ}}
\newcommand{\ldtk}{\textsc{ldtk}}

\newcommand{\kms}{km\,s$^{-1}$}
\newcommand{\ms}{m\,s$^{-1}$}

\newcommand{\mpl}{\mbox{M\textsubscript{p}}}
\newcommand{\rpl}{\mbox{R\textsubscript{p}}}
\newcommand{\mstar}{\mbox{M$_{\star}$}}
\newcommand{\rstar}{\mbox{R$_{\star}$}}

\newcommand{\mjup}{\mbox{M\textsubscript{J}}}
\newcommand{\rjup}{\mbox{R\textsubscript{J}}}
\newcommand{\msun}{\mbox{M$_{\odot}$}}
\newcommand{\rsun}{\mbox{R$_{\odot}$}}

\newcommand{\rearth}{R$_{\oplus}$}
\newcommand{\mearth}{M$_{\oplus}$}
\newcommand{\gcc}{g\,cm$^{-3}$}

\newcommand{\teff}{$T_\text{eff}$}
\newcommand{\teq}{$T_\text{eq}$}

\newcommand{\logg}{$\log g$}

\newcommand{\gc}{g~cm$^{-3}$}

\newcommand{\vsini}{$v \sin i_\star$}   
 
\newcommand{\vmic}{$v_{\rm mic}$}
\newcommand{\vmac}{$v_{\rm mac}$}

\newcommand{\feh}{[Fe/H]}

\newcommand{\logrhk}{$\log\,\mathrm{R}^\prime_\mathrm{HK}$}

\newcommand{\halpha}{H$\alpha$}

\newcommand{\feii}{Fe\,{\sc II}}
\newcommand{\fei}{Fe\,{\sc I}}

\newcommand{\name}{\ngts-38}
\newcommand{\TICID}{65910228}
\newcommand{\GaiaID}{5606317297918628992}
\newcommand{\twoMASSID}{J07145152-2925498}
\newcommand{\TYCID}{6536-3043-1}

\newcommand{\SkyMapperID}{35585731}

\newcommand{\rahms}{$07:14:51.52$}
\newcommand{\dechms}{$-29:25:49.91$}

\newcommand{\pmra}{$13.5664\pm 0.0108$}
\newcommand{\pmdec}{$-21.0834\pm0.0131$}
\newcommand{\parallax}{$3.75592\pm0.01280$}

\newcommand{\Vmag}{$10.230\pm0.020$}
\newcommand{\Bmag}{$10.738\pm0.026$}
\newcommand{\umag}{$11.703\pm0.011$}
\newcommand{\vmag}{$11.300\pm0.011$}
\newcommand{\gmag}{$10.334\pm0.009$}
\newcommand{\rmag}{$10.116\pm0.010$}
\newcommand{\imag}{$10.047\pm0.017$}
\newcommand{\zmag}{$10.051\pm0.012$}
\newcommand{\GaiaGmag}{$10.1103\pm0.0002$}
\newcommand{\GaiaBPmag}{$10.3797\pm0.0003$}
\newcommand{\GaiaRPmag}{$9.6821\pm0.0002$}
\newcommand{\TESSmag}{$9.734\pm0.006$}
\newcommand{\Jmag}{$9.251\pm0.026$}
\newcommand{\Hmag}{$9.010\pm 0.026$}
\newcommand{\Kmag}{$8.938\pm0.023$}
\newcommand{\WOnemag}{$8.909\pm0.023$}
\newcommand{\WTwomag}{$8.947\pm0.021$}
\newcommand{\WThreemag}{$8.912\pm0.027$}

\newcommand{\hostteff}{$6310\pm129$}
\newcommand{\hostlogg}{$4.06\pm0.02$}
\newcommand{\speclogg}{$3.8\pm0.1$}
\newcommand{\hostfeh}{$0.33\pm0.09$}
\newcommand{\hostvmic}{$1.4\pm0.2$}
\newcommand{\hostvmac}{$7.6\pm0.8$}
\newcommand{\hostvsini}{$12\pm3$}

\newcommand{\meanlogrhk}{$-5.06\pm0.07$}

\newcommand{\hostrad}{$1.88\pm0.08$}
\newcommand{\hostmass}{$1.46\pm0.09$}
\newcommand{\hostage}{$2.2\pm0.5$}
\newcommand{\stardist}{$265\pm1$}

\newcommand{\VgalU}{$18.349\pm0.096$}
\newcommand{\VgalV}{$-32.918\pm0.054$}
\newcommand{\VgalW}{$0.526\pm0.019$}
\newcommand{\Vgaltot}{$37.62\pm0.07$}
\newcommand{\eccgal}{0.16}
\newcommand{\Jzgal}{1608.6}

\newcommand{\brr}{$0.05910\pm0.00059$} 
\newcommand{\brsuma}{$0.01330\pm0.00036$} 
\newcommand{\bcosi}{$0.00823\pm0.00032$} 
\newcommand{\bepoch}{$2459209.2282\pm0.0028$} 
\newcommand{\bperiod}{$180.52797\pm0.00036$} 
\newcommand{\bfc}{$0.0788\pm0.0054$} 
\newcommand{\bfs}{$-0.5498\pm0.0095$} 
\newcommand{\bK}{$0.1373\pm0.0017$} 
\newcommand{\hostldcqoneTESS}{$0.2781\pm0.0019$} 
\newcommand{\hostldcqtwoTESS}{$0.388\pm0.022$} 
\newcommand{\hostldcqoneNGTS}{$0.3368_{-0.0029}^{+0.0028}$} 
\newcommand{\hostldcqtwoNGTS}{$0.384\pm0.024$} 
\newcommand{\hostldcqoneLCO}{$0.2829\pm0.0028$} 
\newcommand{\hostldcqtwoLCO}{$0.369\pm0.024$} 
\newcommand{\hostldcqoneElSauce}{$0.3757_{-0.0087}^{+0.0083}$} 
\newcommand{\hostldcqtwoElSauce}{$0.385\pm0.070$} 
\newcommand{\hostldcqoneVignes}{$0.3370\pm0.0084$} 
\newcommand{\hostldcqtwoVignes}{$0.383\pm0.072$} 
\newcommand{\hostldcqoneMallia}{$0.2684\pm0.0074$} 
\newcommand{\hostldcqtwoMallia}{$0.379\pm0.076$} 
\newcommand{\lnerrfluxTESS}{$-7.750\pm0.011$} 
\newcommand{\lnerrfluxNGTS}{$-4.899\pm0.011$} 
\newcommand{\lnerrfluxLCOCTIOL}{$-6.251\pm0.020$} 
\newcommand{\lnerrfluxLCOCTIOS}{$-5.752\pm0.026$} 
\newcommand{\lnerrfluxLCOSSOL}{$-6.664\pm0.023$} 
\newcommand{\lnerrfluxLCOSSOS}{$-5.966\pm0.019$} 
\newcommand{\lnerrfluxElSauce}{$-5.135_{-0.025}^{+0.027}$} 
\newcommand{\lnerrfluxVignes}{$-6.250\pm0.064$} 
\newcommand{\lnerrfluxMallia}{$-6.155\pm0.050$} 
\newcommand{\lnjitterrvCORALIEonefour}{$-6.06_{-1.0}^{+0.95}$} 
\newcommand{\lnjitterrvCORALIEtwofour}{$-6.47_{-0.75}^{+0.86}$} 
\newcommand{\lnjitterrvHARPS}{$-5.46_{-0.42}^{+0.31}$} 
\newcommand{\baselineoffsetrvCORALIEonefour}{$20.0646\pm0.0042$} 
\newcommand{\baselineoffsetrvCORALIEtwofour}{$20.0725\pm0.0027$} 
\newcommand{\baselineoffsetrvHARPS}{$20.0537\pm0.0013$} 
\newcommand{\baselineoffsetfluxNGTS}{$-0.00034\pm0.00011$} 
\newcommand{\baselineoffsetfluxLCOCTIOL}{$-0.000317\pm0.000066$} 
\newcommand{\baselineoffsetfluxLCOCTIOS}{$-0.00032\pm0.00012$} 
\newcommand{\baselineoffsetfluxLCOSSOL}{$0.000017_{-0.000045}^{+0.000043}$} 
\newcommand{\baselineoffsetfluxLCOSSOS}{$-0.000050\pm0.000074$} 
\newcommand{\baselineoffsetfluxElSauce}{$-0.00056\pm0.00020$} 
\newcommand{\baselineoffsetfluxVignes}{$-0.00065_{-0.00016}^{+0.00017}$} 
\newcommand{\baselineoffsetfluxMallia}{$-0.00064\pm0.00015$} 

\newcommand{\bRstarovera}{$0.01255\pm0.00034$} 
\newcommand{\baoverRstar}{$79.7_{-2.0}^{+2.2}$} 
\newcommand{\bRcompanionovera}{$0.000742\pm0.000024$} 
\newcommand{\bRcompanionRearth}{$12.12\pm0.53$} 
\newcommand{\bRcompanionRjup}{$1.081\pm0.047$} 
\newcommand{\baRsun}{$149.8\pm7.5$} 
\newcommand{\baAU}{$0.697\pm0.035$} 
\newcommand{\bi}{$89.529\pm0.018$} 
\newcommand{\be}{$0.3086\pm0.010$} 
\newcommand{\bw}{$278.16\pm0.64$} 
\newcommand{\bq}{$0.00312_{-0.00015}^{+0.00017}$} 
\newcommand{\bMcompanionMearth}{$1520_{-120}^{+130}$} 
\newcommand{\bMcompanionMjup}{$4.77_{-0.37}^{+0.39}$} 
\newcommand{\bMcompanionMsun}{$0.00456_{-0.00036}^{+0.00038}$} 
\newcommand{\bbtra}{$0.8536_{-0.0095}^{+0.0088}$} 
\newcommand{\bTtratot}{$14.86\pm0.13$} 
\newcommand{\bTtrafull}{$9.38_{-0.28}^{+0.26}$} 
\newcommand{\bdensity}{$4.67_{-0.77}^{+0.96}$} 
\newcommand{\bsurfacegravity}{$9550_{-600}^{+680}$} 
\newcommand{\bTeq}{$457\pm11$} 
\newcommand{\bdepthtrundilTESS}{$3.204_{-0.048}^{+0.053}$} 
\newcommand{\bdepthtrundilNGTS}{$3.172_{-0.049}^{+0.054}$} 
\newcommand{\bdepthtrundilLCOSSOL}{$3.210_{-0.052}^{+0.048}$} 

\newcommand{\bdepthtrundilLCOSSOS}{$3.212_{-0.051}^{+0.056}$} 

\newcommand{\bdepthtrundilLCOCTIOL}{$3.210\pm0.052$} 
\newcommand{\bdepthtrundilLCOCTIOS}{$3.211_{-0.051}^{+0.054}$}

\newcommand{\bdepthtrundilElSauce}{$3.153_{-0.066}^{+0.061}$} 
\newcommand{\bdepthtrundilVignes}{$3.176\pm0.062$} 
\newcommand{\bdepthtrundilMallia}{$3.215_{-0.057}^{+0.060}$} 
\newcommand{\hostldcuoneTESS}{$0.409\pm0.024$} 
\newcommand{\hostldcutwoTESS}{$0.118\pm0.024$} 
\newcommand{\hostldcuoneNGTS}{$0.446\pm0.028$} 
\newcommand{\hostldcutwoNGTS}{$0.134\pm0.028$} 
\newcommand{\hostldcuoneLCO}{$0.393\pm0.027$} 
\newcommand{\hostldcutwoLCO}{$0.139\pm0.026$} 
\newcommand{\hostldcuoneElSauce}{$0.472\pm0.087$} 
\newcommand{\hostldcutwoElSauce}{$0.141\pm0.086$} 
\newcommand{\hostldcuoneVignes}{$0.445\pm0.084$} 
\newcommand{\hostldcutwoVignes}{$0.135\pm0.084$} 
\newcommand{\hostldcuoneMallia}{$0.392\pm0.079$} 
\newcommand{\hostldcutwoMallia}{$0.125\pm0.080$} 
\newcommand{\combinedhostdensity}{$0.293_{-0.022}^{+0.025}$} 

\newcommand{\initialhostdensity}{$0.302_{-0.030}^{+0.035}$}

\newcommand{\periastron}{$0.482\pm0.025$}
\newcommand{\apastron}{$0.91\pm0.05$}
\newcommand{\bTeqAp}{$400\pm10$}
\newcommand{\bTeqPeri}{$550\pm14$}

\newcommand{\bbocc}{$0.59\pm0.03$}

\newcommand{\blogg}{$3.98\pm0.03$}

\newcommand{\rocheRJ}{$3.50\pm0.28$}

\newcommand{\hillRJ}{$147\pm9$}

\newcommand{\RosarioRJ}{$44.2\pm2.8$}
\newcommand{\DomingosRJ}{$49.2\pm3.2$}

\newcommand{\Arm}{$14\pm3$}

\title[TIC-\TICID\,b / \name\,b]{TIC-\TICID\,b\,/\,\name\,b, a 180\,day transiting warm super-Jupiter}

\author[T. Rodel et al.]{\parbox{\textwidth}{\Large
Toby Rodel$^{1}$\thanks{Email: trodel01@qub.ac.uk}\orcidlink{0009-0009-2175-7284},
Sol\`ene Ulmer-Moll$^{2,3}$\orcidlink{0000-0003-2417-7006},
Samuel Gill$^{4,5}$\orcidlink{0000-0002-4259-0155},
Christopher. A. Watson$^{1}$\orcidlink{0000-0002-9718-3266},
Yoshi Nike Emilia Eschen$^{4,5,6}$\orcidlink{0009-0006-6397-2503},
Alix V. Freckelton$^{7}$\orcidlink{0009-0007-1053-0004},
Annelies Mortier$^{7}$\orcidlink{0000-0001-7254-4363},
Karen A. Collins$^{8}$\orcidlink{0000-0001-6588-9574},
Diana Dragomir$^{9}$\orcidlink{0000-0003-2313-467X},
Zahra Essack$^{9}$\orcidlink{0000-0002-2482-0180},
Brett Skinner$^{9}$\orcidlink{0009-0004-7366-4841},
Niamh Mallaghan$^{1}$\orcidlink{0009-0004-4749-8173},
Peter J. Wheatley$^{4,5}$\orcidlink{0000-0003-1452-2240},
David R. Anderson$^{10}$\orcidlink{0000-0001-7416-7522},
Ioannis Apergis$^{4,5}$\orcidlink{0009-0004-7473-4573},
Khalid Barkaoui$^{11,12,13}$\orcidlink{0000-0003-1464-9276},
Matthew P. Battley$^{14}$\orcidlink{0000-0002-1357-9774},
Daniel Bayliss$^{4,5}$\orcidlink{0000-0001-6023-1335},
Fran\c{c}ois Bouchy$^{3}$\orcidlink{0000-0002-7613-393X},
Edward M. Bryant$^{4,5}$\orcidlink{0000-0001-7904-4441},
Matthew R. Burleigh$^{15}$\orcidlink{0000-0003-0684-7803},
Benjamin M. J. Cadell$^{1}$\orcidlink{0009-0006-5883-3138},
Samuel J. Carlier$^{15}$\orcidlink{0009-0000-9901-5242},
Yann Carteret$^{3}$\orcidlink{0000-0002-6159-6528},
Sarah L. Casewell$^{15}$\orcidlink{0000-0003-2478-0120},
Alastair B. Claringbold$^{4,5}$\orcidlink{0000-0003-1309-5558},
Jean C. Costes$^{1}$,
Benjamin D. R. Davies$^{4,5}$\orcidlink{0009-0000-5659-9006},
Lauren Doyle$^{4,5}$\orcidlink{0000-0002-9365-2555},
Phil Evans$^{16}$\orcidlink{0000-0002-5674-2404},
Jorge Fern\'andez Fern\'andez$^{4,5}$\orcidlink{0000-0002-1416-2188},
Emile Fontanet$^{3}$\orcidlink{0000-0002-0215-4551},
Edward Gillen$^{14}$\orcidlink{0000-0003-2851-3070},
Michael R. Goad $^{15}$\orcidlink{0000-0002-2908-7360},
George Harvey$^{15}$\orcidlink{0009-0009-1823-7501},
Faith Hawthorn$^{17,4}$\orcidlink{0000-0002-8675-182X},
Katlyn L. Hobbs$^{1}$\orcidlink{0009-0004-4519-5080},
Melissa Hobson$^{3}$\orcidlink{0000-0002-5945-7975},
Giovanni Isopi $^{18,19,20,21}$\orcidlink{0000-0002-8458-0588},
James S. Jenkins$^{22,23}$\orcidlink{0000-0003-2733-8725},
Alicia Kendall$^{15}$\orcidlink{0009-0006-0719-9229},
David Kipping$^{24}$\orcidlink{0000-0002-4365-7366},
Monika Lendl$^{3}$\orcidlink{0000-0001-9699-1459},
Franco Mallia$^{19}$,
Christopher Mann$^{25}$\orcidlink{0000-0002-9312-0073},
James McCormac$^{4,5}$\orcidlink{0000-0003-1631-4170},
Ernst J.W. de Mooij$^{1}$\orcidlink{0000-0001-6391-9266},
Maximiliano Moyano$^{10}$\orcidlink{0000-0002-7927-9555},
Arianna Nigioni$^{3}$\orcidlink{0009-0004-5882-6574},
Mohammad Odeh$^{11}$\orcidlink{0000-0002-8986-6681},
Vera Maria Passegger$^{26,11,27,28}$\orcidlink{0000-0002-8569-7243},
Suman Saha$^{22,23}$\orcidlink{0000-0001-8018-0264},
Richard P. Schwarz$^{8}$\orcidlink{0000-0001-8227-1020},
Amber Sedgley$^{4,5}$\orcidlink{0009-0001-7824-1715},
Avi Shporer$^{29}$\orcidlink{0000-0002-1836-3120},
Abderahmane Soubkiou$^{12}$\orcidlink{0000-0002-0345-2147},
St\'{e}phane Udry$^{3}$\orcidlink{0000-0001-7576-6236},
Dimitri Veras$^{4,5,30}$\orcidlink{0000-0001-8014-6162},
Jean. P. Vignes$^{31}$,
Steven Villanueva Jr.$^{32}$\orcidlink{0000-0001-6213-8804},
Jos\'{e} I. Vin\'{e}s$^{10}$\orcidlink{0000-0002-2135-9018},
Richard West$^{4,5}$\orcidlink{0000-0001-6604-5533},
Thomas G. Wilson$^{4,5}$\orcidlink{0000-0001-8749-1962},
Hannah L. Worters$^{33}$,
Mitchell E. Young$^{1}$\orcidlink{0000-0003-0672-7123},
Aldo Zapparata $^{18,19}$\orcidlink{0000-0002-9428-1573}
}
\vspace{0.2cm}
\\
Authors' institutions are listed at the end of the paper.
}

\date{Accepted XXX. Received YYY; in original form ZZZ}

\pubyear{\the\year{}}

\begin{document}
\label{firstpage}
\pagerange{\pageref{firstpage}--\pageref{lastpage}}
\maketitle

\begin{abstract}
We present the discovery of TIC-\TICID\,b\,/\,\name\,b, a giant exoplanet with a radius of \bRcompanionRjup\,\rjup\ and a mass of \bMcompanionMjup\,\mjup\ on a long-period (\bperiod\,day), moderately eccentric ({\it e}=\be) orbit transiting a bright (V=\Vmag\ mag) metal rich (\feh=\hostfeh\,`dex') F6V-F7V type host star. The planet was initially detected from a single transit in \tess\ Sector 33. A photometric monitoring campaign of 228 nights with \ngts\ detected a transit egress of the planet, which together with spectroscopic radial velocity monitoring with \coralie\ and \harps\ identified an orbital period of 180.5\,d. These radial velocity measurements also showed the mass of the companion to be planetary. Additional transit observations coordinated by the \tess\ follow-up observing program allowed further confirmation and refinement of this period. With its relatively cool equilibrium temperature of \bTeq\,K, \name\,b joins a small but growing population of well characterised transiting warm-Jupiters and has one of the longest periods of any discovered to date. The target is situated in the LOPS2 field of the upcoming \plato\ mission which will allow for greater refinement of the system parameters and potential for the discovery of additional companions too small and/or too long-period to be seen by \tess\ or \ngts. \name\,b's bright host star and wide orbital separation make it an attractive target for further study, including potential measurement of its spin-orbit alignment or targeted exomoon/ring searches.
\end{abstract}

\begin{keywords}
planets and satellites: detection -- techniques: photometric -- techniques: radial velocities -- planets and satellites: individual: TIC-\TICID\,b
\end{keywords}



\section{Introduction}
\label{section:intro}

Despite the large number of confirmed transiting planets \citep[4673 listed on the NASA Exoplanet Archive;][]{Akeson2013, Christiansen2025}\footnote{Accessed 2026 April 28. Available at: \url{https://exoplanetarchive.ipac.caltech.edu}}, periods on the order of tens to hundreds of days remain observationally rare. A relative few of these $(\lessapprox25\%)$ have periods beyond 25 days and even fewer $(\lessapprox5\%)$ have periods $>100$\,days. Many of these $P\geq100$\,days period planets $(\sim90\%)$ were discovered by the \kepler\ mission \citep{Borucki2010} due to its long observational baseline. However, \kepler's relatively narrow field of view meant that many of these targets orbit faint, distant stars. This limits the extent to which they can be characterised in detail through spectroscopic follow-up observations.

Many of the most pressing questions in exoplanet science require a large sample of well characterised long-period ($\sim10-1000$\,days) transiting planets with fully measured masses, radii and orbital solutions. For instance, the existence of close in giant planets, often termed `hot-Jupiters' \citep[e.g. 51 Pegasi\,b;][]{Mayor1995}, challenged pre-existing notions of planet formation and evolution \citep[see review;][]{Dawson2018}. The existence of these planets can either be explained with in-situ formation \citep[e.g;][]{Bodenheimer2000} or ex-situ formation followed by inward migration with the latter usually being preferred. Two main migration channels are proposed: high eccentricity migration \citep{1962Lidov, 1962Kozai, Rasio1996} where a dynamical interaction perturbs the planet into an eccentric orbit that is circularised through tidal interactions over time; or disk migration \citep{Goldreich1980} where interactions with a protoplanetary disk cause a planet to lose angular momentum and migrate inwards.

These migration mechanisms should leave detectable dynamical traces in a planetary system, for instance, the orbital eccentricity and/or the spin-orbit alignment / obliquity between the planet and host star \citep{Naoz2016, Dawson2018, Albrecht2022}. However, for giant planets with orbital periods of 10\,days or shorter, tidal interactions with the star circularise orbits and can dampen obliquities \citep{2012Albrecht, Albrecht2022, Dawson2018} effectively removing any trace of the system's dynamical history. The exception to this being young systems with recently formed planets which may remain eccentric, even on close orbits. Longer-period giant planets (`warm-Jupiters') transiting nearby bright stars are therefore valuable tracers of planet formation and migration as spectroscopic radial velocity observations can constrain their undamped orbital eccentricities and obliquities via the Rossiter-McLaughlin effect (RM; \citet{Rossiter1924, McLaughlin1924}; also see Review by \citet{Triaud2018}). Already, RM surveys have found a greater trend towards alignment in giant planets with periods on the order of 10s to 100s of days compared to those with periods shorter than 10 days \citep{Rice2022, Wang2024, Espinoza-Retamal2025}. Although the numbers are currently small, this could suggest warm-Jupiters may be a distinct sub-population rather than representing a transitional stage towards becoming hot-Jupiters.

Longer-period planets also generally experience less irradiation from their host stars. This drastically reduces the rates of both atmospheric mass loss \citep{Sanz-Forcada2011} and inflation \citep[][]{Dawson2018}, meaning these planets are more likely to retain their primordial atmospheres and interiors. If the host star is bright, the planetary atmosphere can be probed via transmission spectroscopy \citep[e.g;][]{Charbonneau2002}. These observations can provide additional constraints on planet formation and migration mechanisms using tracers such as the C/O and N/O ratios \citep{Oberg2011,Ohno2023a,Ohno2023b}, as well as general atmospheric dynamics and compositions at cooler equilibrium temperatures \citep{Fortney2020}. 

Long period giant planets are also valuable targets in searches for circumplanetary objects such as exomoons and exorings \citep{Szabo2024}, both of which are expected to become more stable with increasing orbital separation \citep{Barnes2002, Barnes2004}. In addition, as we push to longer-periods, wider separations and cooler equilibrium temperatures, the possibility of either directly habitable exoplanets or habitable moons around giant planets becomes greater.

The Transiting Exoplanet Survey Satellite \citep[\tess;][]{ricker2015tess} is specifically designed for the discovery of transiting exoplanets around nearby bright stars. The survey strategy of tiling \tess' wide rectangular field across the sky in `sectors' has led to almost total sky coverage over the primary and extended missions. However, this strategy limits the observational baseline, creating difficulties in detecting long-period planets. Simulation work by \citet{Cooke2018, Cooke2019, Cooke2021, Villanueva2019} and most recently \cite{Rodel2024} has found that the majority of long-period planets detected via \tess\ will have unsolved periods, these being detected either from a single `monotransit' or `duo/triotransits' with two or three events separated by the $\sim2$\,yr gap between \tess\ visits \citep[e.g;][]{Hawthorn2024}. In the former case the period is almost entirely unconstrained, although the duration of transit can give some indication, whilst in the latter case the true period forms part of a sequence of aliases, each an integer fraction of the gap between transits.

This means extensive follow-up observational programs are required to confirm these planets and find their true orbital periods. The Next Generation Transit Survey \citep[\ngts;][]{wheatley2018ngts} is one of the best suited instruments for the photometric follow-up of unsolved \tess\ detections. It has already been used to confirm several long-period transiting giant planets \citep{Gill2020ngts11,Gill2024, Grieves2022, Battley2024, Ulmer-Moll2022, Ulmer-Moll2025, Kendall2025} as well as brown dwarf and late M dwarf companions \citep{Gill2020monofind, Gill2020eblm, Gill2022, Lendl2020, Henderson2024, Rodel2025}. The Characterising ExOPlanet Satellite \citep[\cheops;][]{Benz2021} has similarly been used from space to confirm smaller transiting long-period planets \citep[e.g;][]{Ulmer-Moll2023, Osborn2023, Tuson2023}. Spectroscopic radial velocity campaigns using instruments such as \coralie\ \citep{Queloz2001coralie}, the Fiberfed Extended Range Optical Spectrograph \citep[\feros;][]{Kaufer1999} or the High Accuracy Radial velocity Planet Searcher \citep[\harps;][]{Pepe2000, Mayor2003harps} are also essential in such follow-up efforts. These observations act either to confirm the mass of a planet with a photometrically determined period, or to provide an initial period that can be confirmed more precisely with additional photometry \citep[e.g.][]{Schlecker2020, Hobson2021, Eberhardt2023, Brahm2023, Bieryla2026}.

It is in this context that we present the discovery of TIC-\TICID\,b (which we also refer to as \name\,b): a transiting giant planet on a $180.5$\,day orbit around a bright F type star. This makes it one of the longest period transiting planets ever found and one of the most amenable of these to characterisation. In Section\,\ref{section:obs:phot} we describe the initial detection of the planet from a single \tess\ transit (Section\,\ref{section:obs:phot:tess}), the detection of an egress with \ngts\ (Section\,\ref{section:obs:phot:ngts}), simultaneous observation of the most recent transit from several facilities (Section\,\ref{section:obs:phot:sg1}), and  longer term photometric monitoring (Section\,\ref{section:obs:phot:rot}) used to search for rotational signals. In Section\,\ref{section:obs:spec} we describe the spectroscopic campaigns with \coralie\ (Section\,\ref{section:obs:spec:coralie}) and \harps\ (Section\,\ref{section:obs:spec:harps}) that allowed a mass measurement of the planet via radial velocity and spectroscopic characterisation of the host star, which we describe in Section\,\ref{section:host-analysis}. In Section\,\ref{section:orbit-fit} we describe the joint global model fit we performed on the previously described data in order to obtain a full orbital solution for \name\,b. Then, in Section\,\ref{section:discussion}, we present the parameters derived from our fit alongside some interior structure modelling and discuss their significance. We then discuss the potential for future work on the system in Section\,\ref{section:future} before summarising our work with some concluding remarks in Section\,\ref{section:conclusion}.

\section{Photometric Observations}
\label{section:obs:phot}
The photometric datasets for TIC-\TICID\,\,/\,\name\ are summarised in Table\,\ref{tab:phot_summary} and they are described in the remainder of this section.

\begin{table*}
    \centering
    \caption{Summary of photometric observations of \name.}
    \label{tab:phot_summary}
    \begin{tabular}{ccccccccc}
        \toprule
        Facility &  Night(s) Observed (UTC) &  $N_\text{images}$\textbf{*}&  Exptime (s) & $N_\text{nights}$\textbf{**} & Filter & Pipeline & $N_\text{telescopes}$ & Note(s)\\
        \hline
        \waspsouth & 2011 Nov 2 - 2012 Mar 29 & 2464 & 30 & 77 & \wasp & \wasp & 1 & \\
        \assasn & 2016 Feb 2 - 2018 Sep 17 & 795 & 90 & 276 & V & \assasn & 1 & \\
        \assasn & 2018 Aug 8 - 2025 Nov 23 & 4544 & 90 & 1304 & g & \assasn & 1 & \\
        \tess-S07 & 2019 Jan 8 - 2019 Feb 16 & 1086 & 1800 & 22.6 & \tess & \tess-SPOC & 1 & \\
        \tess-S33 & 2020 Dec 18 - 2021 Jan 13 & 3485 & 600 & 24.2 & \tess & \tess-SPOC & 1 & a \\
        \tess-S34 & 2021 Jan 14 - 2021 Feb 8 & 3358 & 600 & 23.2 & \tess & \tess-SPOC & 1 & \\
        \tess-S61 & 2023 Jan 18 - 2023 Feb 12 & 9503 & 200 & 23.0 & \tess & \tess-SPOC & 1 & \\
        \tess-S87 & 2024 Dec 18 - 2025 Jan 14 & 14576 & 120 & 20.2 &\tess & SPOC & 1 & \\
        \tess-S88 & 2025 Jan 14 - 2025 Feb 11 & 19432 & 120 & 27.0 & \tess & SPOC & 1 &  \\
        \ngts & 2023 Aug 25 - 2024 Feb 4 & 118315 & 10 & 115 &\ngts & \bsproc & 1 & \\
        \ngts & 2024 Apr 16 - 2024 May 12 & 3012 & 10 & 8 & \ngts & \bsproc & 1 &  \\
        \ngts & 2024 Aug 18 - 2024 Nov 21 & 44971 & 10 & 69 & \ngts & \bsproc & 1 & \\
        \ngts & 2024 Nov 22 - 2024 Dec 31 & 131761 & 10 & 36 & \ngts & \bsproc & 2 & b \\
        \ngts & 2025 Dec 2 - 2025 Dec 5 & 22903 & 10 & 4 & \ngts & \bsproc & 4 & \\
        LCO-SSO 1m & 2025 Dec 4 & 900 & 10 & 1 & i$^\prime$ & \aij & 2 & c \\
        LCO-SSO 0.35m & 2025 Dec 4 & 1106 & 27 & 1 & i$^\prime$ & \aij & 2 & c \\
        LCO-CTIO 1m & 2025 Dec 4 & 1068 & 10 & 1 & i$^\prime$ & \aij & 2  & d \\
        LCO-CTIO 0.35 & 2025 Dec 4 & 652 & 27 & 1 & i$^\prime$ & \aij & 1 & d\\
        El Sauce & 2025 Dec 4 & 751 & 32 & 1 & Rc & \aij & 1 & d \\
        DSC0.4m & 2025 Dec 4 & 120 & 135 & 1 & Chroma R & \aij & 1 & d \\
        OACC - CAO (0.6 m) & 2025 Dec 4 & 196 & 100 & 1 & i$^\prime$2 & \aij & 1 & d \\
        \bottomrule
    \end{tabular}
    \begin{tablenotes}
      \small
      \item \textbf{*} For \tess\ $N_\text{images}$ is counted as the number of individual points in the lightcurve with a data quality flag of 0.
      \item \textbf{**} For \tess\ this is calculated by (Exptime (s) $\times N_\text{images})/86400$\,s while for all ground based instruments this is simply the count of nights with observations.
      \item \textbf{a:} Full Transit on UTC 2020 December 25. \textbf{b:} Transit Egress on UTC 2024 December 8. \textbf{c:} Transit Ingress on UTC 2025 December 4. \textbf{d:} Transit Egress on UTC 2025 December 4
    \end{tablenotes}
\end{table*}

\begin{figure}
    \centering
    \includegraphics[width=\columnwidth]{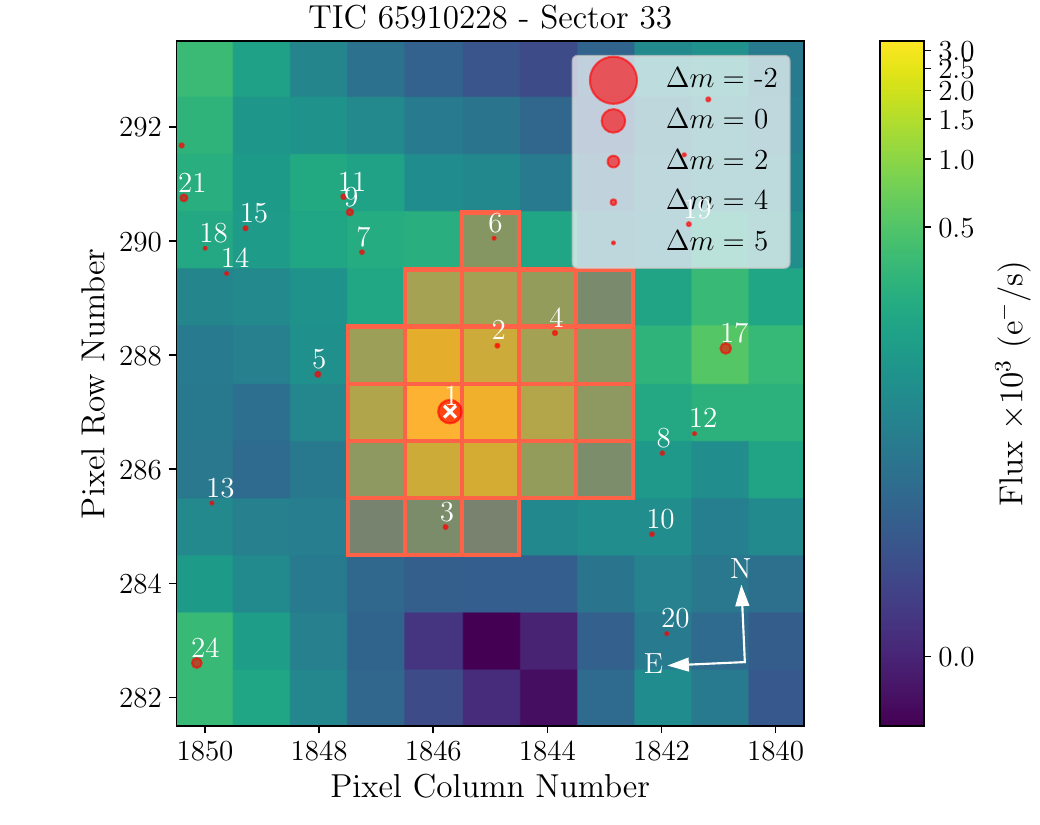}
    \caption{\tess\ target pixel file for \name\ in sector 33. Each red circle represents a \gaia\ source down to a magnitude difference of 6 from the target. The circles are also sized according to the difference in magnitude from the target, which is marked with a white x and numbered 1.}
    \label{fig:tesstpf}
\end{figure}

\begin{figure*}
    \centering
    \begin{subfigure}{\columnwidth}
        \includegraphics[width=\textwidth]{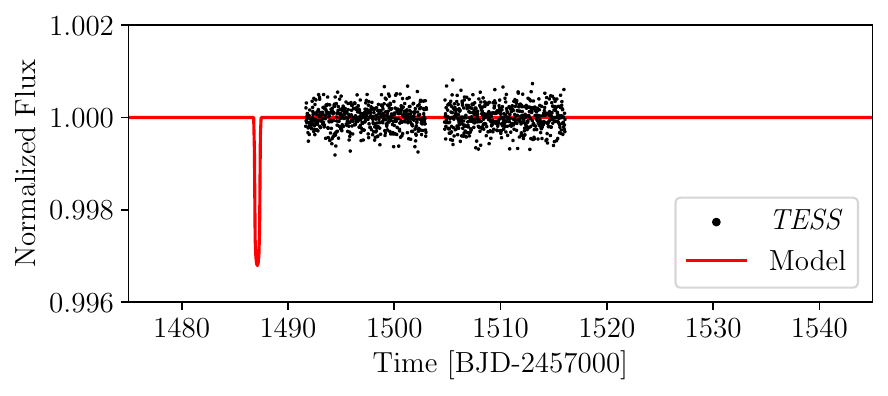}
        \caption{\tess\ Cycle 1 (Sector 7)}
        \label{fig:phot_plot:TESS_Cycle1}
    \end{subfigure}
    \begin{subfigure}{\columnwidth}
        \includegraphics[width=\textwidth]{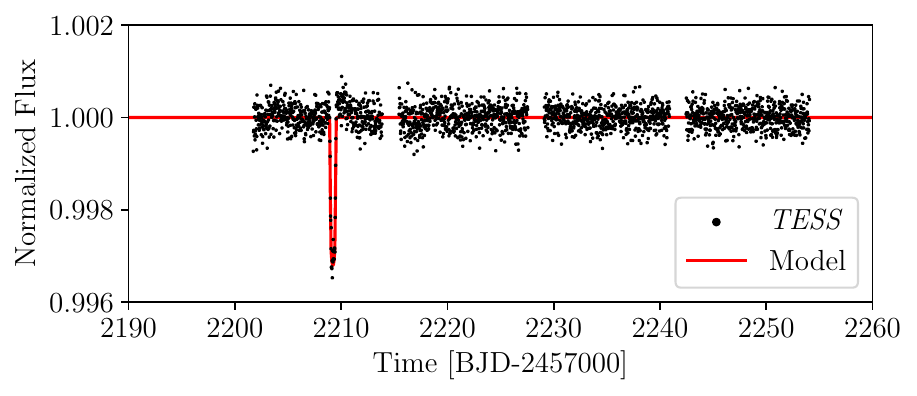}
        \caption{\tess\ Cycle 3 (Sectors 33 and 34)}
        \label{fig:phot_plot:TESS_Cycle3}
    \end{subfigure}
    \begin{subfigure}{\columnwidth}
        \includegraphics[width=\textwidth]{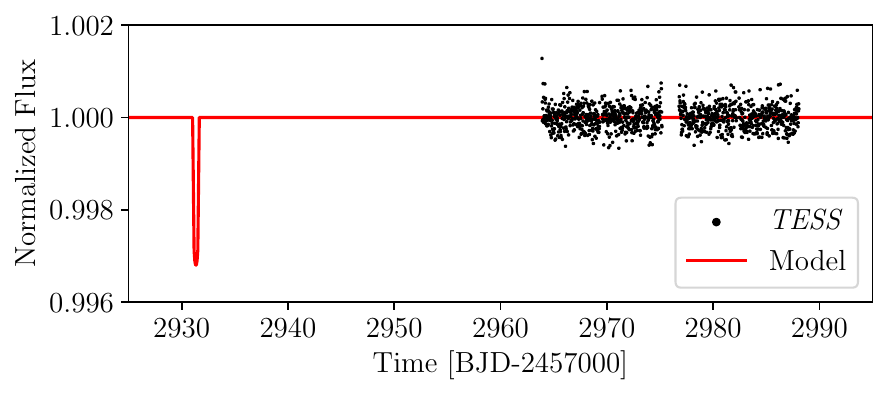}
        \caption{\tess\ Cycle 5 (Sector 61)}
        \label{fig:phot_plot:TESS_Cycle5}
    \end{subfigure}
    \begin{subfigure}{\columnwidth}
        \includegraphics[width=\textwidth]{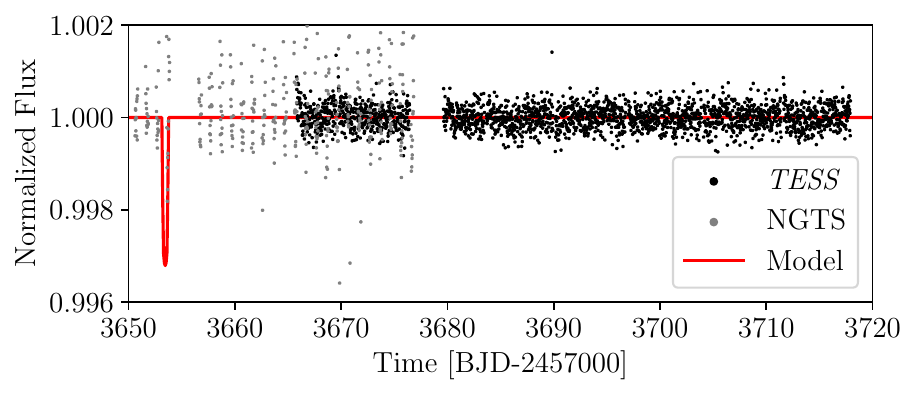}
        \caption{\tess\ Cycle 7 (Sectors 87 and 88)}
        \label{fig:phot_plot:TESS_Cycle7}
    \end{subfigure}
    \begin{subfigure}{\textwidth}
        \includegraphics[width=\textwidth]{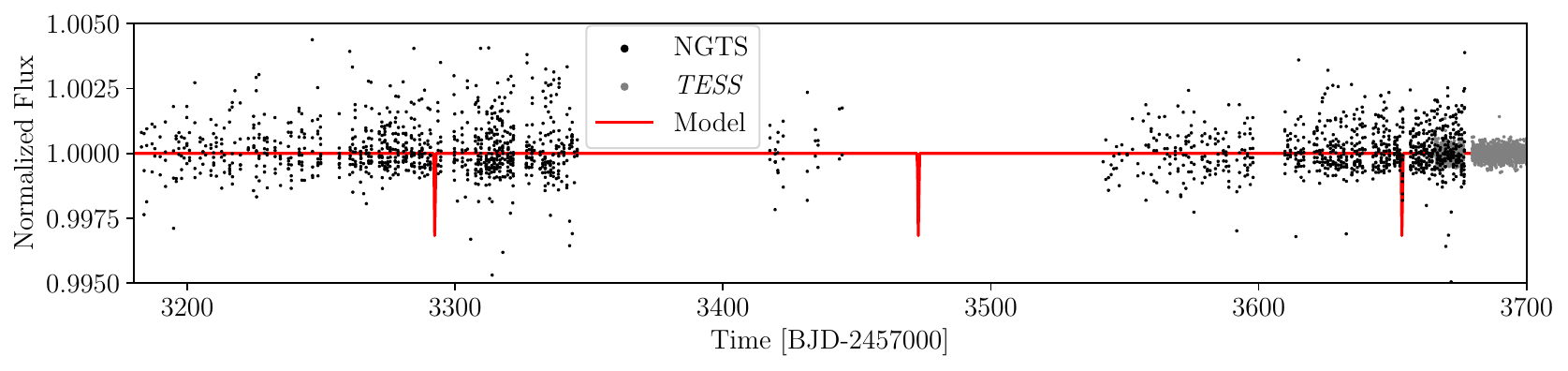}
        \caption{\ngts}
        \label{fig:phot_plot:NGTS}
    \end{subfigure}
    \caption{Discovery photometry for \name\,b. All datasets have been binned to 30 minute cadence for ease of comparison. The top four each panels show 70\,d spans of \tess\ photometry in black with \ngts\ included in gray where the two datasets overlap in the bottom right panel \textbf{(d)}. The bottom panel \textbf{(e)} shows the \ngts\ data in black while \tess\ Sector 87 is shown in gray. In all panels, the transit model is shown as a red line. The transit predicted at BJD-2457000\,$\approx3292$ does fall within the time limits of the first run of \ngts\ data but falls within a data gap.}
    \label{fig:phot_plot}
\end{figure*}

\begin{figure*}
    \centering
    \begin{subfigure}{\columnwidth}
        \includegraphics[width=\textwidth]{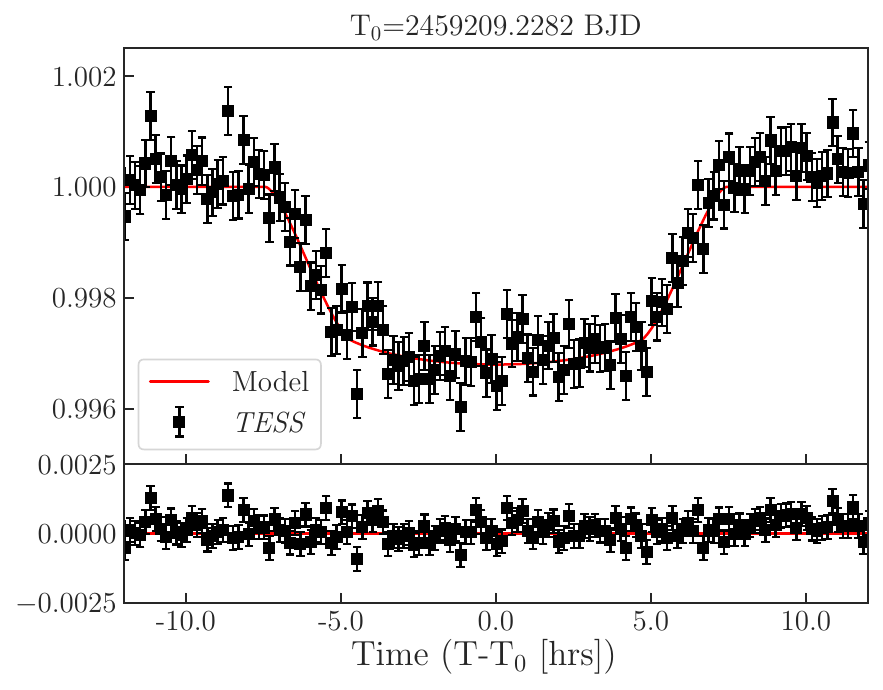}
        \caption{\tess\ Sector 33 transit.}
        \label{fig:transits_TESS}
    \end{subfigure}
    \begin{subfigure}{\columnwidth}
        \includegraphics[width=\textwidth]{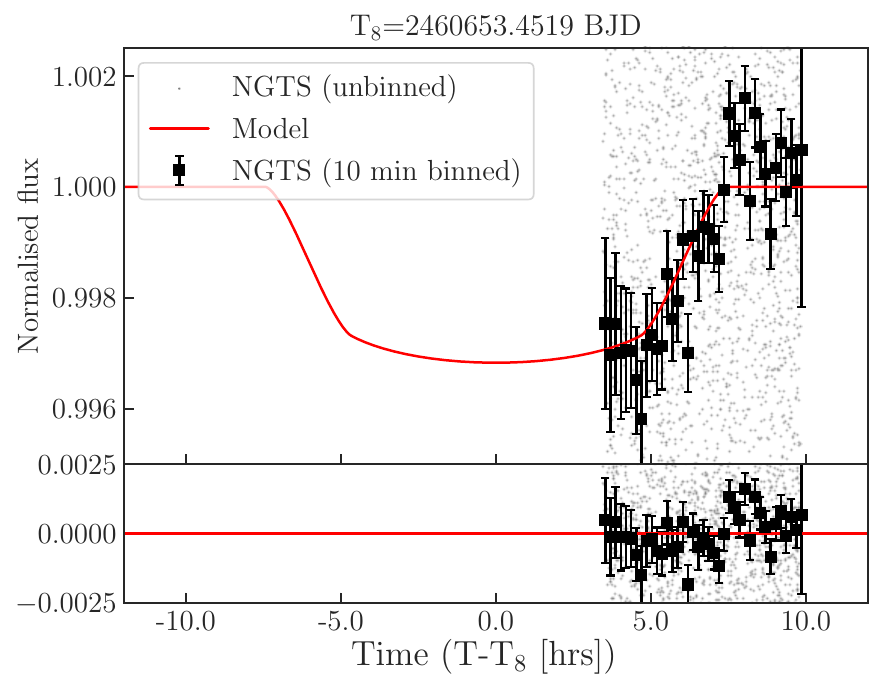}
        \caption{\ngts\ transit.}
        \label{fig:transits_NGTS}
    \end{subfigure}
    \caption{Discovery transit lightcurves of \name\,b. Each panel shows the median transit model (see Section\,\ref{section:orbit-fit}) as a solid red line. The transit data plotted is shown as black square markers with errorbars. For \ngts\ the data are binned to 10-minute cadence to match \tess\ and the unbinned data are shown as gray dots. The top panel of each subfigure shows the data and model while the lower panel of each shows the residuals after the model has been subtracted from the observed data. The TESS data shows a systematic `bump' in flux post transit, this is probably an artefact of the \texttt{PDCSAP\_FLUX} flattening of the lightcurve. Our own spline fit (see Section\,\ref{section:orbit-fit}) failed to remove this but the impact on any fitted parameters is likely to be negligible.}
    \label{fig:transits}
\end{figure*}

\begin{figure*}
    \centering
    \begin{subfigure}{\columnwidth}
        \includegraphics[width=\textwidth]{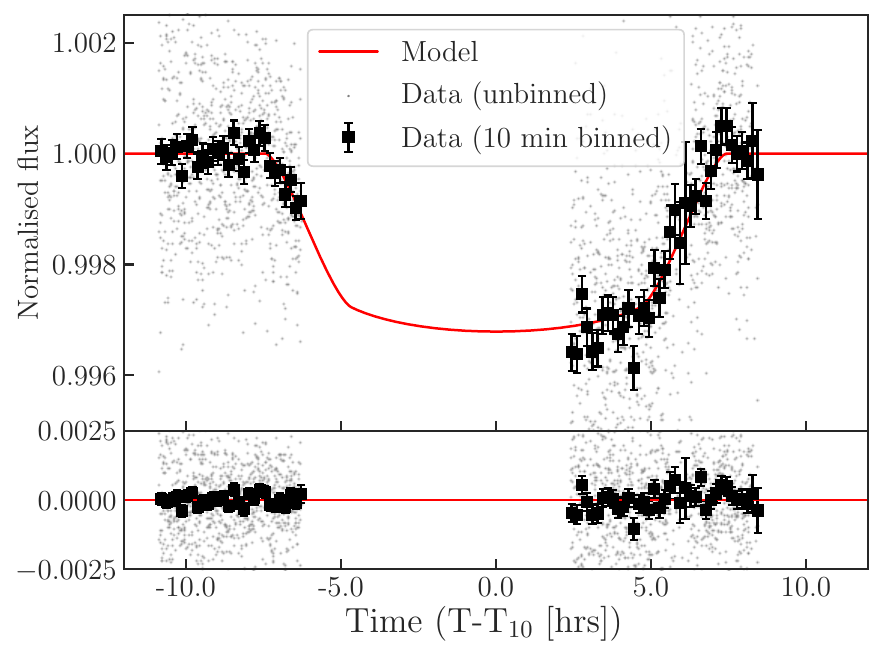}
        \caption{\lco\ 1m}
        \label{fig:transits_LCO-1m}
    \end{subfigure}
    \begin{subfigure}{\columnwidth}
        \includegraphics[width=\textwidth]{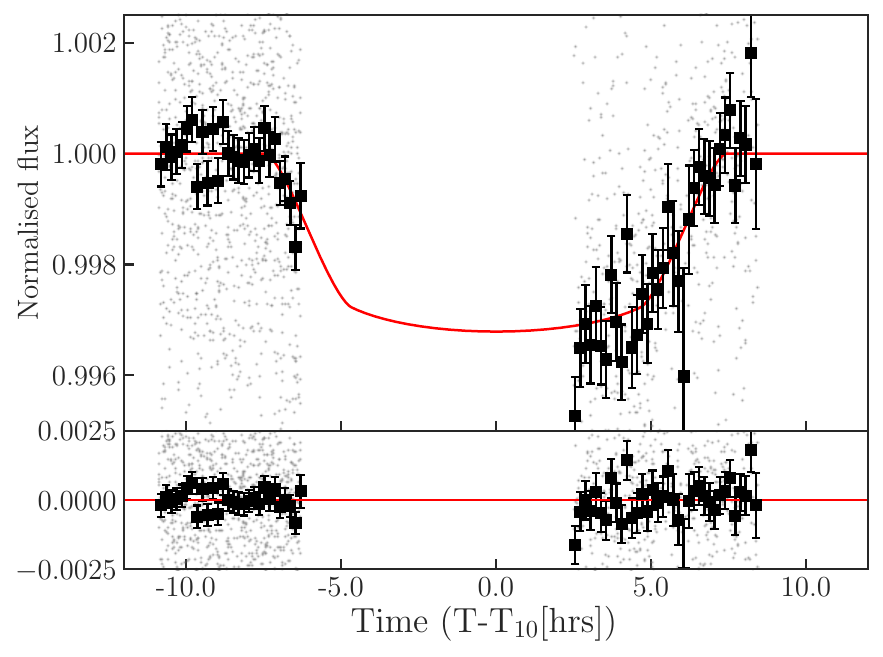}
        \caption{\lco\ 0.35m}
        \label{fig:transits_LCO-35cm}
    \end{subfigure}
    \begin{subfigure}{\columnwidth}
        \includegraphics[width=\textwidth]{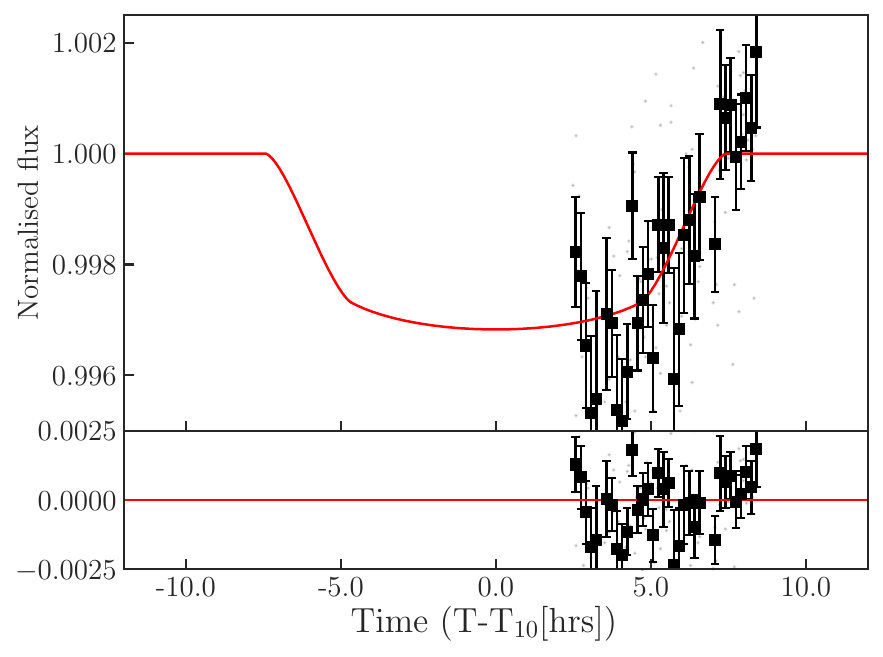}
        \caption{DSC0.4m}
        \label{fig:transits_vignes}
    \end{subfigure}
    \begin{subfigure}{\columnwidth}
        \includegraphics[width=\textwidth]{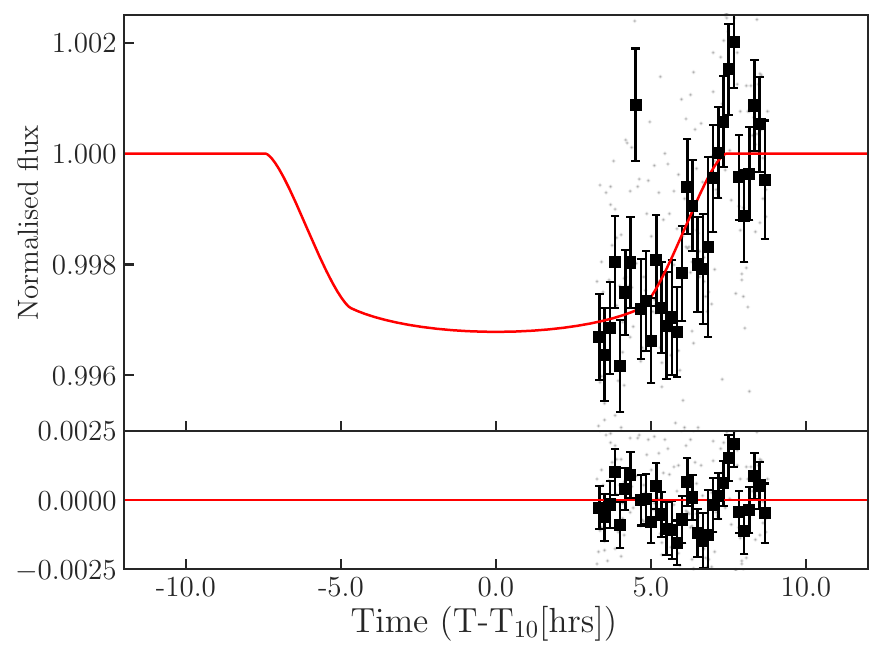}
        \caption{OACC - CAO (0.6m)}
        \label{fig:transits_mallia}
    \end{subfigure}
    \begin{subfigure}{\columnwidth}
        \includegraphics[width=\textwidth]{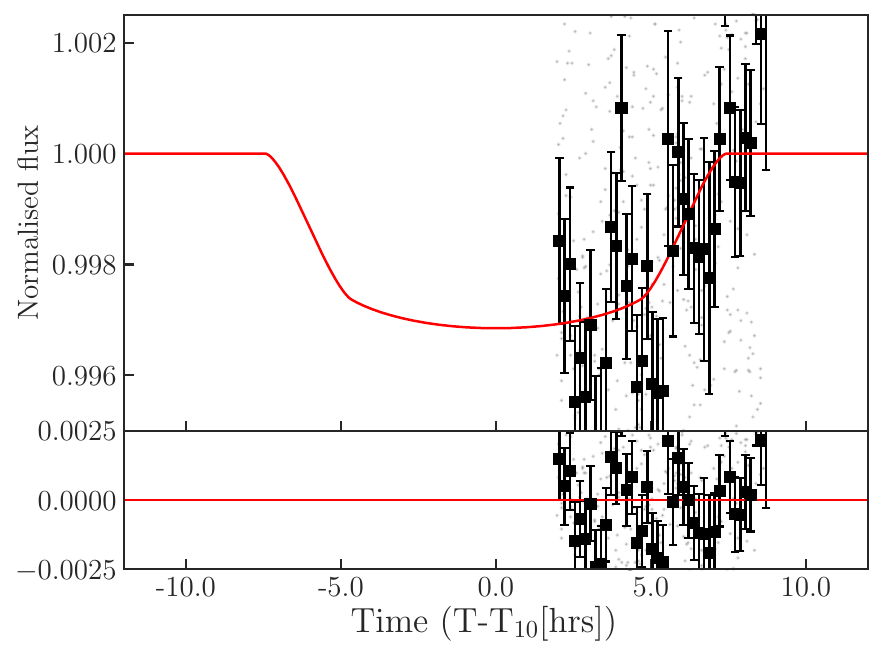}
        \caption{\elsauce}
        \label{fig:transits_elsauce}
    \end{subfigure}
    \caption{Simultaneous transit lightcurves of \name\,b from the night beginning 2025 December 4 UTC (T$_{10}$=2461014.5077\,BJD) Each panel shows the median transit model (see Section\,\ref{section:orbit-fit}) as a solid red line. The top panel of each subfigure shows the data and model while the lower panel of each shows the residuals after the model has been subtracted from the observed data. Unbinned data is shown as gray dots while the black squares with error bars represent the data binned to 10 minute cadence. The \lco\ data for the same diameter telescopes from CTIO and SSO have been stitched together to save space but these data were fitted separately.}
    \label{fig:transits1}
\end{figure*}

\subsection{\tess}
\label{section:obs:phot:tess}

\tess\ has been operating a (near) all-sky survey since 2018 aimed at discovering exoplanets transiting bright host stars. The \tess\ spacecraft carries four $\sim 10$\,cm cameras, each with a custom red-optical filter covering approximately 600\,nm - 1000\,nm and containing four CCDs with an image scale of $21\arcsec$ per pixel. The four cameras are arranged vertically to provide a rectangular $24\degree \times 96\degree$ field of view. \tess\ rotates this field of view around the sky in `sectors', which are each observed for $\sim27$\,days at a time, although overlap between sectors can provide some targets with longer baselines.

\name\ was observed by \tess\ in sectors 7, 33, 34, 61, 87 and 88. We show the sector 33 Target Pixel File created using \textsc{tpfplotter} \citep{Aller2020} in Figure\,\ref{fig:tesstpf}. In the most recent two sectors, lightcurves were produced onboard the spacecraft at two minute cadence by the Science Processing Operations Centre \citep[SPOC;][]{jenkins2016spoc}. The lightcurves for the remaining sectors were produced from the Full Frame Images (FFIs) using the \tess-SPOC pipeline \citep{Caldwell2020} at cadences of 30-minutes for sector 7, 10-minutes for sectors 33 and 34 and 200-seconds for sector 61. We performed quality cuts on all of these lightcurves, only using data points with a quality flag of 0. We used the \texttt{PDCSAP\_FLUX} column for all sectors.
These lightcurves are plotted in Figure\,\ref{fig:phot_plot}.

A single transit of \name\,b was detected at BJD=2459209.23\footnote{This and all other instances of BJD quoted in this paper (including BTJD) use BJD\textsubscript{TDB}.} (see Figures\,\ref{fig:phot_plot} and \ref{fig:transits_TESS}) in the Sector 33 \tess-SPOC FFI lightcurve by the Single Transit Search (STS) pipeline, a custom algorithm from the TESS Single Transit Planet Candidate Working Group (TSTPC WG). The TSTPC WG uses STS to systematically search \textit{TESS}-SPOC lightcurves for single transit events, aiming to increase the yield of TESS long-period planets. The STS algorithm, employing \textsc{astropy} Box-Least Squares method \citep{Kovacs2002, AstropyCollab2022}, identified the event by searching for periods near or beyond the TESS sector length across a thousand trial durations from 1 hour to 1 day. The transit signal was then compared against a sinusoidal model to exclude the possibility of stellar variability. The resulting data validation report was twice independently vetted, followed by a group evaluation session. The field is crowded (seen in Figure \ref{fig:tesstpf}), with numerous faint sources within the photometric aperture. However, pixel-level lightcurves generated with \textsc{eleanor} \citep{Feinstein2019} confirmed that the transit source is on-target on \name. The contaminating sources are too faint ($\Delta \text{m}_{\tess} > 4$) to produce the observed $\sim3$ parts per thousand (ppt) transit depth, and no significant centroid motion was detected during the event, ruling them out as the event's origin. The validated planet candidate parameters were uploaded to ExoFOP-TESS as the final step of TSTPC's processes. The event in Sector 33 was also detected by the \textsc{monofind} algorithm \citep{Gill2020monofind, Hawthorn2024}, independently from and contemporaneously with the TSTPC WG detection.

\subsection{\ngts}
\label{section:obs:phot:ngts}

\ngts\ is a ground based photometric facility made up of twelve individual robotic 20cm telescopes located in Paranal, Chile. Like \tess, each \ngts\ telescope is fitted with a custom broad band red optical filter spanning approximately 520\,nm - 850\,nm. However, \ngts\ has a finer image scale of approximately $5\arcsec$ per pixel. Despite being ground based, the precision of \ngts\ is able to rival or exceed that of \tess\ \citep{Bryant2020, O'Brien2022, Bayliss2022}, making it very well suited to photometric follow-up of \tess\ single transit candidates \citep{Bayliss2020} such as \name.

After the initial transit detection in \tess, \name\ was observed by \ngts\ in four separate targeted observing campaigns for a total of 228 nights between August of 2023 and January of 2025, these are summarised in Table\,\ref{tab:phot_summary}. The first campaign consisted of 115 nights of observation between 2023 August 25 and 2024 February 4. The second brief campaign consisted of 8 nights between 2024 April 16 and 2024 May 12, this sparse coverage was due to technical issues with the \ngts\ roof control system and cloudy weather. The third campaign began on 2024 August 18 with 69 nights before 2024 November 21. These first three campaigns all used a single telescope while the fourth and final campaign used two simultaneously for 36 nights of observations beginning immediately after the previous on 2024 November 22 and lasting until 2024 December 31. 

These data were reduced using the \bsproc\ pipeline as described in \cite{Bryant2020} using fixed comparison stars and apertures across each run to ensure night to night photometric stability. A sample of the reduced lightcurve data can be found in Table\,\ref{tab:ngts_data}. Before any analysis was done, we median normalised the data across each campaign.

We searched the \ngts\ monitoring data for a second planetary transit using the algorithm described by \citet{Gill2020monofind}, which uses the initial \tess\ detection as a template in a cross-correlation with the \ngts\ photometry. This revealed a candidate egress event on the night of 2024 December 8, which was confirmed with a visual inspection of the data (see Figure\,\ref{fig:transits_NGTS}). No other transit events were identified in the 228 nights of \ngts\ photometry. 

With two transits of \name\,b detected, the orbital period of the planet was constrained to a finite set of aliases, each an integer fraction of the 1444\,d time separation between the two events. The majority of these aliases were ruled out by either the \tess\ or \ngts\ photometry, including all alias periods shorter than 72\,d.

The egress event on 2024 December 8 showed a strong systematic increase in flux, comparable to the transit depth, at the beginning of the night, which was coincident with observations at high airmass ($\gtrsim1.4$). To correct for this, we fit a linear model of flux, $f$, against airmass, $\sec z$, to the residuals from the template match model against the data from the night of the transit (2024 December 8). The function took the form $f=C_1\sec z + C_0$ with the fit resulting in coefficients of $C_1=0.0024\pm0.0005$ and $C_0=-0.0038\pm0.0006$.

We launched a short fifth \ngts\ campaign from 2025 December 2 to 2025 December 5 with four telescopes to target the 180.5\,d alias, which was the best match to our radial velocity observations (see Section\,\ref{section:obs:spec}). However, due to high cloud cover over Paranal at the time we were unable to obtain any usable data.

\subsection{TFOP SG1 Observations}
\label{section:obs:phot:sg1}

After identifying the most likely orbital period for \name\,b of 180.5\,d using \tess\ and \ngts\ photometry (Sects.\,\ref{section:obs:phot:tess}\,\&\,\ref{section:obs:phot:ngts}) and radial velocity observations (Sect.\,\ref{section:obs:spec}), we sent out a general call to the \tess\ Follow-up Observing Program Sub Group 1 \citep[TFOP SG1;][]{Collins2019}\footnote{\url{https://tess.mit.edu/followup}} to target the predicted transit ingress on the night beginning 2025 December 4 UTC and the egress on 2025 December 5 UTC. Observations were contributed by the Las Cumbres Observatory Global Telescope Network \citep[\LCOGT; Section\,\ref{section:obs:phot:lco};][]{Brown2013} \tess\ Follow-up Key Project (PI: Shporer), as well as three private observatories: The Deep Sky Chile 0.4 m (DSC 0.4 m; Section\,\ref{section:obs:phot:vignes}), the Osservatorio Astronomico di Campo Catino (OACC; Section\,\ref{section:obs:phot:mallia}) and the El Sauce Observatory (Section\,\ref{section:obs:phot:elsauce}. 

Photometric extraction was carried out for all of these instruments with the \aij\ package \citep{Collins2017}. The comparison star ensemble for each lightcurve was selected by minimizing the RMS of the residuals of a simultaneous fit to a transit model and a linear fit to airmass using \aij. 

The resulting transit light curves are all shown in Figure\,\ref{fig:transits1}. The detection of this transit ruled out the majority of the remaining alias periods allowed by the \tess\ and \ngts\ transits (Sect.\,\ref{section:obs:phot:ngts} and Fig.\,\ref{fig:transits}). The full set of aliases allowed by the combined photometry are: 72, 90, 120, 181 and 361\,d. Only the 180.5\,d alias is also consistent with our radial velocity observations (see Section\,\ref{section:obs:spec}).

\subsubsection{\LCOGT}
\label{section:obs:phot:lco}

A transit of \name\,b beginning on 2025 December 4 UTC was observed from two Las Cumbres Observatory Global Telescope network nodes. The transit ingress was observed simultaneously with two 1.0\,m and two 0.35\,m telescopes at the Siding Spring Observatory (SSO) node near Coonabarabran, Australia and the egress was observed simultaneously with two 1.0\,m and one 0.35\,m telescopes at the Cerro Tololo Inter-American Observatory (CTIO) node in Chile. The 1.0\,m telescopes are equipped with 4096$\times$4096 SINISTRO Cameras, with an image scale of $0.389\arcsec$ per pixel and a field of view (FOV) of $26^{\prime} \times 26^{\prime}$. The 0.35\,m Planewave Delta Rho 350 telescopes are equipped with $9576\times6388$ QHY600 CMOS cameras having an image scale of $0\farcs73$ per pixel, resulting in a $114\arcmin\times72\arcmin$ full field of view. We used the optional $30\arcmin\times30\arcmin$ sub field of view for a faster detector read-out. All observations were carried out in the Sloan-i$^\prime$ filter with exposure times of 10\,s and 27\,s for the 1.0\,m and 0.35\,m telescopes, respectively. The image data were calibrated with the standard \lco\ \textsc{banzai} pipeline \citep{McCully2018}, and photometric extraction was performed with \aij\ using uncontaminated photometric apertures of 6--9\arcsec. We combined all 1.0\,m datasets in Figure\,\ref{fig:transits_LCO-1m} and all 0.35\,m datasets in Figure\,\ref{fig:transits_LCO-35cm}.

\subsubsection{DSC 0.4m}
\label{section:obs:phot:vignes}
The 2025 December 5 UTC egress was also observed from near Ovalle, Chile, with the Deep Sky Chile 0.4\,m (DSC 0.4\,m) Planewave CDK400 telescope, which was equipped with a QHYCCD600M camera and Chroma R filter (650\,nm $\pm$ 50\,nm). The 9576 x 6388 pixel detector has an image scale of 0.4" per pixel and a FOV of $62^{\prime} \times 42^{\prime}$. The images were calibrated and photometric data were extracted with \aij\ using circular 5" photometric apertures. The egress lightcurve is shown in Figure \ref{fig:transits_vignes}.

\subsubsection{OACC - CAO 0.6m}
\label{section:obs:phot:mallia}
The 2025 December 5 UTC egress was also observed from the Osservatorio Astronomico di Campo Catino (OACC) Campocatino Austral Observatory (CAO) 0.6\,m telescope (El Sauce, Atacama desert, Chile) equipped with a Moravian C3-61000EC PRO camera that provides a FOV of $30^{\prime} \times 20^{\prime}$ with a pixel scale of 0.2" per pixel. The images were aqcuired with a standard photometric Astrodon i$^\prime$2 filter. The time series, composed of 196 x 100s images, was calibrated and photometric data were extracted with \aij\ using circular 6.8" photometric apertures. The egress lightcurve is shown in Figure \ref{fig:transits_mallia}.

\subsubsection{\elsauce}
\label{section:obs:phot:elsauce}
The transit egress was also observed from \elsauce\ observatory in Coquimbo Province, Chile on UTC 2025/12/05 with a CDK500 0.5\,m robotic telescope equipped with an Rc filter and a Moravian C3-26000 camera with 6252 x 4176 pixels binned 2 x 2 in-camera resulting in an image scale of 0.449"/pixel. The 751 x 32\,s images were calibrated and photometric data were extracted with \aij\ using circular 5.8" photometric apertures. The egress lightcurve is shown in Figure \ref{fig:transits_elsauce}.

\subsection{Archival Photometry}
\label{section:obs:phot:rot}

\subsubsection{\wasp}
\label{section:obs:phot:wasp}
The Wide Angle Search for Planets \citep[\wasp;][]{Pollacco2006} was a ground based transit survey with a northern facility (often simply referred to as \wasp) at the Observatorio del Roque de los Muchachos on the island of La Palma in the Canary Islands and a Southern facility (\waspsouth) at the Sutherland Station of the South African Astronomical Observatory. Both \wasp\ instruments consisted of eight Canon Telephoto lenses with a $\sim11$\,cm aperture with custom $2048\times2048$ pixel CCD cameras on a fixed robotic mount, resulting in a wide 482 square degree field of view. \name\ was observed by \waspsouth\ from 2011 November 2 to 2012 March 29. The dataset contains no transit coverage, and the transit is shallower than \wasp's $\sim1\%$ detection limit. However, the \wasp\ lightcurve was still used to attempt to measure a photometric rotation period of the star which we present in Section\,\ref{section:hoststarrot}.

\subsubsection{\assasn}
\label{section:obs:phot:assasn}
The All Sky Automated Survey for SuperNovae \citep[\assasn][]{Kochanek2017} is made up of several stations hosted at various locations around the world by the \lco\ network (see Section\,\ref{section:obs:phot:sg1} and \ref{section:obs:phot:lco}). Each station consists of a single Nikon telephoto lens with a 14\,cm aperture and an approximately 4.5 square degree field of view. \assasn\ is a transient survey and so is not well suited for transit detection given the low number of observations per target on a given night, however, its long baseline of nightly coverage does allow for monitoring rotation and/or activity in stars \citep[e.g;][]{Eschen2025}. \name\ was observed by \assasn\ from 2016 February 2 to 2018 September 17 in the V band and 2018 August 8 to 2025 November 23 in the g band. We apply the same quality cuts as in \cite{Eschen2025} to our dataset in order to use the \assasn\ lightcurve to search for a stellar rotation signal, which we present in Section\,\ref{section:hoststarrot}.

\section{Spectroscopic Observations}
\label{section:obs:spec}

\begin{figure}
    \centering
    \begin{subfigure}{\columnwidth}
        \centering
        \includegraphics[width=\textwidth]{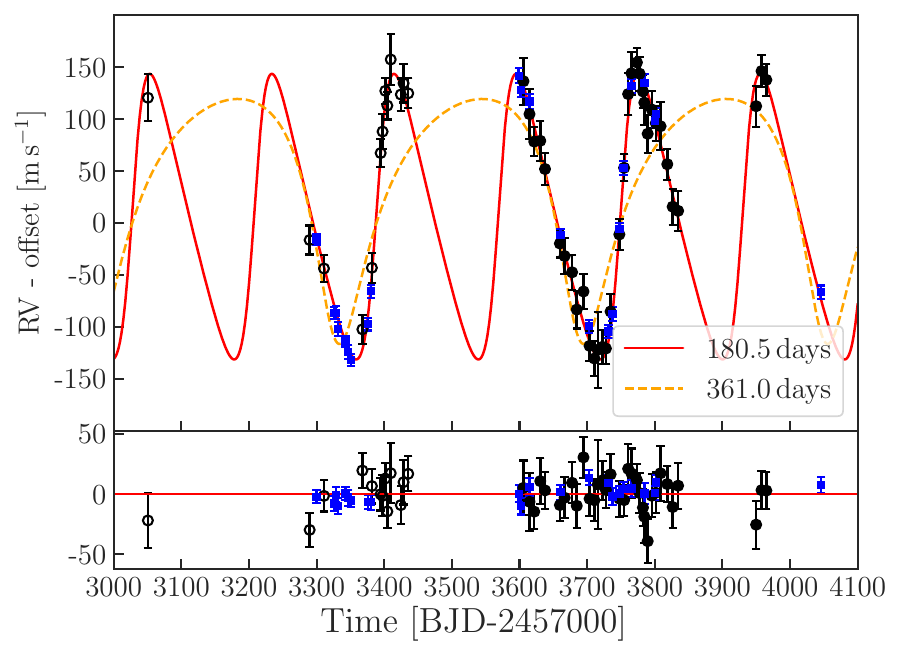}
        \caption{Radial velocities as a function of time.}
        \label{fig:RV_plot_time}
    \end{subfigure}
    \begin{subfigure}{\columnwidth}
        \centering
        \includegraphics[width=\textwidth]{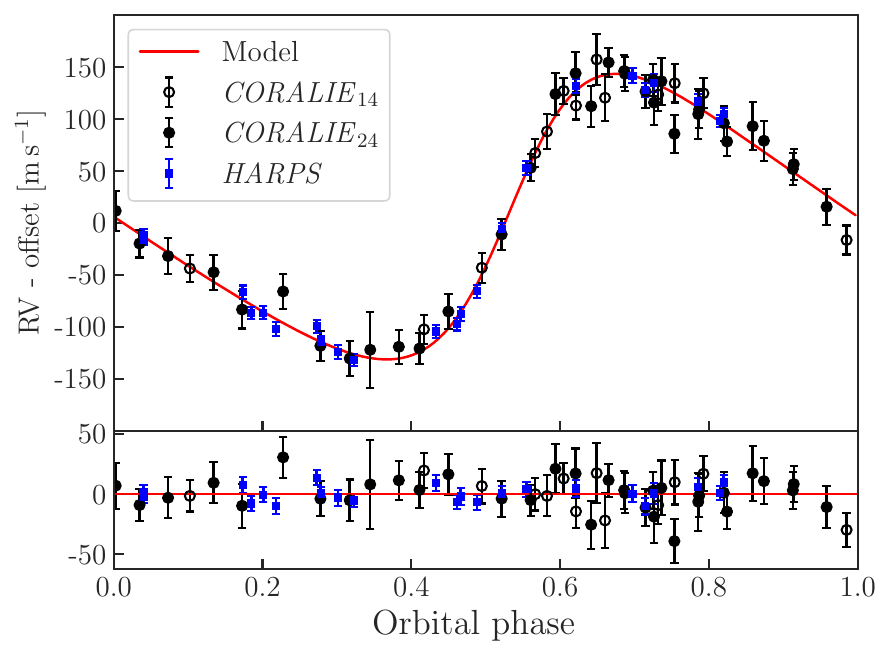}
        \caption{Radial velocities after phase-folding on the orbital period.}
        \label{fig:RV_plot_phase}
    \end{subfigure}
    \caption[Radial velocity measurements of \name.]{Radial velocity measurements of \name\ with the fitted radial velocity baseline offset values subtracted (see Section\,\ref{section:orbit-fit} and Table\,\ref{tab:ns_table}). Data points from \coralie\ are denoted with black circles with those from the first epoch unfilled and those from the second epoch filled. \harps\ data points are shown as blue squares. The median fitted RV model is also overplotted with a red solid  while the best fitting model to the 361.0\,day period alias is shown with an orange dashed line in the top panel. Subfigure \textbf{(a)} shows the data plotted versus time while \textbf{(b)} shows the same data as a function of orbital phase. For each subfigure, the top panel shows the data and model, while the bottom panel shows the residuals with the model subtracted.}
    \label{fig:RV_plot}
\end{figure}

\begin{figure}
    \centering
    \includegraphics[width=\columnwidth]{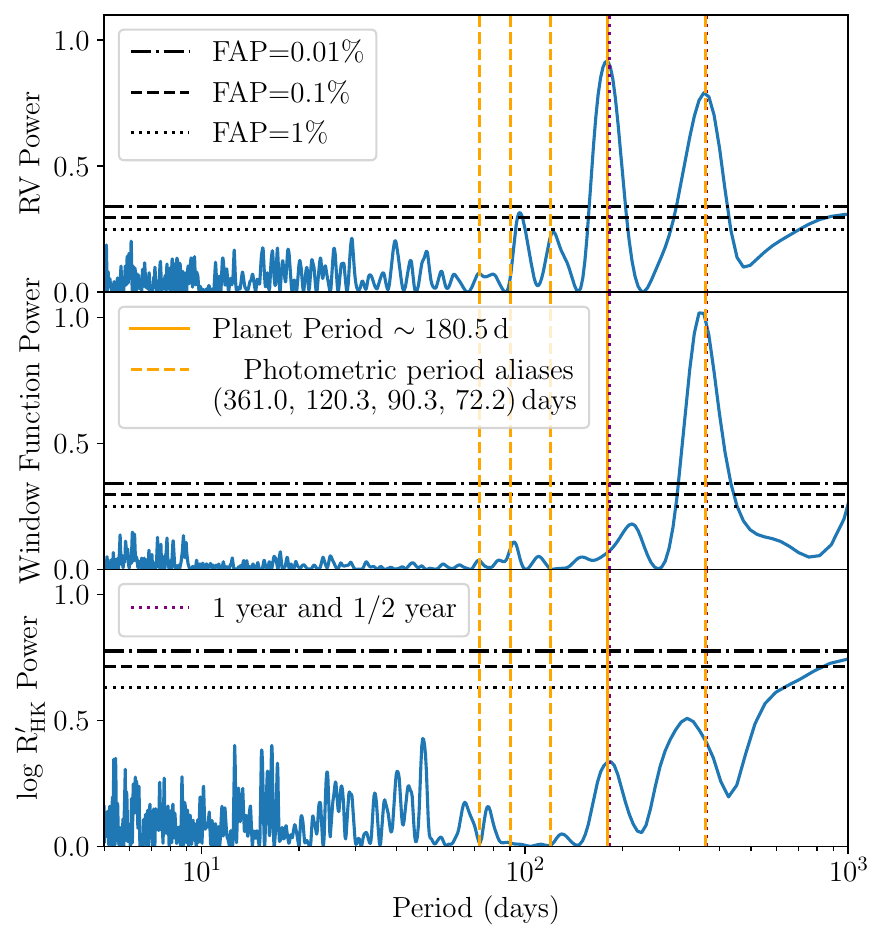}
    \caption{Spectroscopic Lomb-Scargle periodograms of the radial velocity timeseries (\textbf{top}), the combined \harps\ and \coralie\ window function (\textbf{middle}) and \logrhk\ measurements from \harps\ (\textbf{bottom}). The window function was created by running a Lomb-Scargle periodogram on the RV timeseries with a constant y value of 1. False alarm probabilities of 1\%, 0.1\% and 0.01\% are represented by horizontal black lines with dotted, dashed and dash-dotted styles, respectively. A solid orange vertical line represents the orbital period of the planet at $\sim180.5$\,d while dashed vertical orange lines represent the photometric period aliases at 361.0, 120.3, 90.3 and 72.2 days. Additionally purple dotted vertical lines are shown to represent periods of 1 year and half a year.}
    \label{fig:rv_periodogram}
\end{figure}

In this section we describe the spectroscopic follow-up campaigns, which began after the initial \tess\ single transit detection of TIC-\TICID\,b\,/\,\name\,b, using \coralie\ \citep[][Section\,\ref{section:obs:spec:coralie}]{Queloz2001coralie} and \harps\ \citep[][Section\,\ref{section:obs:spec:harps}]{Pepe2000, Mayor2003harps}. These spectra were used to produce time-series radial velocity measurements shown in Tables\,\ref{tab:rvs_coralie}\,and\,\ref{tab:rvs_harps}, respectively, and are presented together in Figure\,\ref{fig:RV_plot}. Before any models were fitted to the data, we performed an initial search for periodic signals by creating a generalised Lomb-Scargle periodogram \citep{Lomb1976, Scargle1982, Zechmeister2009} of the combined \coralie\ and \harps\ radial velocity data using \textsc{astropy} \citep{AstropyCollab2022}. To combine these datasets we subtracted the median from each set of RVs before concatenating them together. The resulting periodogram is shown in Figure\,\ref{fig:rv_periodogram}. The periodogram shows two peaks at $\sim180$ and $\sim360$\,days which effectively rules out all but the 361\,day and 180.5\,day photometric period aliases (see Section\,\ref{section:obs:phot}). The relative strength of the 180.5\,day peak compared to to the 361\,day signal as well as visually inspecting and comparing the best fits at both periods (see Figure\,\ref{fig:RV_plot_time} and Section\,\ref{section:orbit-fit}) led us to conclude that the 180.5\,day alias was in fact the true period of the system.

\subsection{\coralie}
\label{section:obs:spec:coralie}
\coralie\ is a fiber-fed echelle spectrograph with a resolution of $\text{R}\sim60\,000$, mounted on the 1.2-m Leonhard Euler Telescope at ESO La Silla Observatory in Chile. \coralie\ is capable of achieving radial velocity precision on the order of $\sim10$\ms \citep{Queloz2001coralie}, which makes it well suited to the detection of giant planets such as \name\,b. 

On the night of 2023 April 16 a single spectrum of \name\ was taken followed by an additional twelve points between 2023 December 10 and 2024 May 5, with a further 31 observations performed between 2024 October 21 and 2025 June 8 with 44 total spectra taken. All the observations of \name\ taken with \coralie\ had an exposure time of 1200-s and were taken at airmasses between 1.00 and 1.74 with the resulting spectra having signal-to-noise ratios (SNRs) between 19.6 and 40.4 in spectral order 62. The spectra were reduced using the standard \coralie\ Data Reduction System (DRS) 3.3.12 with a G2 mask used for cross correlation that produced the radial velocity measurements shown in Table\,\ref{tab:rvs_coralie} and Figure\,\ref{fig:RV_plot}. Due to an instrument update for \coralie\ between the 2nd and 3rd set of observations described above, we treat the data from before and after this point as two separate instruments (\coralie$_{14}$ and \coralie$_{24}$, respectively) for the purpose of fitting instrumental offsets and jitter terms (see Section\,\ref{section:orbit-fit}).

\subsection{\harps}
\label{section:obs:spec:harps}
Alongside the observations made with \coralie, we also observed \name\ with \harps; a fiber-fed echelle spectrograph with a spectral resolution of $\text{R}\sim115000$, which is mounted on the 3.6-m telescope at the ESO La Silla Observatory. The greater telescope diameter and spectral resolution of \harps\ allow for more precise radial velocity measurements than \coralie, with \harps\ capable of reaching $\sim1$\ms\ precision. 

The \harps\ observations of \name\ were taken in a very similar time-frame to the \coralie\ data. Nine spectra were taken between 2023 December 20 and 2024 March 11, with a further 13 between 2024 October 14 and 2025 May 6. An additional \harps\ spectrum, taken on 2026 January 5, was incorporated into the RV dataset but not used in the determination of the host star's spectral parameters (see Section\,\ref{section:PAWS}). The first seven of these observations used an exposure time of 1500-s and were taken at airmasses between 1.02 and 1.59 with a resulting SNR of 51.4-78.4 in order 64. The remaining spectra were all taken with an exposure time of 1200-s at airmasses between 1.00 and 1.49 with resulting SNRs between 35.3 and 56.1 in order 64. These data were reduced with DRS 3.3.6 using a G2 mask. The resultant radial velocity values as well as activity indicators, including the S index derived \logrhk\ value are shown in Table\,\ref{tab:rvs_harps} and Figure\,\ref{fig:RV_plot}.

\section{Host star analysis}
\label{section:host-analysis}

\subsection{Catalogue parameters}
\label{section:catalog-params}
Catalogue parameters for TIC-\TICID\,/\,\name\ are shown in Table~\ref{tab:host_properties}. These were taken from the 2 Micron All Sky Survey \citep[2MASS;][]{skrutskie20062mass}, \gaia\ data release 3 \citep[\gaia\ DR3;][]{gaia2021dr3}, the \tess\ Input Catalogue v8 \citep[TIC 8;][]{Stassun2019tic}, the AAVSO Photometric All Sky Survey \citep[APASS;][]{Henden2014APASS}, the Wide field Infrared Survey Explorer \citep[WISE;][]{wright2010wise}, \tycho\,\citep{Hog2000} and SkyMapper \citep{Keller2007SkyMapper} Data Release 4 \citep[DR4;][]{Onken2024} catalogues where specified. We use the \textsc{gaiadr3-zeropoint} package to calculate a zero point offset to the parallax as described in \citet{Lindegren2020, Lindegren2021}, finding a value of of -0.01592\,mas which we apply to the \gaia\ DR3 parallax of $3.7400\pm0.0128$\,mas resulting in the final value we quote in Table\,\ref{tab:host_properties}.

\begin{figure}
    \centering
    \includegraphics[width=\columnwidth]{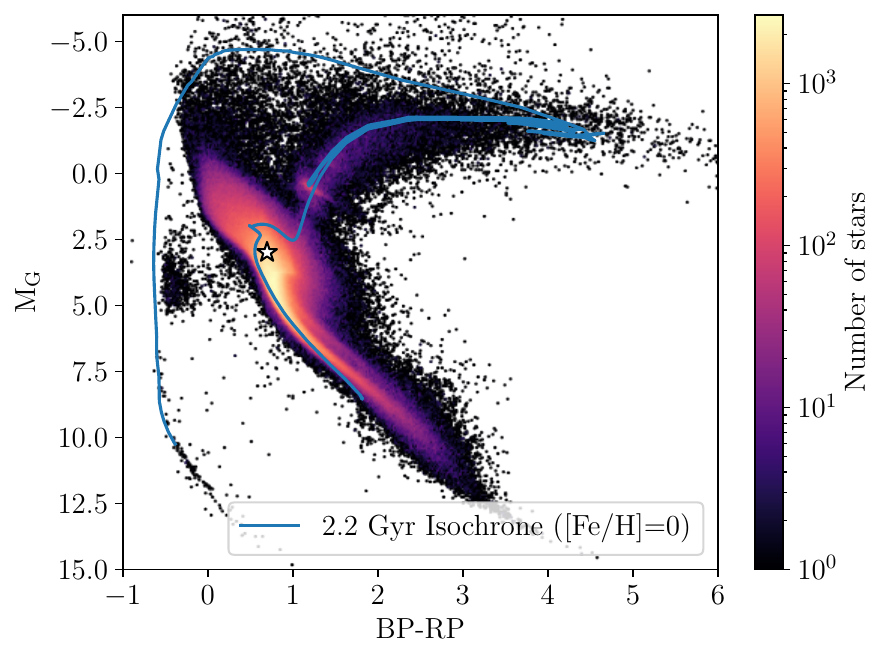}
    \caption{\gaia\ DR3 Hertzsprung-Russel diagram for all \tess-SPOC target stars from \citet{Doyle2024}. \name\ is overplotted as a white star with black borders. A solar metallicity 2.2\,Gyr MIST isochrone \citep{Dotter2016, Choi2016} is overplotted as a solid blue line. The star sits significantly above the isochrone due to its enhanced metallicity giving it a larger radius (see Sections\,\ref{section:isochrones} and \ref{section:overview}).}
    \label{fig:gaia_HRD}
\end{figure}

To aid in visualising the colour and magnitude information from these catalogues we overplot \name\ on a \gaia\,DR3 Hertzsprung-Russell diagram \citep[HRD;][]{Hertzsprung1913, Russell1913} of all \tess-SPOC target stars from \citet{Doyle2024} in Figure\,\ref{fig:gaia_HRD}. The star's position on the HRD shows a colour and luminosity broadly consistent with a main sequence F6-F7 type star, albeit a mildly over-luminous one.

\begin{table}
    \centering
    \caption{\name\ stellar parameters.}
    \label{tab:host_properties}
    \begin{tabular}{ccc}
        \toprule
         Property&  Value & Source\\ \hline
         \hline
         \multicolumn{3}{c}{\textit{Identifiers}} \\ 
         \hline
         \tycho\ ID & \TYCID & \tycho (\S\ref{section:catalog-params}) \\
         SkyMapper DR4 ID & \SkyMapperID & SkyMapper (\S\ref{section:catalog-params}) \\
         2MASS ID & \twoMASSID & 2MASS (\S\ref{section:catalog-params}) \\
         \gaia\ Source ID&  \GaiaID & \gaia\ DR3 (\S\ref{section:catalog-params}) \\
         TIC ID&   \TICID & TIC 8 (\S\ref{section:catalog-params}) \\
         \hline
         \multicolumn{3}{c}{\textit{Coordinates}} \\ 
         \hline
         RA (hh:mm:ss.ss)&  \rahms & \gaia\ DR3 (\S\ref{section:catalog-params}) \\
         DEC (dd:mm:ss.ss)&  \dechms & \gaia\ DR3 (\S\ref{section:catalog-params}) \\
         \hline
         \multicolumn{3}{c}{\textit{Proper motion and parallax}} \\ 
         \hline
         $\mu_\text{RA}$ (mas y$^{-1}$)&  \pmra & \gaia\ DR3 (\S\ref{section:catalog-params}) \\
         $\mu_\text{DEC}$ (mas y$^{-1}$)&  \pmdec & \gaia\ DR3 (\S\ref{section:catalog-params}) \\
         Parallax (mas)&  \parallax & \gaia\ DR3 (\S\ref{section:catalog-params}) \\
         Systemic velocity (\kms) & \baselineoffsetrvHARPS & This work (\S\ref{section:orbit-fit}) \\
         \hline
         \multicolumn{3}{c}{\textit{Galactic Kinematics}} \\ 
         \hline
         $U$ (\kms)& \VgalU & This work (\S\ref{section:toomre}) \\
         $V$ (\kms)& \VgalV & This work (\S\ref{section:toomre}) \\
         $W$ (\kms)& \VgalW & This work (\S\ref{section:toomre}) \\
         $v_\text{tot}$ (\kms)& \Vgaltot & This work (\S\ref{section:toomre}) \\
         $e_\text{gal}$& \eccgal & This work (\S\ref{section:toomre}) \\
         $J_z$ (kpc \kms)& \Jzgal & This work (\S\ref{section:toomre}) \\
         \hline
         \multicolumn{3}{c}{\textit{Magnitudes}} \\ 
         \hline
         V (mag) & \Vmag & APASS (\S\ref{section:catalog-params}) \\
         B (mag) & \Bmag & APASS (\S\ref{section:catalog-params}) \\
         u (mag) & \umag & SkyMapper (\S\ref{section:catalog-params}) \\
         v (mag) & \vmag & SkyMapper (\S\ref{section:catalog-params}) \\
         g (mag) & \gmag & SkyMapper (\S\ref{section:catalog-params}) \\
         r (mag) & \rmag & SkyMapper (\S\ref{section:catalog-params}) \\
         i (mag) & \imag & SkyMapper (\S\ref{section:catalog-params}) \\
         z (mag) & \zmag & SkyMapper (\S\ref{section:catalog-params}) \\
         G (mag) & \GaiaGmag & \gaia\ DR3 (\S\ref{section:catalog-params}) \\
         BP (mag) & \GaiaBPmag & \gaia\ DR3 (\S\ref{section:catalog-params}) \\
         RP (mag) & \GaiaRPmag & \gaia\ DR3 (\S\ref{section:catalog-params}) \\
         \tess\ (mag) & \TESSmag & TIC 8 (\S\ref{section:catalog-params}) \\
         J (mag) & \Jmag & 2MASS (\S\ref{section:catalog-params}) \\
         H (mag) & \Hmag & 2MASS (\S\ref{section:catalog-params}) \\
         K (mag) & \Kmag & 2MASS (\S\ref{section:catalog-params}) \\
         W1 (mag) & \WOnemag & WISE (\S\ref{section:catalog-params}) \\
         W2 (mag) & \WTwomag & WISE (\S\ref{section:catalog-params}) \\
         W3 (mag) & \WThreemag & WISE (\S\ref{section:catalog-params}) \\
         \hline
         \multicolumn{3}{c}{\textit{Spectral Parameters}}\\
         \hline
         \vmic\ (\kms) & \hostvmic & This Work (\S\ref{section:PAWS})\\
         \vmac\ (\kms) & \hostvmac & This Work (\S\ref{section:PAWS})\\
         \vsini\ (\kms) & \hostvsini & This Work (\S\ref{section:PAWS})\\
         \teff\ (K) & \hostteff & This Work (\S\ref{section:PAWS}) \\
         \feh\ (`dex') & \hostfeh & This Work (\S\ref{section:PAWS}) \\
         \logg\ ($\log(\text{cgs})$) & \hostlogg & This Work (\S\ref{section:isochrones}) \\
         \logrhk & \meanlogrhk & This Work (\S\ref{section:logrhk}) \\
         \hline
         \multicolumn{3}{c}{\textit{Derived parameters}} \\ 
         \hline
         $\rho_\star$ (\gcc) & \initialhostdensity & This Work (\S\ref{section:orbit-fit}) \\
         \rstar\ (\rsun) & \hostrad & This Work (\S\ref{section:isochrones}) \\
         \mstar\ (\msun) & \hostmass & This Work (\S\ref{section:isochrones}) \\
         Age (Gyr) & \hostage & This Work (\S\ref{section:isochrones}) \\
         Distance (pc) & \stardist & This Work (\S\ref{section:isochrones}) \\
         \bottomrule
    \end{tabular}
\end{table}

\subsection{Spectral fit}
\label{section:PAWS}
To fit spectral parameters for the host star we co-added the first 22 \harps\ and 44 \coralie\ spectra into two combined spectra with SNR values of 151 and 112, respectively.

We then analysed these spectra using Parameters Approximated With Synthesis \citep[\paws;][]{Freckelton2024, Freckelton2025}. \paws\ uses the functions provided in the \textsc{iSpec} package \citep{Blanco-Cuaresma2014, Blanco-Cuaresma2019} to determine the atmospheric parameters via the equivalent widths (EWs) of the \fei\ and \feii\ lines, assuming ionization and excitation equilibrium. The resulting values from this method were then used as priors for a spectral synthesis fit to obtain the final stellar parameters. 

We calculated an inverse variance weighted mean of the results from both the \coralie\ and \harps\ spectra, obtaining values of \teff, \feh, \vmic, \vmac, and \vsini\,that we adopt in Table\,\ref{tab:host_properties}. Note that we do {\it not} adopt the spectroscopically derived \logg\ of \speclogg\ from \paws\ as such values from spectra are known to be unreliable due to strong correlations with \teff\ and \feh\ \citep{Torres2012, Mortier2014}.

\subsection{Isochrone fit}
\label{section:isochrones}

To determine the final set of stellar parameters we adopt in Table\,\ref{tab:host_properties}, we fitted the stellar models from the MESA \citep[Modules for Experiments in Stellar Astrophysics;][]{Paxton2010} Isochrones and Stellar Tracks \citep[MIST;][]{Dotter2016, Choi2016} using the \isochrones\ package \citep{Morton2015}. In order to derive radius, mass (and thus isochronal \logg), and age, we used the optical B and V band magnitudes as well as the J,H,K,W1,W2 and W3 magnitudes to cover a wider range of the spectral energy distribution. All of these magnitudes are listed in Table\,\ref{tab:host_properties}. We furthermore used the values of \teff\,and \feh\ from \paws\ (see Section\,\ref{section:PAWS}) and the zero point corrected \gaia\,DR3 parallax applied listed in Table\,\ref{tab:host_properties}. Additionally, we also included the orbit derived host density (\initialhostdensity\,\gcc). This value was derived from a preliminary global model fit with no stellar parameters included and wide priors on the limb-darkening parameters (see Section\,\ref{section:orbit-fit}). We note that the final result did not depend, within errors, on the inclusion or exclusion of the stellar density derived from the transit. Final values and precision uncertainties were extracted as the median value and the 16th and 84th percentiles of the posterior distributions.

We adopt the recommendations of \cite{Tayar2022} to account for the inherent uncertainties between stellar models; adding an additional 0.042\% error term in quadrature to the host radius. In addition we made use of the \textsc{kiauhoku} package \citep{Claytor2020} to compare fits of the spectroscopic \teff, \logg\ and \feh\ against MIST, Dartmouth \citep{Dotter2008}, Yale Rotating stellar Evolution Code \citep[YREC;][]{Demarque2008} and Garching STellar Evolution Code \citep[GARSTEC;][]{Weiss2008} stellar models. We calculated the standard deviation across the values fitted against each model and added these in quadrature to our values. This error term for \logg\ was so small as to be negligible $(\sim10^{-7})$. However, we used terms for the stellar mass and age of 0.08\,\msun\ and 0.4\,Gyr, respectively. With these error terms added this resulted in the final stellar parameters we adopt in Table\,\ref{tab:host_properties}.

\subsection{Activity and rotation}
\label{section:activity}

\subsubsection{Rotation}
\label{section:hoststarrot}

\begin{figure}
    \centering
    \includegraphics[width=\columnwidth]{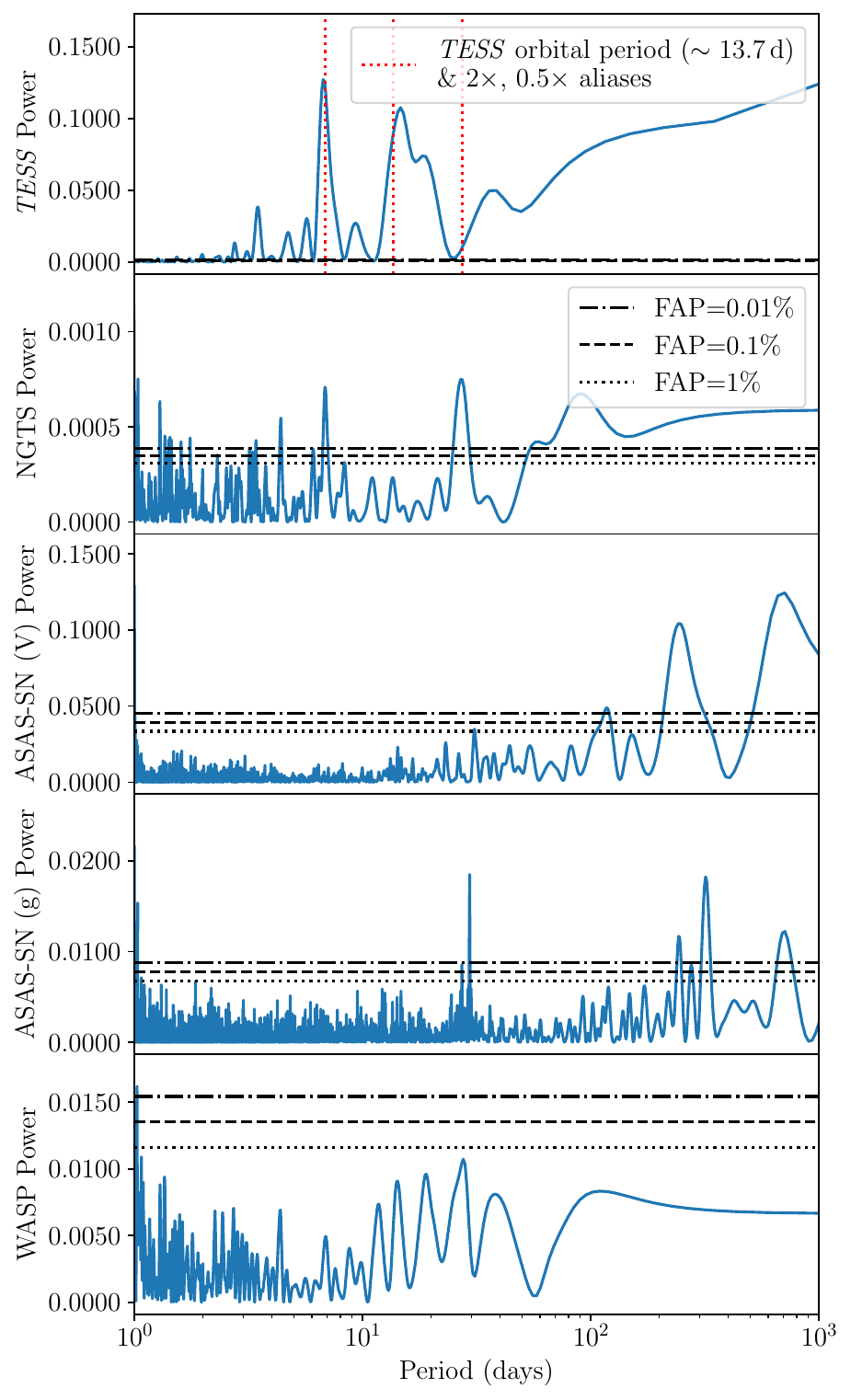}
    \caption{Lomb Scargle periodograms of \tess\ sectors 87 and 88 (\textbf{top}), the first run of \ngts\ data (\textbf{upper middle}), \assasn\ in both the V band (\textbf{middle}) and g band (\textbf{lower middle}) and \wasp\ (\textbf{bottom}) for \name. False alarm probabilities of 1\%, 0.1\% and 0.01\% are represented by horizontal black lines with dotted, dashed and dash-dotted styles respectively. For \tess, its 13.7\,d orbital period is marked with a vertical red dashed line as well as aliases at $2\times$ (a \tess\ sector duration) and $0.5\times$ this period (which coincides with the mid-orbit gap in newer sectors).}
    \label{fig:LombScarglephot}
\end{figure}

Using equation\,1 from \cite{Watson2010} and our value of \vsini\ from \paws\ (see Section\,\ref{section:PAWS}), we estimate an upper limit on the stellar rotation period of $8\pm2$\,d. We attempt to measure a photometric rotation period by creating Lomb-Scargle periodograms on the largest consecutive runs of each of our datasets. For \ngts\ we use the first run between 2023 August 25 - 2024 February 4 and for \tess\ we separately median normalised and then stitched together the SPOC lightcurves (using the \texttt{SAP\_FLUX} column) for sectors 87 and 88. We also used the entire \wasp\ and \assasn\ datasets (See Section\,\ref{section:obs:phot:wasp} and \ref{section:obs:phot:assasn}), separating the latter by V and g bands. The resulting periodograms are shown in Figure\,\ref{fig:LombScarglephot}.

The \tess\ periodogram shows two peaks of interest, one at 6.7\,d and another broad peak around $\sim20-30$\,d. However, these are close to the mid orbit gap and mid sector gap, respectively, and should be interpreted cautiously. The \ngts\ periodogram is nosier than \tess\ but still shows a few peaks of interest. The peak at $\sim1$\,d is almost certainly due to the nightly cadence of the data. There is a peak at 6.9\,d similar to that found in \tess, with another at 26.9\,d which coincides with the lunar period of 27.3\,d and thus is likely caused by dilution from moonlight modulating with the lunar phases. Another peak appears at 93\,d, however given this is approaching the baseline it should also be interpreted cautiously. A peak at $\sim7$\,d appears in both \tess\ and \ngts\ photometry and is in agreement with the \vsini\ derived rotation period -- lending some credence that this may be the true rotation period of the star. However, phase folding on this period doesn't show a clear sinusoidal signal for either dataset. Additionally, we do not see this peak in either \assasn\ bandpass, seeing only a $\sim300$\,d peak in both likely due to the Earth's orbit and another at 29.5\,d in the g band likely due to the moon. The \wasp\ periodogram also shows no convincing peaks at any period beyond $\sim1$\,d. Hence we conclude that, in spite of a tentative $\sim7$\,d signal, we do not recover any significant rotation period for \name\ photometrically.

\subsubsection{\logrhk}
\label{section:logrhk}

We calculated an inverse variance weighted mean and standard deviation of all the individual values of \logrhk\ calculated by the DRS from each \harps\ spectrum (see Table\,\ref{tab:rvs_harps}), finding of \logrhk=\meanlogrhk. This indicates a low level of stellar activity, albeit not unusually so for an F6-F7 star \citep[See \logrhk\ catalogues;][]{Jenkins2006, Jenkins2008, Jenkins2011}. 

A potential long-term downward trend in \logrhk\ is visible. This trend is, however, unlikely to be genuine as the lower \logrhk\ data points have systematically poorer precision and come from lower signal to noise spectra. We still attempted to investigate this by producing a periodogram of the \logrhk\ data, which we present in Figure\,\ref{fig:rv_periodogram}. The periodogram shows a potential long term trend with a high plateau in power beyond $10^3$\,d, although this could be driven by the lower SNR spectra taken later in the campaign. Also notable are the peaks that appear to coincide with those in the RV periodogram at $\sim180$ and $\sim360$\,. There is also a weak negative correlation between the RVs and \logrhk. Although, it is highly unlikely that the RV signal is activity induced due to its high amplitude. These peaks are instead likely an artefact of the spectroscopic window function. This is further supported by the fact that we see no similar peaks or RV correlations in other activity indicators such as the Full Width Half Maximum (FWHM) and Bisector Inverse Slope (BIS).

\subsection{Galactic Kinematics and population membership}
\label{section:toomre}
Using the proper motions and parallax reported in Gaia DR3 and the \harps\ systemic RV (see Sections\,\ref{section:obs:spec},\ref{section:orbit-fit} and Table\,\ref{tab:host_properties}), we compute the kinematic properties of \name\ following \citet{Johnson1987kinematics, Bensby2003thinthickdisk} and report U, V and W in Table\,\ref{tab:host_properties}. As laid out in \cite{Bensby2003thinthickdisk, Bensby2014thinthickdisk, Reddy2006thinthickdisk}, we compute the kinematic membership probability by correcting for the Local Standards of Rest reported in \citet{Koval_LSR, Schoenrich_LSR,Coskunoglu_LSR,Bobylev_LSR,Francis_LSR,Tian_LSR} and \citet{Almeida_Fernandes_LSR}, also using the velocity dispersions and stellar fractions reported in \citet{Bensby2014thinthickdisk,Chen2021thinthickdisk}. This results in a weighted kinematic thin disc probability of 98.2\% for \name. This is consistent with the star's enriched metallicity as thin disc stars are typically more metal rich than those in the thick disc \citep{Bensby2003thinthickdisk, Bensby2014thinthickdisk}.

\section{Orbital solution}
\label{section:orbit-fit}

We performed a joint model fit to the combined photometric and spectroscopic radial velocity dataset using the \alles\ package \citep{allesfitter-code, allesfitter-paper}, which packages together various fitting code suites with the \textsc{ellc} modelling package \citep{Maxted2016ellc}. We opted to use nested sampling \citep{Skilling2004, Skilling2006nestedsampling} via the \textsc{dynesty} package \citep{Speagle2020} contained within \alles.

\alles\ uses 8 main fitted astrophysical parameters to describe the orbital solution of a system. The first and simplest of these is the orbital period, $P$, in days. Transit timings are described largely by this as well as the transit epoch, $T_0$. The transit shape comes from the radius ratio between the planet and host star, \rpl/\rstar, the combined host and planet radius relative to the orbital semi-major axis, $(\rstar+\rpl)/a$, and the cosine of the orbital inclination, $\cos i$. Radial velocity data solely determines the semi-amplitude, $K$, in \kms\ and also largely controls the orbital eccentricity ($e$) and angular argument of periastron ($\omega$), parametrized as $\sqrt{e}\sin\omega$ and $\sqrt{e}\cos\omega$.

On top of these astrophysical parameters, a wide array of instrumental parameters were also fitted to the data. For photometry we included logarithmic error terms, $\log\sigma$ per instrument, which we initially set with wide uniform priors between -14 and 0. For all ground based photometry instruments we fit a simple flux baseline offset term with a uniform prior between $\pm0.001$. For \tess\ we performed a least squares minimisation spline fit to the baseline at each sampling step using \alles's built in \texttt{hyrbrid\_spline} method. We modelled limb darkening using a quadratic law, for which \alles\ uses the $q_1,q_2$ parametrization from \citet{Kipping2013}, varying the priors between 0 and 1 for all instruments in all fits with the exception of the final fits described in Section\,\ref{section:orbit-fit:final}. The limb-darkening parameters were coupled together for the four \lco\ instruments. For the radial velocity data, a logarithmic jitter term, $\log\sigma$, was fitted per instrument with initial uniform priors, as well as an RV baseline offset.

\subsection{Preliminary fits}
\label{section:orbit-fit:prelim}

The first fit we performed was only to the \tess\ and radial velocity data. We allowed the period to vary uniformly between 179 and 181 days, based on the most likely peak in the RV periodogram (see Figure\,\ref{fig:rv_periodogram}). Similarly, the transit epoch had a uniform prior distribution of a 2 day range centred on the reported \tess\ epoch on BJD=2459209.228. Wide uniform priors of $\mathcal{U}(0, 0.1)$, $\mathcal{U}(0, 0.05)$ were placed on \rpl/\rstar and $(\rstar+\rpl)/a$, respectively, based on visual inspection of the \tess\ transit. The radial velocity semi-amplitude was allowed to vary between 0 and 500\,\ms. $\cos i$ was allowed to vary across its full possible range between 0 and 1 and the parameters of $\sqrt{e}\sin\omega$ and $\sqrt{e}\cos\omega$ were similarly sampled across their full possible ranges between -1 and 1. The RV baseline priors were set between 19.9 and 20.01\,\kms\ for all RV instruments based on a visual inspection of the data. All RV instrumental jitter terms had uniform priors between -7.6 and -4.6 $\ln$\,\kms. The transit parameters resulting from this fit were used for the template matching used to detect the \ngts\ egress described in Section\,\ref{section:obs:phot:ngts}.

After the detection of the \ngts\ egress (see Section\,\ref{section:obs:phot:ngts}), the detrended \ngts\ lightcurve was incorporated into a fit. The period of $180.531\pm0.001$\,d and epoch of $\text{BJD}=2459209.2277\pm0.0032$ from this fit were used to schedule the simultaneous set of additional transit observations described in Section\,\ref{section:obs:phot:sg1}. We then performed a preliminary global fit, incorporating the entire dataset which we performed to determine the transit derived stellar density used in the host star characterisation (see Section\,\ref{section:isochrones}). As with all the previous fits, no stellar parameters were inputted to \alles\ to avoid any biasing of the resulting stellar density value.

\subsection{Final fits}
\label{section:orbit-fit:final}

All further fits were conducted with the addition of the stellar radius, mass and effective temperature into the fit, which allowed the determination of derived parameters such as the planet mass and radius. We also allowed \alles\ to use these stellar parameters to set an external prior on the stellar density. Solutions where the derived host density did not match this prior were penalised. In addition, we used the Limb Darkening Tool Kit \citep[\ldtk;][]{Parviainen2015} package to determine priors on the limb darkening coefficients. Since \ldtk\ returns limb darkening coefficients in the usual $u_1,u_2$ parametrization, we converted these values back to the \citet{Kipping2013} $q_1,q_2$ parameters in order to be usable in \alles. \ldtk\ uses values of the host star \teff, \feh\ and \logg\ along with filter response functions for each instrument. For \tess, \ngts\ and \lco\ exact filter response functions were available while for the \elsauce, DSC0.4m and OACC-CAO instruments we used generic Cousins R, Johnson R, and Sloan i$^\prime$ filters respectively from the Spanish Virtual Observatory filter profile service,\footnote{Available at \url{https://svo2.cab.inta-csic.es/theory/fps/}} assuming these to be close enough approximations. In these cases we widened the spread of the normal priors on both limb darkening coefficients by a factor of 3 to account for the approximation. It should be noted that in the preliminary fits with uniform priors, the limb darkening parameters remained largely unconstrained. The only exception to this being those for \tess\ which are in disagreement with the final values derived from \ldtk. However, given that the solutions were otherwise extremely similar and the physically motivated nature of this prior, we still consider its use appropriate.

We used this setup to perform two fits; one with a $\mathcal{U}(179,181)$\,d prior on the period used in all previous fits and another with a prior on the period of $\mathcal{U}(360,361)$\,d to test the other remaining photometric alias. Whilst the $\mathcal{U}(179,181)$\,d period prior fit provided a good fit to the data, the $\mathcal{U}(360,361)$\,day fit was much poorer (see Figure\,\ref{fig:RV_plot_time}), effectively allowing us to rule out this period. The prior and posterior distributions of all parameters from the final fit to the 180.5\,d alias are shown in Table\,\ref{tab:ns_table}. Also presented are the derived parameters using the inputted stellar parameters, shown in Table\,\ref{tab:ns_derived_table}.

\section{Results}
\label{section:discussion}

\subsection{The TIC-\TICID\,/\,\name\ system}
\label{section:overview}

\begin{table}
    \centering
    \caption{\name\,b properties}
    \label{tab:planet_properties}
    \begin{tabular}{ccc}
        \toprule
        Parameter & Value & Source \\
        \hline
        Period\,(days) & \bperiod & \alles \\
        \hline
        \multicolumn{3}{c}{\it Transit Parameters} \\
        \hline
        T$_0$\,(BJD) & \bepoch & \alles \\
        \rpl/\rstar & \brr & \alles \\
        \rstar/a & \bRstarovera & \alles \\
        $b_\text{transit}$ & \bbtra & \alles \\
        $b_\text{occulation}$ & \bbocc & This work (\S\ref{section:atmosphere}) \\
        T$_{14}$\,(hrs) & \bTtratot & \alles \\
        T$_{23}$\,(hrs) & \bTtrafull & \alles \\
        \hline
        \multicolumn{3}{c}{\it RV parameters} \\
        \hline
        $K$\,(\kms) & \bK & \alles \\
        $e$ & \be & \alles \\
        $\omega$\,(degrees) & \bw & \alles \\
        \hline
        \multicolumn{3}{c}{\it Derived parameters} \\
        \hline
        \rpl\,(\rjup) & \bRcompanionRjup & \alles \\
        \mpl\,(\mjup) & \bMcompanionMjup & \alles \\
        $\rho_\text{p}$\,(\gcc) & \bdensity & \alles \\
        \logg\,($\log(\text{cgs})$) & \blogg & \alles \\
        $a$\,(au) & \baAU & \alles \\
        $r_\text{periastron}$ (au) & \periastron & $a(1-e)$ \\
        $r_\text{apastron}$ (au) & \apastron & $a(1+e)$ \\
        \teq\,(K) & \bTeq & \alles \\
        \teq\textsubscript{;apastron}\,(K) & \bTeqAp & This work (\S\ref{section:overview}) \\
        \teq\textsubscript{;periastron}\,(K) & \bTeqPeri & This work (\S\ref{section:overview}) \\
        $R_\text{Hill}$ (\rjup) & \hillRJ & This work (\S\ref{section:moonsnrings}) \\
        $d_\text{Roche}$ (\rjup) & \rocheRJ & This work (\S\ref{section:moonsnrings}) \\
        \bottomrule
    \end{tabular}
\end{table}

TIC-\TICID\,/\,\name\ is a bright (Vmag=\Vmag) star with a temperature of \hostteff\,K, which is consistent with an F6V-F7V type star \citep{Pecaut2013}. While the star may superficially appear evolved due to its large radius (\hostrad\,\rsun) compared to its mass (\hostmass\,\msun) and resulting low surface gravity (\logg=\hostlogg), the star is still on the main sequence. In evolutionary models of F type stars \citep[e.g.][]{Dotter2016, Choi2016}, the stellar radius is fairly dependent on metallicity as a higher metallicity leads to higher opacities in the upper stellar atmosphere. Hence, a metal rich star like \name\ (\feh=\hostfeh) can have an apparently inflated radius before leaving the main sequence (See the isochrone in Figure\,\ref{fig:gaia_HRD}. This is reflected in the isochrone derived stellar age of \hostage\,Gyr we find for \name, which is expected of an F6-F7 type star on the main sequence.

The system is host to a transiting planet companion, \name\,b, for which we list some properties in Table\,\ref{tab:planet_properties}. We find the planet to have a radius similar to that of Jupiter (\rpl=\bRcompanionRjup\,\rjup) but almost five times the mass (\mpl=\bMcompanionMjup\,\mjup), placing it firmly among the rare `super-Jupiter' class of exoplanets. The planet is on a long-period (P=\bperiod\,d), wide separation ($a=$\baAU\,au) orbit with a moderate eccentricity ($e=$\be). We investigated the timescales for both tidal locking and orbital circularisation using equations 1 and 4 from \citet{Guillot1996} and \citet{Jackson2008} respectively, finding both to be far in excess of a Hubble time. This is unsurprising for a planet as widely separated as \name\,b. This wide orbital separation also gives \name\,b a relatively cool equilibrium temperature for a transiting exoplanet with \teq=\bTeq\,K at its semi-major axis. It should be noted, however, that a planet with as long a period and as eccentric an orbit as \name\,b will experience temperature fluctuations throughout its orbit. Using the same assumptions used by \alles\ to estimate \teq\ at $a$ (those being a Jupiter like albedo of 0.3 and emissivity of 1) we calculate the equilibrium temperature at periastron and apastron, which we list in Table\,\ref{tab:planet_properties}.

\subsection{\name\,b in a population context}
\label{section:pop}
\name\,b is one of the longest period transiting planets ever discovered being in the 97.5th percentile of all transiting planets in terms of orbital period \citep[NASA exoplanet archive;][]{Akeson2013, Christiansen2025}. Within this 97.5th period percentile it is also one of the most amenable to characterisation with its host star belonging to the 90th percentile in V band magnitude. \name\,b is also one of only fifteen planets with orbital periods >100\,d to be discovered by the \tess\ mission at the time of writing, being the 7th longest period of these.

The mass of \name\,b is just within the sparsely populated mass regime between $\sim4$\,\mjup\ and $\sim10$\,\mjup, proposed to represent the transition from core accretion to gravitational instability as the primary mechanism of planet formation \citep{Santos2017, Schlaufman2018, Narang2018}. \name\,b's presence in this mass regime may be more intriguing given the enriched metallicity of its host star since \citet{Narang2018} find the occurrence rate of planets with $\mpl\geq4\mjup$ decreases with increasing host metallicity (although the occurrence of planets with periods longer than 10\,days increases). Additionally the moderate eccentricity of \name\,b ($e$=\be) is consistent with the peak in the eccentricity of super-Jupiters at $e\sim0.3$ identified by \citet{Blunt2026}.

\subsubsection{Archive sample}
\label{section:archive}
\begin{figure}
    \begin{subfigure}{\columnwidth}
        \centering
        \includegraphics[width=\columnwidth]{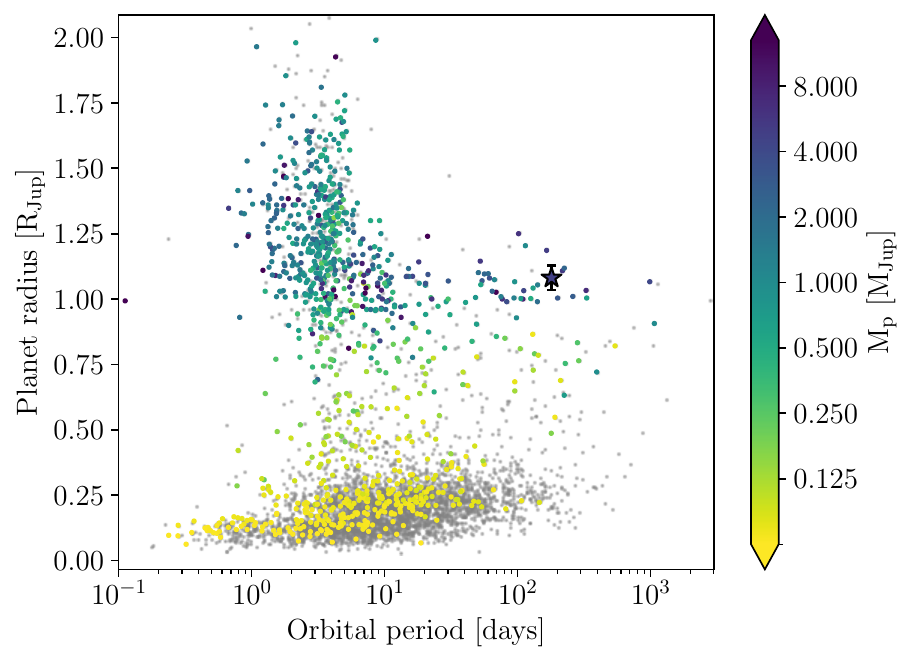}
        \caption{Period-radius}
        \label{fig:pop_plots:period_rad}
    \end{subfigure}
    \begin{subfigure}{\columnwidth}
        \centering
        \includegraphics[width=\columnwidth]{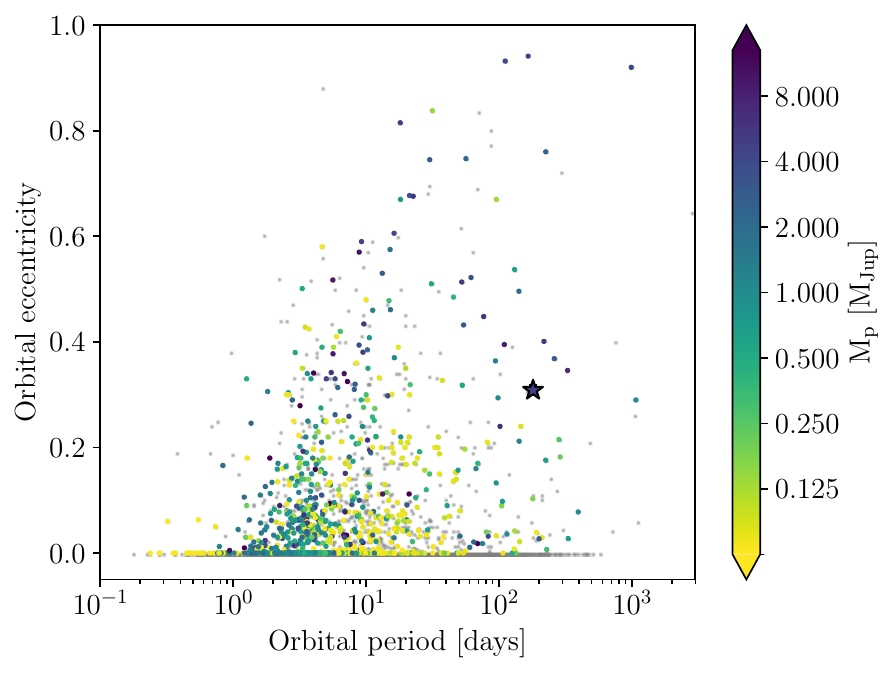}
        \caption{Period-eccentricity}
        \label{fig:pop_plots:period_ecc}
    \end{subfigure}
    \caption{Exoplanet population plots with \name\,b overplotted as a black-bordered star. Orbital period on a base 10 logarithmic scale is plotted against planet radius (top) and eccentricity (bottom). The general transiting sample is shown in grey while the `well-constrained' subset is shown coloured by mass. Both these samples are described in Section\,\ref{section:archive}.}
    \label{fig:pop_plots}
\end{figure}

We retrieved a sample of transiting exoplanets with well defined parameters from the NASA exoplanet archive on 2026 April 28 to compare against \name\,b. We used the planetary systems composite data table, containing 6271 entries, to avoid repeated entries of the same planet. We then eliminated dubious planets by setting the controversial flag to 0 and limited our sample to only transiting planets by setting the detected by transits flag to 1. This left us with 4673 confirmed transiting planets. We then created a subset of this population with `well-constrained' parameters as follows. First we set the limit flags on radius, mass (eliminating M$\sin i$ only constraints) and orbital periods to 0 to only allow constrained values on these parameters. This left us with 1477 planets. We then made a precision cut of 25\% on mass, 10\% on radius and 1\% on orbital period, leaving 968 planets in the final comparison sample. We show \name\,b plotted with this sample in Figure\,\ref{fig:pop_plots}. Of these, \name\,b has the 16th longest period.

\subsection{Interior}
\label{section:interior_results}

\begin{figure}
    \centering
    \includegraphics[width=\columnwidth]{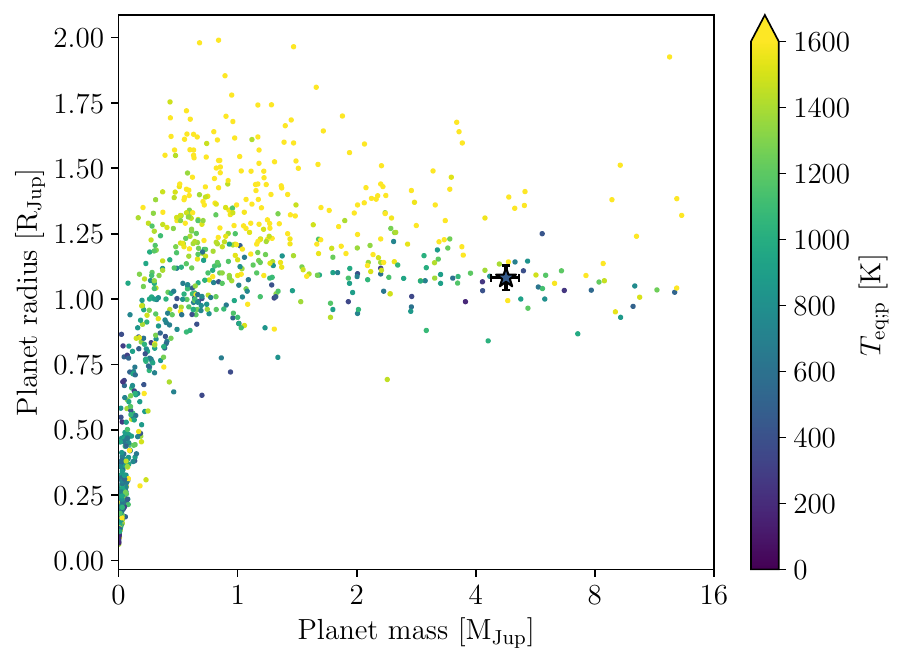}
    \caption{Planet mass in Jupiter masses (on a $\log_2$ scale) plotted against planet radius in Jupiter radii for the well-constrained transiting planet population described in Section\,\ref{section:archive}. Planets are coloured by their equilibrium temperature. \name\,b is shown as a black bordered star with errorbars.} 
    \label{fig:mass-radius}
\end{figure}

As can be seen in Figure\,\ref{fig:mass-radius}, \name\,b appears to have a normal radius and density for a temperate planet of its mass, unaffected by inflation like many hotter planets. 

We used the planetary code \completo\ \citep{Mordasini2012} to estimate the heavy element content of \name\,b. We built a grid of interior models assuming that planets are described with a layered model including a core with a mass comprised between 0 and 10\,\mearth\ and an envelope coupled with a semi-grey atmospheric model. The heavy elements are modelled as water with the AQUA equation of state (EoS) from \citet{Haldemann2020} and they are homogeneously mixed in the H/He envelope are described with EoS from \citet{Chabrier2021}.

The grid of evolution models was then coupled to a Bayesian inference model to retrieve the heavy element content that is compatible with the observed planetary mass, radius, equilibrium temperature, and age of the system. We find that the heavy element mass ($\rm M_{Z}$) is equal to $\rm 110^{+90}_{-80}$ $\rm M_{\oplus}$. The overall stellar metallicity is approximated with a scaling law with the iron abundance $\rm Z_{\star} = 0.0142 \times 10^{[Fe/H]}$ \citep{Asplund2009}. The planetary heavy element enrichment ($\rm Z_{p} / Z_{\star} $, with $\rm Z_{p} = M_{Z} / M_{P} $,) is equal to $2.4^{+2.0}_{-1.7}$. This value is consistent with no metal enrichment of \name\,b relative to its host star at $2\,\sigma$. This result is in line with other high-mass ($\rm M>2\,M_{J}$) and weakly irradiated giant planets orbiting metal-rich stars, such as Kepler-1704\,b \citep{Dalba2021} or TOI-2180 b \citep{Dalba2022}, which also show relatively low bulk metal enrichment.

We note that the results of the interior structure modelling are dependent on several assumptions, in particular regarding the chosen interior structure and equations of state. The planetary metal enrichment can vary depending on these model assumptions but also on the interior modelling codes used to explore the planet's composition.

\section{Discussion}
\label{section:future}

\subsection{Alignment}
\label{section:obliq}
We do not find a photometric rotation period for TIC-\TICID\ / \name\ to compare against the spectroscopic \vsini\ derived upper limit (see Section\,\ref{section:hoststarrot}). This makes any constraints on the alignment of \name\,b impossible without measuring the sky projected obliquity by observing the Rossiter-McLaughlin effect via a spectroscopic transit. If this was done, it would make \name\,b the third longest period planet with a measured obliquity behind HIP 41378 d and f \citep{Grouffal2022, Grouffal2025} and one of only five such systems with periods beyond 100\,d. The planet's moderate eccentricity would also make it stand out among this sample, with the other four systems having either effectively circular or highly eccentric orbits \citep[Source: NASA exoplanet archive;][Accessed 2026/01/16]{Akeson2013, Christiansen2025}. Of additional interest, is the fact that the effective temperature of the host star lies close to the Kraft break \citep{Kraft1967}, which has been linked to a bimodal distribution in observed giant planet obliquities \citep{Winn2010, Albrecht2012, Winn2015, Triaud2018, Albrecht2022, Knudstrup2024, Wang2024}. 

Using equation\,1 from \cite{Triaud2018} we estimate an RM amplitude of \Arm\,\ms\ for \name\,b. This is within the detection limits of high precision radial velocity spectrographs such as the Echelle SPectrograph for Rocky Exoplanets and Stable Spectroscopic Observations \citep[\espresso;][]{Pepe2021} and potentially \harps, albeit marginally. Although, the long transit duration (T$_{14}=$\bTtratot\,hrs) of \name\,b will make such observations challenging from the ground. However, using multiple facilities across the globe it may be possible to obtain an effective observation baseline long enough to view the entire spectroscopic transit. This approach has already been effectively used for RM observations of the long-period planets in the HIP 41378 system \citep{Grouffal2022, Grouffal2025}.

\subsection{Atmosphere}
\label{section:atmosphere}
\begin{figure}
    \centering
    \includegraphics[width=\columnwidth]{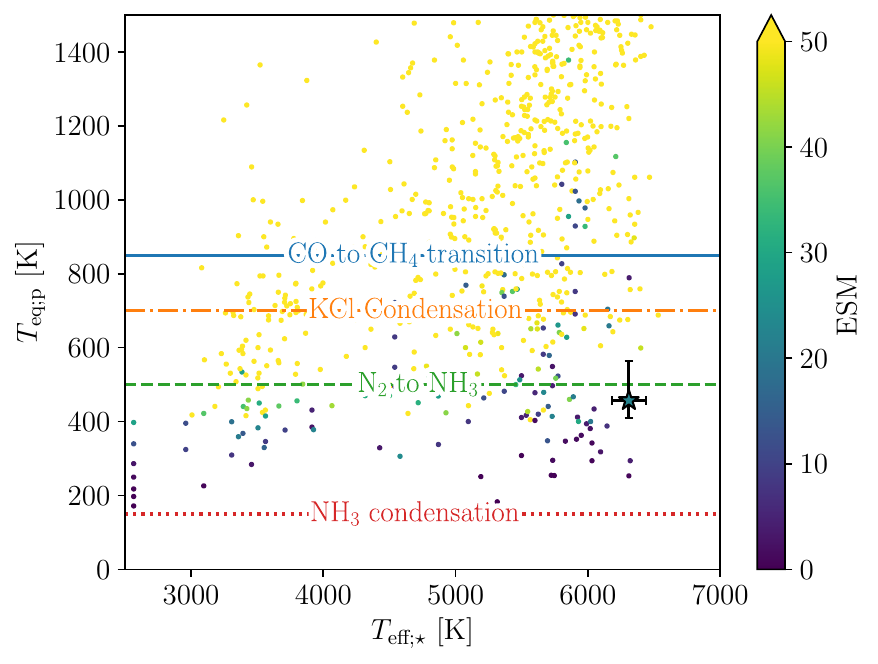}
    \caption{Stellar effective temperature, \teff, plotted against planetary equilibrium temperature, \teq, for the well-constrained transiting planet population described in Section\,\ref{section:archive}. \name\,b is overplotted as a black bordered star with errorbars sized to reflect the range of equilibrium temperatures throughout the planet's orbit. Points are coloured according to their Emission Spectroscopy Metric \citep[ESM;][]{Kempton2018}. Horizontal lines are plotted at 850, 700, 500 and 150 K to indicate the CO to CH$_4$ transition, KCl condensation, the N$_2$ to NH$_3$ transition and NH$_3$ condensation, respectively, for a 1\mjup\,planet \citep{Fortney2020}.}
    \label{fig:Teff-Teq}
\end{figure}

As can be seen in Figure\,\ref{fig:Teff-Teq}, \name\,b has one of the coolest equilibrium temperatures out of our sample of transiting exoplanets with well defined parameters (see Section\,\ref{section:archive}). The \teff\ of its host star is also among the hottest of all the planets with \teq$\lessapprox500$\,K in this sample - potentially leading to unique photochemistry in its upper atmosphere.

Such temperate transiting giant planets allow for the potential detection of molecular nitrogen and therefore measurement of both N/O and C/O ratios, with both of these being key tracers of formation and migration \citep{Oberg2011, Ohno2023a, Ohno2023b}. One key nitrogen species is NH$_3$, which is well documented in temperate Y dwarfs \citep[e.g;][]{Beiler2023, Lew2024}. Comparatively few temperate giant planets have had their atmospheres similarly characterised, although NH$_3$ has been detected in the directly imaged planet HR 8799\,b \citep{Xuan2026}. The exact temperature at which transitions occur between chemical species in exoplanet atmospheres is dependent on surface gravity, pressure and mixing ratios (among other factors). Although the N$_2$ to NH$_3$ transition, where the atmospheric nitrogen content switches between being dominated by diatomic Nitrogen to the much more detectable NH$_3$, should occur at $\sim500$\,K for a gaseous giant planet \citep{Fortney2020}. The equilibrium temperature of \name\,b lies below this transition, making NH$_3$ potentially observable. Although it should be noted that the equilibrium temperature at periastron (\bTeqPeri\,K) is above this transition.

In addition, the wide orbital separation of the planet means the rotation is unlikely to be tidally locked (see Section\,\ref{section:overview}). The lack of a permanent day or night side will almost certainly have significant effects on the heat distribution and circulation dynamics of the planet. Similarly, the lack of tidal locking means the planetary rotation period could be much shorter than most hot-Jupiters, leading to further differences in atmospheric dynamics. For reviews into the atmospheric dynamics of giant planets, including rotational and tidal locking effects see: \cite{Pierrehumbert2019} and \cite{Showman2020}.

To test the observability of \name\,b's atmosphere, we calculate the Transmission Spectroscopy Metric \citep[TSM;][]{Kempton2018} for \name\,b, finding a low value of $\sim2.1$. The planet's high mass results in a low atmospheric scale height and thus lower signal-to-noise compared to a less massive planet. This makes the planet a poor prospect for transmission spectroscopy follow-up.

Because emission SNR is largely independent of planet mass we also investigate emission spectroscopy as an alternative approach to probing the planet's atmosphere. Given the planet's large orbital separation and relatively high transit impact parameter ($b_\text{transit}$=\bbtra) a secondary eclipse would seem relatively unlikely. However, using equation\,8 from \cite{Winn2014}, we find a value of $b_\text{occultation}$=\bbocc, meaning the planet is occulted by its host and emission spectroscopy may be possible. We calculate a value of the Emission Spectroscopy Metric \citep[ESM;][]{Kempton2018} - although we choose to use the equilibrium temperature for \name\,b instead of estimating a daytime temperature, since the planet is extremely unlikely to be tidally locked at its wide orbital separation. Using this method we find a value of $\sim22.5$, which, while modest, is relatively high for a planet with as low an equilibrium temperature as \name\,b (see Figure\,\ref{fig:Teff-Teq}).

\subsection{Additional planetary companions}
\label{section:extra_planets}
Multiplicity is common in exoplanetary systems with 2628 out of 6271 $(\sim42\%)$ confirmed planets being in systems with other planets \citep[Source: NASA exoplanet archive;][Accessed 2026 April 28]{Akeson2013, Christiansen2025}. A number of transiting warm Jupiter systems have already been found to host smaller inner companions \citep[e.g. \ngts-11\,b and c;][]{Gill2020ngts11, anderson2025ngts11ctransitingneptunemass} and warm-Jupiters appear to be more likely to have close planetary neighbours than their hot-Jupiter counterparts \citep{Huang2016}. Given the wide orbital separation of \name\,b it is fairly likely that additional interior planets could have stable orbits. With this in mind we performed a transit least squares search on the \tess\ dataset using the \alles\ package which acts as a wrapper for the \textsc{transitleastsquares} package \citep{Hippke2019tls} alongside \textsc{wotan} \citep{Hippke2019wotan} for detrending. However, this search returned no convincing results. Additionally a generalised Lomb-Scargle periodogram of the RV residuals showed no significant peaks. Thus we detect no additional companions apart from \name\,b, although this does not necessarily rule out the existence of any other companions in the system.

\subsubsection{Sensitivity maps}
\label{section:tiara}

\begin{figure}
    \begin{subfigure}{\columnwidth}
        \centering
        \includegraphics[width=\columnwidth]{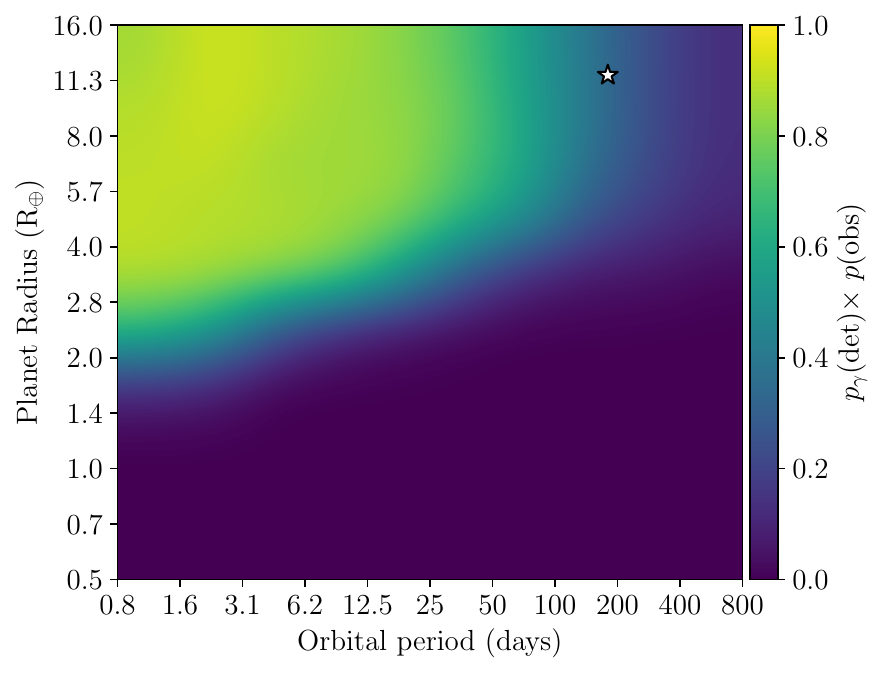}
        \caption{\tess}
        \label{fig:tiara_tess}
    \end{subfigure}
    \begin{subfigure}{\columnwidth}
        \centering
        \includegraphics[width=\columnwidth]{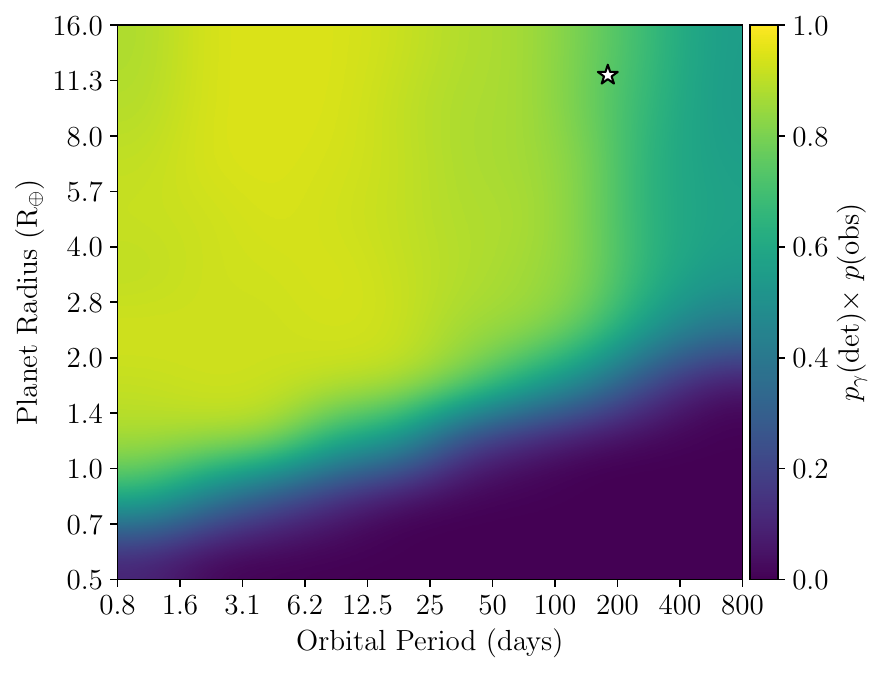}
        \caption{\plato}
        \label{fig:tiara_plato}
    \end{subfigure}
    \caption{\tiara\ sensitivity maps for \name\ in \tess\ (top) and \plato\ (bottom) with \name\,b overplotted as a white star with black borders.}
    \label{fig:tiara}
\end{figure}

To ascertain the detection limits of the \tess\ data for \name\ we made use of the Transit Investigation and Recoverability Application \citep[\tiara;][]{Rodel2024}. \tiara\ takes photometric data as inputs, specifically lightcurve timestamps, contamination values and measured photometric precision with the latter two in this instance being taken from the lightcurve \texttt{fits} file headers 1-\texttt{CROWDSAP} and \texttt{CDPP2\_0}, respectively. The sector 61 lightcurve lacked a recorded noise value so we instead used the sector 34 noise value in its place. \tiara\ then uses the lightcurve properties to calculate the signal-to-noise and detection probability of generated signals producing a sensitivity map which we show with a smoothing function from \cite{Eschen2024} in Figure\,\ref{fig:tiara_tess}. The resulting sensitivity map shows that planets smaller than 2-4\,\rearth (depending on orbital period) are unlikely to be detectable in the \tess\ data, at least at periods shorter than $\sim100$\,days. This allows us to effectively rule out larger transiting companions than this, although, the aforementioned lack of any residual RV signal makes any large non-transiting companions additionally unlikely.

Potential small inner transiting companions around \name\ could be detected with the upcoming PLAnetary Transits and Oscillations of Stars \citep[\plato;][]{Rauer2025} mission. This mission will observe a field spanning $\sim$5\% of the sky. For at least the first two years of its mission, \plato\ will observe a field in the Southern Hemisphere, LOPS2 \citep[][]{Nascimbeni2025}. Unlike \tess\ which observes a target with one of its four cameras, \plato\ has four sets of six cameras that overlap in the centre of its field. Hence a target can be observed with 6, 12, 18 or 24 cameras resulting in a higher precision of the photometric data. This and its longer observations enable \plato\ to detect planets smaller than \tess\ \citep[][]{Rauer2025,Eschen2024}. \name\ lies within the \plato\ LOPS2 field and will be monitored by \plato\ with 6 cameras during its observation of the Southern field. 
These observations will not only enable us to determine the period and ephemeris more precisely but also to search for smaller inner companions. Following \citet{Eschen2024}, we apply \tiara\ to 2 years of simulated \plato\ data using the noise \citep[][]{Börner2024} reported in the \plato\ Input Catalogue \citep[][]{Montalto2021}. These simulations show that \plato\ is predicted to detect potential inner companions down to sub 1\,\rearth\ radii at short periods as we show in Figure\,\ref{fig:tiara_plato}. 

\subsection{Potential for the detection of exomoons and exorings}
\label{section:moonsnrings}

We find no observational evidence of exomoons or exorings around \name\,b. However, its large mass and wide orbital separation make it a potentially high value target in future searches for such circumplanetary systems. In the remainder of this section we assess the ability of the planet to form and retain moons and/or rings and the potential to detect them.

\subsubsection{Formation and retention}
\label{section:moonsnrings:survivability}

The planet's current position is well interior to any primordial ice line of the system making the formation of icy moons or rings like those in the solar system unlikely if the planet formed in situ. If the planet formed beyond the ice line and migrated inwards any icy moons or rings would have to survive this migration. The theoretical lifespans of rings are too short to allow survival of such a ring system to the present. Moons could potentially survive such a migration, although dynamically violent migration mechanisms could potentially be more likely to disrupt any moons present \citep{Gong2013}. Alternatively, moons or rings could be formed from the volatile poor silicate material found in the inner region of the disk.

\cite{Barnes2002} find no meaningful upper limit on the mass of moons able to survive for planets orbiting their hosts beyond $\sim0.6$\,au. \name\,b sits on the edge of this limit with its semi major axis of \baAU\,au, although its moderate eccentricity means the periastron distance falls below this limit. However, based on figure\,2 of \cite{Barnes2002} this still allows the survivability of a $\lessapprox1$\,\mearth\ satellite. 

To further investigate the survivability and possible extent of circumplanetary systems around \name\,b, we calculate the Hill sphere radius as follows:

\begin{equation}
    R_\text{Hill} = a\sqrt\frac{\mpl}{3(\mpl+\mstar)},
    \label{eq:hillradius}
\end{equation}

\noindent finding a value of \hillRJ\,\rjup. This is much smaller than that of Jupiter $(\sim700\,\rjup)$ but still relatively large compared to most transiting exoplanets. \cite{Rosario-Franco2020} find that moons are stable out to $\lessapprox0.3$\,R\textsubscript{Hill}. For \name\,b this is equivalent to a distance of \RosarioRJ\,\rjup. \citet{Domingos2006} similarly find an outer limit of stability for a moon on a circular prograde orbit of $R_\text{Hill}\times0.4895(1.0000-1.0305e)$. For \name\,b we find a value of \DomingosRJ\,\rjup, which we show plotted against host \gaia\ G-band magnitude in Figure\,\ref{fig:moon}. Both these theoretical stable regions are greater than the separation between Jupiter and Callisto, the outermost Galilean moon. This means a Galilean moon system is tentatively within the stability limits of \name\,b, however given the planet's location interior to the ice line it is unclear whether such a system could form.

We also calculate the Roche limit radius for \name\,b as follows:

\begin{equation}
    d_\text{Roche; p} = 2.44\rpl\left(\frac{\rho_\text{b}}{\rho_\text{sat}}\right)^\frac{1}{3},
    \label{eq:roche_lim}
\end{equation}

\noindent where $\rho_\text{sat}$ is the density of a satellite, which we set to 2\gc (similar to many asteroids and planetesimals). We calculate a value of \rocheRJ\,\rjup, which is considerably larger than the limits for Jupiter (2.13\,\rjup) and Saturn (1.39\,\rjup). The Roche limit not only represents the innermost separation any potential exomoon could have without being disrupted but can also inform the extent of possible exorings. The ring systems around giant planets in the solar system tend to extend to 1-2 Roche limit radii, although proposed exoring candidates \citep[e.g. J1407\,b;][]{Kenworthy2015} are much larger and fill the entire hill sphere of their theoretical host planets. However, if such a ring system existed around \name\,b it would have almost certainly already been detectable in the existing \tess\ and \ngts\ photometry.

\subsubsection{Detectability of moons and rings}
\label{section:moonsnrings:detection}

\begin{figure}
    \centering
    \includegraphics[width=\columnwidth]{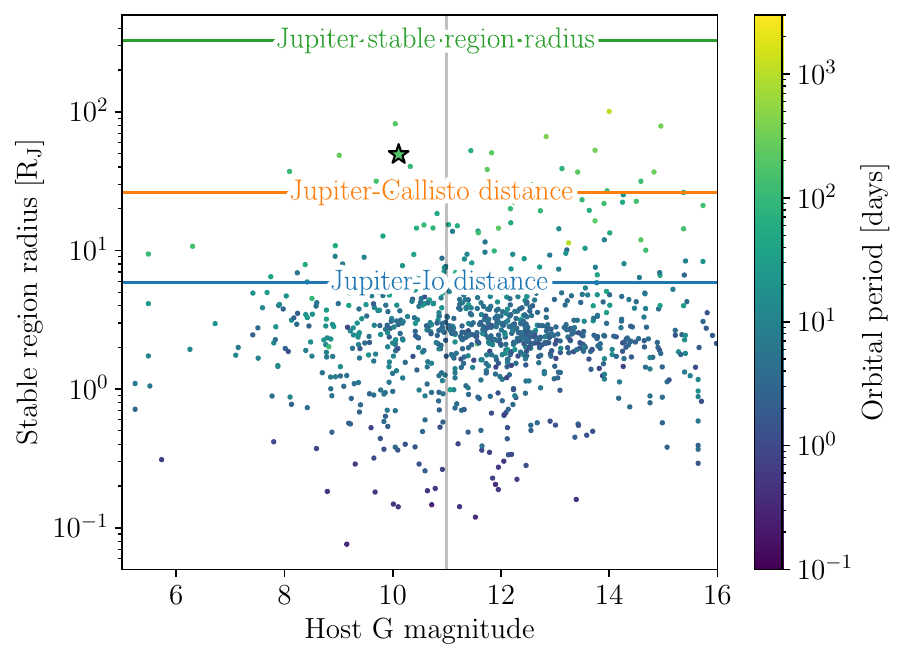}
    \caption{\gaia\ G-band magnitude plotted against stable region radius from \citet{Domingos2006} for all planets in the `well-constrained' sample described in Section\,\ref{section:archive}. \name\,b is plotted as a black bordered star. Points are coloured according to their orbital period. A grey vertical line is plotted at G=11 to represent the brightness cutoff from \citet{Szabo2024}. In addition, horizontal lines are plotted at 329\,\rjup, 26.3\,\rjup and 5.9\,\rjup\ to represent the radius of the stable region for Jupiter, the distance between Jupiter and Callisto (the most widely separated Galilean moon), and the distance between Jupiter and the closest in Galilean moon; Io.}
    \label{fig:moon}
\end{figure}

\citet{Szabo2024} identify planets with periods $>100$\,days transiting stars brighter than a \gaia\ G band magnitude of 11 as high priority targets for exomoon searches. \name\,b comfortably meets both of these criteria (see Figure\,\ref{fig:moon}). We can refine this prioritisation further by only allowing planets where the outermost prograde circular orbit radius from \citet{Domingos2006} is larger than the separation between Jupiter and the outermost Galilean moon (Callisto). This leaves only 6 high priority targets, including \name\,b (see Figure\,\ref{fig:moon}). The other five planets in this regime are HD-114082\,b \citep{Zakhozhay2022}, TIC-172900988\,b \citep{Kostov2021}, TOI-2010\,b \citep{Mann2023}, TOI-2180\,b \citep{Dalba2022} and TOI-4465 \citep{Essack2025}. This makes \name\,b one of the most exciting prospects for future exomoon surveys, although it should be noted that while the brightness of its host star will improve SNR, the large radius of the host will make detection of exomoons via the transit method more difficult. Jupiter's largest moon; Ganymede, would have a transit depth of just over 14\,ppm across the sun and only $\sim4$\,ppm for \name. 

We also considered the potential TTV signal of a Ganymede and an Earth mass moon on a circular orbit around \name\,b at equal separation to that between Ganymede and Jupiter. We estimated the RMS amplitude of a TTV signal for both these cases using equation\,3 from \citet{Kipping2009}, finding signals of $0.39\pm0.04$\,s and $15.7\pm1.5$\,s respectively. A signal this small would be extremely challenging to detect, making TTV exomoon detection for \name\, b unlikely.

It could be possible to detect exorings around the planet by searching for abnormalities in the planet's transit shape or reflected light signals \citep{Barnes2004, Aizawa2017, Akinsanmi2018, Heller2018}. We used the \exorings\ software package \citep{Kenworthy_software, Kenworthy2015} to model an exoring with a radius equal to $d_\text{Roche}$ around \name\,b, and find that this model is indistinguishable from a ringless transiting planet in \tess. However, with the greater photometric precision available from instruments such as \plato, \jwst, or the \elt\ it may be possible to detect such a system. Alternatively a spectroscopic transit observation of the RM effect, which would already be of scientific interest for constraining the orbital obliquity of the system (see Section\,\ref{section:obliq}), could be used to detect an exoring system \citep{deMooij2017}. 

\section{Summary and Conclusions}
\label{section:conclusion}
TIC-\TICID\,b\,/\,\name\,b was first detected from a single \tess\ transit in Sector 33. We then initiated parallel campaigns of spectroscopic and photometric follow-up that yielded an additional transit egress from \ngts\ and a full radial velocity solution from \harps\ and \coralie. The spectra obtained also allowed a characterisation of the host star, revealing it to have a temperature consistent with an F6V-F7V type star (\hostteff\,K), and a low surface gravity (\logg=\hostlogg) resulting from a large radius (\hostrad\,\rsun) compared to its mass (\hostmass\,\msun). This due to the enhanced metalicity (\feh=\hostfeh\,`dex') of the star. A combined fit with the photometric and spectroscopic data confirmed the planet to be a \bMcompanionMjup\,\mjup, \bRcompanionRjup\,\rjup\ giant planet on a \bperiod\,d orbit with a moderate eccentricity of \be.

The planet's wide orbital separation gives it a low equilibrium temperature compared to other transiting planets, varying between \bTeqPeri\,K and \bTeqAp\,K throughout its eccentric orbit. On account of this cooler temperature, the atmosphere is likely to contain molecular nitrogen species not found in the atmospheres of hotter giant planets. In addition, this wide separation also makes \name\,b one of the most likely transiting planets to host stable exomoon and/or ring systems which could be detectable with targeted observations from high precision facilities like \plato, \jwst\ or the \elt.

While the planet is unsuitable for transmission spectroscopy (TSM$\sim2.1$) with current generation instruments, there is a better chance of detecting an atmosphere through emission spectroscopy of the secondary eclipse (ESM$\sim22.5$). However, perhaps the most pertinent additional observation of \name\,b would be to constrain its orbital obliquity through spectroscopic transit observations. This will require extensive international collaboration to obtain longitudinal coverage from multiple facilities worldwide to observe the full $\sim15$\,hr transit and baseline. Although, with an estimated RM amplitude of \Arm\,\ms\ this should be possible.

\name\,b's position within the LOPS2 field of the upcoming \plato\ mission allows for the potential discovery of additional companions within the system. Either by directly detecting additional companions with transits too shallow and/or infrequent to be seen in current datasets or inferring outer companions through measuring transit timing variations of \name\,b with its long baseline. 

\section*{Acknowledgements}

This manuscript has been prepared independently from and simultaneously with a paper on the same system from the Warm gIaNts with tEss (WINE) team who detected the same planet in a separate set of follow-up observations. We would like to extend our thanks to Felipe Rojas, Dr Rafael Brahm and the rest of the WINE team for agreeing to separate independent publications and for coordinating submission with us.

The author list and affiliations in this paper were created using \textsc{PyTeXAuthors}. The code is available at \url{https://github.com/TobyRodel/pytexauthors}.

This paper includes data collected by the \TESS\ mission. Funding for the \TESS\ mission is provided by the NASA Explorer Program. Resources supporting this work were provided by the NASA High-End Computing (HEC) Program through the NASA Advanced Supercomputing (NAS) Division at Ames Research Center for the production of the SPOC data products. The \tess\ team shall assure that the masses of fifty (50) planets with radii less than 4\rearth\ are determined.

We acknowledge the use of public \TESS\ Alert data from pipelines at the \TESS\ Science Office and at the \TESS\ Science Processing Operations Center.

This work makes use of data from the European Space Agency (ESA) mission \gaia\ (\url{https://www.cosmos.esa.int/gaia}), processed by the \gaia\ Data Processing and Analysis Consortium (DPAC, \url{https://www.cosmos.esa.int/web/gaia/dpac/consortium}). Funding for the DPAC has been provided by national institutions, in particular the institutions participating in the \gaia\ Multilateral Agreement.

This research makes use of the Exoplanet Follow-up Observation Program website, which is operated by the California Institute of Technology, under contract with the National Aeronautics and Space Administration under the Exoplanet Exploration Program.

This paper includes data collected by the TESS mission that are publicly available from the Mikulski Archive for Space Telescopes (MAST).

This work is based in part on data collected under the NGTS project at the ESO La Silla Paranal Observatory. The NGTS facility is operated by a consortium of institutes with support from the UK Science and Technology Facilities Council (STFC) under projects ST/M001962/1, ST/S002642/1 and ST/W003163/1.


This work makes use of observations from the LCOGT network. Part of the LCOGT telescope time was granted by NOIRLab through the Mid-Scale Innovations Program (MSIP). MSIP is funded by NSF.


This paper uses observations made with the Las Cumbres Observatory’s education network telescopes that were upgraded through generous support from the Gordon and Betty Moore Foundation.


This research has made use of the NASA Exoplanet Archive, which is operated by the California Institute of Technology, under contract with the National Aeronautics and Space Administration under the Exoplanet Exploration Program.

TR is supported by an STFC studentship. C.A.W., E.dM., M.E.Y and J.C.C would like to acknowledge support from the UK Science and Technology Facilities Council (STFC, grant number ST/X00094X/1). JSJ gratefully acknowledges support by FONDECYT grant 1240738 and from the ANID BASAL project FB210003. ML acknowledges support of the Swiss National Science Foundation under grant number PCEFP2\_194576. The contributions of ML, FB, SU and SUM have been carried out within the framework of the NCCR PlanetS supported by the Swiss National Science Foundation under grants 51NF40\_182901 and 51NF40\_205606. D.D., Z.E. and B.S. acknowledge support from the TESS Guest Investigator Program grant 80NSSC23K0769. D.D. also acknowledges support from TESS Guest Investigator Program grant 80NSSC22K1353. A.M. acknowledges funding from a UKRI Future Leader Fellowship, grant number MR/X033244/1 and a UK Science and Technology Facilities Council (STFC) small grant ST/Y002334/1. E.G. gratefully acknowledges support from UK Research and Innovation (UKRI) under the UK government’s Horizon Europe funding guarantee for an ERC Starting Grant [grant number EP/Z000890/1]. Funding for K.B. was provided by the European Union (ERC AdG SUBSTELLAR, GA 101054354). A.N. acknowledges support from the Swiss National Science Foundation (SNSF) under grant PZ00P2\_208945.

\section*{Data Availability}

The \tess\ data is accessible via the MAST (Mikulski Archive for Space Telescopes) portal at \url{https://mast.stsci.edu/portal/Mashup/Clients/Mast/Portal.html}.




Any code used for analysis or in producing the plots in this paper can be made available upon reasonable request to the author(s).

\section*{Affiliations}
$^{1}$Astrophysics Research Centre, School of Mathematics and Physics, Queen’s University Belfast, Belfast, BT7 1NN, UK\\
$^{2}$Leiden Observatory, Leiden University, P.O. Box 9513, 2300 RA Leiden, The Netherlands\\
$^{3}$Observatoire de Gen\`{e}ve, Universit\'{e} de Gen\`{e}ve, 51 Ch. des Maillettes, CH-1290 Sauverny, Switzerland\\
$^{4}$Department of Physics, University of Warwick, Gibbet Hill Road, Coventry CV4 7AL, UK\\
$^{5}$Centre for Exoplanets and Habitability, University of Warwick, Gibbet Hill Road, Coventry CV4 7AL, UK\\
$^{6}$Isaac Newton Group of Telescopes, Apartado de correos 321, E-38700 Santa Cruz de La Palma, Canary Islands; Spain\\
$^{7}$School of Physics \& Astronomy, University of Birmingham, Edgbaston, Birmingham B15 2TT, UK\\
$^{8}$Center for Astrophysics \textbar \ Harvard \& Smithsonian, 60 Garden Street, Cambridge, MA 02138, USA\\
$^{9}$Department of Physics and Astronomy, The University of New Mexico, 210 Yale Blvd NE, Albuquerque, NM 87106, USA\\
$^{10}$Instituto de Astronom\'{i}a, Universidad Cat\'{o}lica del Norte, Angamos 0610, 1270709, Antofagasta, Chile\\
$^{11}$Instituto de Astrof\'{i}sica de Canarias, c/ V\'{i}a L\'{a}ctea s/n, 38205 La Laguna, Tenerife, Spain\\
$^{12}$Astrobiology Research Unit, Université de Liège, 19C Allée du 6 Août, 4000 Liège, Belgium\\
$^{13}$Department of Earth, Atmospheric and Planetary Science, Massachusetts Institute of Technology, 77 Massachusetts Avenue, Cambridge, MA 02139, USA\\
$^{14}$Astronomy Unit, Queen Mary University of London, Mile End Road, London E1 4NS, UK\\
$^{15}$School of Physics and Astronomy, University of Leicester, Leicester LE1 7RH, UK\\
$^{16}$El Sauce Observatory, Coquimbo Province, Chile\\
$^{17}$Rugby School, Lawrence Sheriff Street, Rugby, Warwickshire CV22 5EH, UK\\
$^{18}$Sapienza Universit\`{a} di Roma, Piazzale Aldo Moro, 5, 00185, Rome (RM), Italy\\
$^{19}$Campo Catino Astronomical Observatory, Regione Lazio, Guarcino (FR), 03010 Italy\\
$^{20}$INFN Sezione Roma1, Piazzale Aldo Moro, 2, 00185, Rome (RM), Italy\\
$^{21}$INAF OAC, Via della Scienza, 5, 09047, Selargius (CA), Italy\\
$^{22}$Instituto de Estudios Astrofísicos, Facultad de Ingeniería y Ciencias, Universidad Diego Portales, Av. Ejército Libertador 441, Santiago, Chile\\
$^{23}$Centro de Excelencia en Astrofísica y Tecnologías Afines (CATA), Camino El Observatorio 1515, Las Condes, Santiago, Chile\\
$^{24}$Columbia University, 550 W 120th Street, New York NY 10027, USA\\
$^{25}$National Research Council Canada, Herzberg Astronomy \& Astrophysics Research Centre, 5071 West Saanich Road, Victoria, BC V9E 2E7, Canada\\
$^{26}$Subaru Telescope, National Astronomical Observatory of Japan, 650 North A‘ohoku Place, Hilo, HI 96720, USA\\
$^{27}$Departamento de Astrof\'{i}sica, Universidad de La Laguna, 38206 La Laguna, Tenerife, Spain\\
$^{28}$Hamburger Sternwarte, Gojenbergsweg 112, 21029 Hamburg, Germany\\
$^{29}$Massachusetts Institute of Technology, Department of Physics and Kavli Institute for Astrophysics and Space Research, Cambridge, MA 02139, USA\\
$^{30}$Centre for Space Domain Awareness, University of Warwick, Gibbet Hill Road, Coventry CV4 7AL, UK\\
$^{31}$American Association of Variable Star Observers, 49 Bay State Road, Cambridge, MA 02138, USA\\
$^{32}$NASA Goddard Space Flight Center, 8800 Greenbelt Road, Greenbelt, MD 20771, USA\\
$^{33}$South African Astronomical Observatory, P.O Box 9, Observatory 7935, Cape Town, South Africa\\



\bibliographystyle{mnras}
\bibliography{refs} 

@ARTICLE{Feinstein2019,
       author = {{Feinstein}, Adina D. and {Montet}, Benjamin T. and {Foreman-Mackey}, Daniel and {Bedell}, Megan E. and {Saunders}, Nicholas and {Bean}, Jacob L. and {Christiansen}, Jessie L. and {Hedges}, Christina and {Luger}, Rodrigo and {Scolnic}, Daniel and {Cardoso}, Jos{\'e} Vin{\'\i}cius de Miranda},
        title = "{eleanor: An Open-source Tool for Extracting Light Curves from the TESS Full-frame Images}",
      journal = {\pasp},
         year = 2019,
        month = sep,
       volume = {131},
       number = {1003},
        pages = {094502},
          doi = {10.1088/1538-3873/ab291c},
archivePrefix = {arXiv},
       eprint = {1903.09152},
 primaryClass = {astro-ph.IM},
       adsurl = {https://ui.adsabs.harvard.edu/abs/2019PASP..131i4502F}
}

@ARTICLE{AstropyCollab2022,
       author = {{Astropy Collaboration} and {Price-Whelan}, Adrian M. and {Lim}, Pey Lian and {Earl}, Nicholas and {Starkman}, Nathaniel and {Bradley}, Larry and {Shupe}, David L. and {Patil}, Aarya A. and {Corrales}, Lia and {Brasseur}, C.~E. and {N{\"o}the}, Maximilian and {Donath}, Axel and {Tollerud}, Erik and {Morris}, Brett M. and {Ginsburg}, Adam and {Vaher}, Eero and {Weaver}, Benjamin A. and {Tocknell}, James and {Jamieson}, William and {van Kerkwijk}, Marten H. and {Robitaille}, Thomas P. and {Merry}, Bruce and {Bachetti}, Matteo and {G{\"u}nther}, H. Moritz and {Aldcroft}, Thomas L. and {Alvarado-Montes}, Jaime A. and {Archibald}, Anne M. and {B{\'o}di}, Attila and {Bapat}, Shreyas and {Barentsen}, Geert and {Baz{\'a}n}, Juanjo and {Biswas}, Manish and {Boquien}, M{\'e}d{\'e}ric and {Burke}, D.~J. and {Cara}, Daria and {Cara}, Mihai and {Conroy}, Kyle E. and {Conseil}, Simon and {Craig}, Matthew W. and {Cross}, Robert M. and {Cruz}, Kelle L. and {D'Eugenio}, Francesco and {Dencheva}, Nadia and {Devillepoix}, Hadrien A.~R. and {Dietrich}, J{\"o}rg P. and {Eigenbrot}, Arthur Davis and {Erben}, Thomas and {Ferreira}, Leonardo and {Foreman-Mackey}, Daniel and {Fox}, Ryan and {Freij}, Nabil and {Garg}, Suyog and {Geda}, Robel and {Glattly}, Lauren and {Gondhalekar}, Yash and {Gordon}, Karl D. and {Grant}, David and {Greenfield}, Perry and {Groener}, Austen M. and {Guest}, Steve and {Gurovich}, Sebastian and {Handberg}, Rasmus and {Hart}, Akeem and {Hatfield-Dodds}, Zac and {Homeier}, Derek and {Hosseinzadeh}, Griffin and {Jenness}, Tim and {Jones}, Craig K. and {Joseph}, Prajwel and {Kalmbach}, J. Bryce and {Karamehmetoglu}, Emir and {Ka{\l}uszy{\'n}ski}, Miko{\l}aj and {Kelley}, Michael S.~P. and {Kern}, Nicholas and {Kerzendorf}, Wolfgang E. and {Koch}, Eric W. and {Kulumani}, Shankar and {Lee}, Antony and {Ly}, Chun and {Ma}, Zhiyuan and {MacBride}, Conor and {Maljaars}, Jakob M. and {Muna}, Demitri and {Murphy}, N.~A. and {Norman}, Henrik and {O'Steen}, Richard and {Oman}, Kyle A. and {Pacifici}, Camilla and {Pascual}, Sergio and {Pascual-Granado}, J. and {Patil}, Rohit R. and {Perren}, Gabriel I. and {Pickering}, Timothy E. and {Rastogi}, Tanuj and {Roulston}, Benjamin R. and {Ryan}, Daniel F. and {Rykoff}, Eli S. and {Sabater}, Jose and {Sakurikar}, Parikshit and {Salgado}, Jes{\'u}s and {Sanghi}, Aniket and {Saunders}, Nicholas and {Savchenko}, Volodymyr and {Schwardt}, Ludwig and {Seifert-Eckert}, Michael and {Shih}, Albert Y. and {Jain}, Anany Shrey and {Shukla}, Gyanendra and {Sick}, Jonathan and {Simpson}, Chris and {Singanamalla}, Sudheesh and {Singer}, Leo P. and {Singhal}, Jaladh and {Sinha}, Manodeep and {Sip{\H{o}}cz}, Brigitta M. and {Spitler}, Lee R. and {Stansby}, David and {Streicher}, Ole and {{\v{S}}umak}, Jani and {Swinbank}, John D. and {Taranu}, Dan S. and {Tewary}, Nikita and {Tremblay}, Grant R. and {de Val-Borro}, Miguel and {Van Kooten}, Samuel J. and {Vasovi{\'c}}, Zlatan and {Verma}, Shresth and {de Miranda Cardoso}, Jos{\'e} Vin{\'\i}cius and {Williams}, Peter K.~G. and {Wilson}, Tom J. and {Winkel}, Benjamin and {Wood-Vasey}, W.~M. and {Xue}, Rui and {Yoachim}, Peter and {Zhang}, Chen and {Zonca}, Andrea and {Astropy Project Contributors}},
        title = "{The Astropy Project: Sustaining and Growing a Community-oriented Open-source Project and the Latest Major Release (v5.0) of the Core Package}",
      journal = {\apj},
         year = 2022,
        month = aug,
       volume = {935},
       number = {2},
          eid = {167},
        pages = {167},
          doi = {10.3847/1538-4357/ac7c74},
archivePrefix = {arXiv},
       eprint = {2206.14220},
 primaryClass = {astro-ph.IM},
       adsurl = {https://ui.adsabs.harvard.edu/abs/2022ApJ...935..167A}
}

@ARTICLE{gaia2021dr3,
       author = {{Gaia Collaboration} and {Brown}, A.~G.~A. and {Vallenari}, A. and {Prusti}, T. and {de Bruijne}, J.~H.~J. and {Babusiaux}, C. and {Biermann}, M. and {Creevey}, O.~L. and {Evans}, D.~W. and {Eyer}, L. and {Hutton}, A. and {Jansen}, F. and {Jordi}, C. and {Klioner}, S.~A. and {Lammers}, U. and {Lindegren}, L. and {Luri}, X. and {Mignard}, F. and {Panem}, C. and {Pourbaix}, D. and {Randich}, S. and {Sartoretti}, P. and {Soubiran}, C. and {Walton}, N.~A. and {Arenou}, F. and {Bailer-Jones}, C.~A.~L. and {Bastian}, U. and {Cropper}, M. and {Drimmel}, R. and {Katz}, D. and {Lattanzi}, M.~G. and {van Leeuwen}, F. and {Bakker}, J. and {Cacciari}, C. and {Casta{\~n}eda}, J. and {De Angeli}, F. and {Ducourant}, C. and {Fabricius}, C. and {Fouesneau}, M. and {Fr{\'e}mat}, Y. and {Guerra}, R. and {Guerrier}, A. and {Guiraud}, J. and {Jean-Antoine Piccolo}, A. and {Masana}, E. and {Messineo}, R. and {Mowlavi}, N. and {Nicolas}, C. and {Nienartowicz}, K. and {Pailler}, F. and {Panuzzo}, P. and {Riclet}, F. and {Roux}, W. and {Seabroke}, G.~M. and {Sordo}, R. and {Tanga}, P. and {Th{\'e}venin}, F. and {Gracia-Abril}, G. and {Portell}, J. and {Teyssier}, D. and {Altmann}, M. and {Andrae}, R. and {Bellas-Velidis}, I. and {Benson}, K. and {Berthier}, J. and {Blomme}, R. and {Brugaletta}, E. and {Burgess}, P.~W. and {Busso}, G. and {Carry}, B. and {Cellino}, A. and {Cheek}, N. and {Clementini}, G. and {Damerdji}, Y. and {Davidson}, M. and {Delchambre}, L. and {Dell'Oro}, A. and {Fern{\'a}ndez-Hern{\'a}ndez}, J. and {Galluccio}, L. and {Garc{\'\i}a-Lario}, P. and {Garcia-Reinaldos}, M. and {Gonz{\'a}lez-N{\'u}{\~n}ez}, J. and {Gosset}, E. and {Haigron}, R. and {Halbwachs}, J. -L. and {Hambly}, N.~C. and {Harrison}, D.~L. and {Hatzidimitriou}, D. and {Heiter}, U. and {Hern{\'a}ndez}, J. and {Hestroffer}, D. and {Hodgkin}, S.~T. and {Holl}, B. and {Jan{\ss}en}, K. and {Jevardat de Fombelle}, G. and {Jordan}, S. and {Krone-Martins}, A. and {Lanzafame}, A.~C. and {L{\"o}ffler}, W. and {Lorca}, A. and {Manteiga}, M. and {Marchal}, O. and {Marrese}, P.~M. and {Moitinho}, A. and {Mora}, A. and {Muinonen}, K. and {Osborne}, P. and {Pancino}, E. and {Pauwels}, T. and {Petit}, J. -M. and {Recio-Blanco}, A. and {Richards}, P.~J. and {Riello}, M. and {Rimoldini}, L. and {Robin}, A.~C. and {Roegiers}, T. and {Rybizki}, J. and {Sarro}, L.~M. and {Siopis}, C. and {Smith}, M. and {Sozzetti}, A. and {Ulla}, A. and {Utrilla}, E. and {van Leeuwen}, M. and {van Reeven}, W. and {Abbas}, U. and {Abreu Aramburu}, A. and {Accart}, S. and {Aerts}, C. and {Aguado}, J.~J. and {Ajaj}, M. and {Altavilla}, G. and {{\'A}lvarez}, M.~A. and {{\'A}lvarez Cid-Fuentes}, J. and {Alves}, J. and {Anderson}, R.~I. and {Anglada Varela}, E. and {Antoja}, T. and {Audard}, M. and {Baines}, D. and {Baker}, S.~G. and {Balaguer-N{\'u}{\~n}ez}, L. and {Balbinot}, E. and {Balog}, Z. and {Barache}, C. and {Barbato}, D. and {Barros}, M. and {Barstow}, M.~A. and {Bartolom{\'e}}, S. and {Bassilana}, J. -L. and {Bauchet}, N. and {Baudesson-Stella}, A. and {Becciani}, U. and {Bellazzini}, M. and {Bernet}, M. and {Bertone}, S. and {Bianchi}, L. and {Blanco-Cuaresma}, S. and {Boch}, T. and {Bombrun}, A. and {Bossini}, D. and {Bouquillon}, S. and {Bragaglia}, A. and {Bramante}, L. and {Breedt}, E. and {Bressan}, A. and {Brouillet}, N. and {Bucciarelli}, B. and {Burlacu}, A. and {Busonero}, D. and {Butkevich}, A.~G. and {Buzzi}, R. and {Caffau}, E. and {Cancelliere}, R. and {C{\'a}novas}, H. and {Cantat-Gaudin}, T. and {Carballo}, R. and {Carlucci}, T. and {Carnerero}, M.~I. and {Carrasco}, J.~M. and {Casamiquela}, L. and {Castellani}, M. and {Castro-Ginard}, A. and {Castro Sampol}, P. and {Chaoul}, L. and {Charlot}, P. and {Chemin}, L. and {Chiavassa}, A. and {Cioni}, M. -R.~L. and {Comoretto}, G. and {Cooper}, W.~J. and {Cornez}, T. and {Cowell}, S. and {Crifo}, F. and {Crosta}, M. and {Crowley}, C. and {Dafonte}, C. and {Dapergolas}, A. and {David}, M. and {David}, P. and {de Laverny}, P. and {De Luise}, F. and {De March}, R. and {De Ridder}, J. and {de Souza}, R. and {de Teodoro}, P. and {de Torres}, A. and {del Peloso}, E.~F. and {del Pozo}, E. and {Delbo}, M. and {Delgado}, A. and {Delgado}, H.~E. and {Delisle}, J. -B. and {Di Matteo}, P. and {Diakite}, S. and {Diener}, C. and {Distefano}, E. and {Dolding}, C. and {Eappachen}, D. and {Edvardsson}, B. and {Enke}, H. and {Esquej}, P. and {Fabre}, C. and {Fabrizio}, M. and {Faigler}, S. and {Fedorets}, G. and {Fernique}, P. and {Fienga}, A. and {Figueras}, F. and {Fouron}, C. and {Fragkoudi}, F. and {Fraile}, E. and {Franke}, F. and {Gai}, M. and {Garabato}, D. and {Garcia-Gutierrez}, A. and {Garc{\'\i}a-Torres}, M. and {Garofalo}, A. and {Gavras}, P. and {Gerlach}, E. and {Geyer}, R. and {Giacobbe}, P. and {Gilmore}, G. and {Girona}, S. and {Giuffrida}, G. and {Gomel}, R. and {Gomez}, A. and {Gonzalez-Santamaria}, I. and {Gonz{\'a}lez-Vidal}, J.~J. and {Granvik}, M. and {Guti{\'e}rrez-S{\'a}nchez}, R. and {Guy}, L.~P. and {Hauser}, M. and {Haywood}, M. and {Helmi}, A. and {Hidalgo}, S.~L. and {Hilger}, T. and {H{\l}adczuk}, N. and {Hobbs}, D. and {Holland}, G. and {Huckle}, H.~E. and {Jasniewicz}, G. and {Jonker}, P.~G. and {Juaristi Campillo}, J. and {Julbe}, F. and {Karbevska}, L. and {Kervella}, P. and {Khanna}, S. and {Kochoska}, A. and {Kontizas}, M. and {Kordopatis}, G. and {Korn}, A.~J. and {Kostrzewa-Rutkowska}, Z. and {Kruszy{\'n}ska}, K. and {Lambert}, S. and {Lanza}, A.~F. and {Lasne}, Y. and {Le Campion}, J. -F. and {Le Fustec}, Y. and {Lebreton}, Y. and {Lebzelter}, T. and {Leccia}, S. and {Leclerc}, N. and {Lecoeur-Taibi}, I. and {Liao}, S. and {Licata}, E. and {Lindstr{\o}m}, E.~P. and {Lister}, T.~A. and {Livanou}, E. and {Lobel}, A. and {Madrero Pardo}, P. and {Managau}, S. and {Mann}, R.~G. and {Marchant}, J.~M. and {Marconi}, M. and {Marcos Santos}, M.~M.~S. and {Marinoni}, S. and {Marocco}, F. and {Marshall}, D.~J. and {Martin Polo}, L. and {Mart{\'\i}n-Fleitas}, J.~M. and {Masip}, A. and {Massari}, D. and {Mastrobuono-Battisti}, A. and {Mazeh}, T. and {McMillan}, P.~J. and {Messina}, S. and {Michalik}, D. and {Millar}, N.~R. and {Mints}, A. and {Molina}, D. and {Molinaro}, R. and {Moln{\'a}r}, L. and {Montegriffo}, P. and {Mor}, R. and {Morbidelli}, R. and {Morel}, T. and {Morris}, D. and {Mulone}, A.~F. and {Munoz}, D. and {Muraveva}, T. and {Murphy}, C.~P. and {Musella}, I. and {Noval}, L. and {Ord{\'e}novic}, C. and {Orr{\`u}}, G. and {Osinde}, J. and {Pagani}, C. and {Pagano}, I. and {Palaversa}, L. and {Palicio}, P.~A. and {Panahi}, A. and {Pawlak}, M. and {Pe{\~n}alosa Esteller}, X. and {Penttil{\"a}}, A. and {Piersimoni}, A.~M. and {Pineau}, F. -X. and {Plachy}, E. and {Plum}, G. and {Poggio}, E. and {Poretti}, E. and {Poujoulet}, E. and {Pr{\v{s}}a}, A. and {Pulone}, L. and {Racero}, E. and {Ragaini}, S. and {Rainer}, M. and {Raiteri}, C.~M. and {Rambaux}, N. and {Ramos}, P. and {Ramos-Lerate}, M. and {Re Fiorentin}, P. and {Regibo}, S. and {Reyl{\'e}}, C. and {Ripepi}, V. and {Riva}, A. and {Rixon}, G. and {Robichon}, N. and {Robin}, C. and {Roelens}, M. and {Rohrbasser}, L. and {Romero-G{\'o}mez}, M. and {Rowell}, N. and {Royer}, F. and {Rybicki}, K.~A. and {Sadowski}, G. and {Sagrist{\`a} Sell{\'e}s}, A. and {Sahlmann}, J. and {Salgado}, J. and {Salguero}, E. and {Samaras}, N. and {Sanchez Gimenez}, V. and {Sanna}, N. and {Santove{\~n}a}, R. and {Sarasso}, M. and {Schultheis}, M. and {Sciacca}, E. and {Segol}, M. and {Segovia}, J.~C. and {S{\'e}gransan}, D. and {Semeux}, D. and {Shahaf}, S. and {Siddiqui}, H.~I. and {Siebert}, A. and {Siltala}, L. and {Slezak}, E. and {Smart}, R.~L. and {Solano}, E. and {Solitro}, F. and {Souami}, D. and {Souchay}, J. and {Spagna}, A. and {Spoto}, F. and {Steele}, I.~A. and {Steidelm{\"u}ller}, H. and {Stephenson}, C.~A. and {S{\"u}veges}, M. and {Szabados}, L. and {Szegedi-Elek}, E. and {Taris}, F. and {Tauran}, G. and {Taylor}, M.~B. and {Teixeira}, R. and {Thuillot}, W. and {Tonello}, N. and {Torra}, F. and {Torra}, J. and {Turon}, C. and {Unger}, N. and {Vaillant}, M. and {van Dillen}, E. and {Vanel}, O. and {Vecchiato}, A. and {Viala}, Y. and {Vicente}, D. and {Voutsinas}, S. and {Weiler}, M. and {Wevers}, T. and {Wyrzykowski}, {\L}. and {Yoldas}, A. and {Yvard}, P. and {Zhao}, H. and {Zorec}, J. and {Zucker}, S. and {Zurbach}, C. and {Zwitter}, T.},
        title = "{Gaia Early Data Release 3. Summary of the contents and survey properties}",
      journal = {\aap},
         year = 2021,
        month = may,
       volume = {649},
          eid = {A1},
        pages = {A1},
          doi = {10.1051/0004-6361/202039657},
archivePrefix = {arXiv},
       eprint = {2012.01533},
 primaryClass = {astro-ph.GA},
       adsurl = {https://ui.adsabs.harvard.edu/abs/2021A&A...649A...1G}
}

@ARTICLE{skrutskie20062mass,
       author = {{Skrutskie}, M.~F. and {Cutri}, R.~M. and {Stiening}, R. and {Weinberg}, M.~D. and {Schneider}, S. and {Carpenter}, J.~M. and {Beichman}, C. and {Capps}, R. and {Chester}, T. and {Elias}, J. and {Huchra}, J. and {Liebert}, J. and {Lonsdale}, C. and {Monet}, D.~G. and {Price}, S. and {Seitzer}, P. and {Jarrett}, T. and {Kirkpatrick}, J.~D. and {Gizis}, J.~E. and {Howard}, E. and {Evans}, T. and {Fowler}, J. and {Fullmer}, L. and {Hurt}, R. and {Light}, R. and {Kopan}, E.~L. and {Marsh}, K.~A. and {McCallon}, H.~L. and {Tam}, R. and {Van Dyk}, S. and {Wheelock}, S.},
        title = "{The Two Micron All Sky Survey (2MASS)}",
      journal = {\aj},
         year = 2006,
        month = feb,
       volume = {131},
       number = {2},
        pages = {1163-1183},
          doi = {10.1086/498708},
       adsurl = {https://ui.adsabs.harvard.edu/abs/2006AJ....131.1163S}
}

@ARTICLE{Stassun2019tic,
       author = {{Stassun}, Keivan G. and {Oelkers}, Ryan J. and {Paegert}, Martin and {Torres}, Guillermo and {Pepper}, Joshua and {De Lee}, Nathan and {Collins}, Kevin and {Latham}, David W. and {Muirhead}, Philip S. and {Chittidi}, Jay and {Rojas-Ayala}, B{\'a}rbara and {Fleming}, Scott W. and {Rose}, Mark E. and {Tenenbaum}, Peter and {Ting}, Eric B. and {Kane}, Stephen R. and {Barclay}, Thomas and {Bean}, Jacob L. and {Brassuer}, C.~E. and {Charbonneau}, David and {Ge}, Jian and {Lissauer}, Jack J. and {Mann}, Andrew W. and {McLean}, Brian and {Mullally}, Susan and {Narita}, Norio and {Plavchan}, Peter and {Ricker}, George R. and {Sasselov}, Dimitar and {Seager}, S. and {Sharma}, Sanjib and {Shiao}, Bernie and {Sozzetti}, Alessandro and {Stello}, Dennis and {Vanderspek}, Roland and {Wallace}, Geoff and {Winn}, Joshua N.},
        title = "{The Revised TESS Input Catalog and Candidate Target List}",
      journal = {\aj},
         year = 2019,
        month = oct,
       volume = {158},
       number = {4},
          eid = {138},
        pages = {138},
          doi = {10.3847/1538-3881/ab3467},
archivePrefix = {arXiv},
       eprint = {1905.10694},
 primaryClass = {astro-ph.SR},
       adsurl = {https://ui.adsabs.harvard.edu/abs/2019AJ....158..138S}
}

@ARTICLE{Henden2014APASS,
       author = {{Henden}, A. and {Munari}, U.},
        title = "{The APASS all-sky, multi-epoch BVgri photometric survey}",
      journal = {Contributions of the Astronomical Observatory Skalnate Pleso},
         year = 2014,
        month = mar,
       volume = {43},
       number = {3},
        pages = {518-522},
       adsurl = {https://ui.adsabs.harvard.edu/abs/2014CoSka..43..518H}
}

@ARTICLE{wright2010wise,
       author = {{Wright}, Edward L. and {Eisenhardt}, Peter R.~M. and {Mainzer}, Amy K. and {Ressler}, Michael E. and {Cutri}, Roc M. and {Jarrett}, Thomas and {Kirkpatrick}, J. Davy and {Padgett}, Deborah and {McMillan}, Robert S. and {Skrutskie}, Michael and {Stanford}, S.~A. and {Cohen}, Martin and {Walker}, Russell G. and {Mather}, John C. and {Leisawitz}, David and {Gautier}, Thomas N., III and {McLean}, Ian and {Benford}, Dominic and {Lonsdale}, Carol J. and {Blain}, Andrew and {Mendez}, Bryan and {Irace}, William R. and {Duval}, Valerie and {Liu}, Fengchuan and {Royer}, Don and {Heinrichsen}, Ingolf and {Howard}, Joan and {Shannon}, Mark and {Kendall}, Martha and {Walsh}, Amy L. and {Larsen}, Mark and {Cardon}, Joel G. and {Schick}, Scott and {Schwalm}, Mark and {Abid}, Mohamed and {Fabinsky}, Beth and {Naes}, Larry and {Tsai}, Chao-Wei},
        title = "{The Wide-field Infrared Survey Explorer (WISE): Mission Description and Initial On-orbit Performance}",
      journal = {\aj},
         year = 2010,
        month = dec,
       volume = {140},
       number = {6},
        pages = {1868-1881},
          doi = {10.1088/0004-6256/140/6/1868},
archivePrefix = {arXiv},
       eprint = {1008.0031},
 primaryClass = {astro-ph.IM},
       adsurl = {https://ui.adsabs.harvard.edu/abs/2010AJ....140.1868W}
}

@ARTICLE{Keller2007SkyMapper,
       author = {{Keller}, S.~C. and {Schmidt}, B.~P. and {Bessell}, M.~S. and {Conroy}, P.~G. and {Francis}, P. and {Granlund}, A. and {Kowald}, E. and {Oates}, A.~P. and {Martin-Jones}, T. and {Preston}, T. and {Tisserand}, P. and {Vaccarella}, A. and {Waterson}, M.~F.},
        title = "{The SkyMapper Telescope and The Southern Sky Survey}",
      journal = {\pasa},
         year = 2007,
        month = may,
       volume = {24},
       number = {1},
        pages = {1-12},
          doi = {10.1071/AS07001},
archivePrefix = {arXiv},
       eprint = {astro-ph/0702511},
 primaryClass = {astro-ph},
       adsurl = {https://ui.adsabs.harvard.edu/abs/2007PASA...24....1K}
}

@ARTICLE{wheatley2018ngts,
       author = {{Wheatley}, Peter J. and {West}, Richard G. and {Goad}, Michael R. and {Jenkins}, James S. and {Pollacco}, Don L. and {Queloz}, Didier and {Rauer}, Heike and {Udry}, St{\'e}phane and {Watson}, Christopher A. and {Chazelas}, Bruno and {Eigm{\"u}ller}, Philipp and {Lambert}, Gregory and {Genolet}, Ludovic and {McCormac}, James and {Walker}, Simon and {Armstrong}, David J. and {Bayliss}, Daniel and {Bento}, Joao and {Bouchy}, Fran{\c{c}}ois and {Burleigh}, Matthew R. and {Cabrera}, Juan and {Casewell}, Sarah L. and {Chaushev}, Alexander and {Chote}, Paul and {Csizmadia}, Szil{\'a}rd and {Erikson}, Anders and {Faedi}, Francesca and {Foxell}, Emma and {G{\"a}nsicke}, Boris T. and {Gillen}, Edward and {Grange}, Andrew and {G{\"u}nther}, Maximilian N. and {Hodgkin}, Simon T. and {Jackman}, James and {Jord{\'a}n}, Andr{\'e}s and {Louden}, Tom and {Metrailler}, Lionel and {Moyano}, Maximiliano and {Nielsen}, Louise D. and {Osborn}, Hugh P. and {Poppenhaeger}, Katja and {Raddi}, Roberto and {Raynard}, Liam and {Smith}, Alexis M.~S. and {Soto}, Maritza and {Titz-Weider}, Ruth},
        title = "{The Next Generation Transit Survey (NGTS)}",
      journal = {\mnras},
         year = 2018,
        month = apr,
       volume = {475},
       number = {4},
        pages = {4476-4493},
          doi = {10.1093/mnras/stx2836},
archivePrefix = {arXiv},
       eprint = {1710.11100},
 primaryClass = {astro-ph.EP},
       adsurl = {https://ui.adsabs.harvard.edu/abs/2018MNRAS.475.4476W}
}

@ARTICLE{ricker2015tess,
       author = {{Ricker}, George R. and {Winn}, Joshua N. and {Vanderspek}, Roland and {Latham}, David W. and {Bakos}, G{\'a}sp{\'a}r {\'A}. and {Bean}, Jacob L. and {Berta-Thompson}, Zachory K. and {Brown}, Timothy M. and {Buchhave}, Lars and {Butler}, Nathaniel R. and {Butler}, R. Paul and {Chaplin}, William J. and {Charbonneau}, David and {Christensen-Dalsgaard}, J{\o}rgen and {Clampin}, Mark and {Deming}, Drake and {Doty}, John and {De Lee}, Nathan and {Dressing}, Courtney and {Dunham}, Edward W. and {Endl}, Michael and {Fressin}, Francois and {Ge}, Jian and {Henning}, Thomas and {Holman}, Matthew J. and {Howard}, Andrew W. and {Ida}, Shigeru and {Jenkins}, Jon M. and {Jernigan}, Garrett and {Johnson}, John Asher and {Kaltenegger}, Lisa and {Kawai}, Nobuyuki and {Kjeldsen}, Hans and {Laughlin}, Gregory and {Levine}, Alan M. and {Lin}, Douglas and {Lissauer}, Jack J. and {MacQueen}, Phillip and {Marcy}, Geoffrey and {McCullough}, Peter R. and {Morton}, Timothy D. and {Narita}, Norio and {Paegert}, Martin and {Palle}, Enric and {Pepe}, Francesco and {Pepper}, Joshua and {Quirrenbach}, Andreas and {Rinehart}, Stephen A. and {Sasselov}, Dimitar and {Sato}, Bun'ei and {Seager}, Sara and {Sozzetti}, Alessandro and {Stassun}, Keivan G. and {Sullivan}, Peter and {Szentgyorgyi}, Andrew and {Torres}, Guillermo and {Udry}, Stephane and {Villasenor}, Joel},
        title = "{Transiting Exoplanet Survey Satellite (TESS)}",
      journal = {Journal of Astronomical Telescopes, Instruments, and Systems},
         year = 2015,
        month = jan,
       volume = {1},
          eid = {014003},
        pages = {014003},
          doi = {10.1117/1.JATIS.1.1.014003},
       adsurl = {https://ui.adsabs.harvard.edu/abs/2015JATIS...1a4003R}
}

@INPROCEEDINGS{jenkins2016spoc,
       author = {{Jenkins}, Jon M. and {Twicken}, Joseph D. and {McCauliff}, Sean and {Campbell}, Jennifer and {Sanderfer}, Dwight and {Lung}, David and {Mansouri-Samani}, Masoud and {Girouard}, Forrest and {Tenenbaum}, Peter and {Klaus}, Todd and {Smith}, Jeffrey C. and {Caldwell}, Douglas A. and {Chacon}, A.~D. and {Henze}, Christopher and {Heiges}, Cory and {Latham}, David W. and {Morgan}, Edward and {Swade}, Daryl and {Rinehart}, Stephen and {Vanderspek}, Roland},
        title = "{The TESS science processing operations center}",
    booktitle = {Software and Cyberinfrastructure for Astronomy IV},
         year = 2016,
       editor = {{Chiozzi}, Gianluca and {Guzman}, Juan C.},
       series = {Society of Photo-Optical Instrumentation Engineers (SPIE) Conference Series},
       volume = {9913},
        month = aug,
          eid = {99133E},
        pages = {99133E},
          doi = {10.1117/12.2233418},
       adsurl = {https://ui.adsabs.harvard.edu/abs/2016SPIE.9913E..3EJ}
}

@ARTICLE{Gill2020monofind,
       author = {{Gill}, Samuel and {Bayliss}, Daniel and {Cooke}, Benjamin F. and {Wheatley}, Peter J. and {Nielsen}, Louise D. and {Lendl}, Monika and {McCormac}, James and {Bryant}, Edward M. and {Acton}, Jack S. and {Anderson}, David R. and {Belardi}, Claudia and {Bouchy}, Fran{\c{c}}ois and {Burleigh}, Matthew R. and {Collier Cameron}, Andrew and {Casewell}, Sarah L. and {Chaushev}, Alexander and {Goad}, Michael R. and {G{\"u}nther}, Maximilian N. and {Hellier}, Coel and {Jackman}, James A.~G. and {Jenkins}, James S. and {Moyano}, Maximiliano and {Pollacco}, Don and {Raynard}, Liam and {Smith}, Alexis M.~S. and {Tilbrook}, Rosanna H. and {Turner}, Oliver and {Udry}, St{\'e}phane and {West}, Richard G.},
        title = "{NGTS and WASP photometric recovery of a single-transit candidate from TESS}",
      journal = {\mnras},
         year = 2020,
        month = jan,
       volume = {491},
       number = {2},
        pages = {1548-1553},
          doi = {10.1093/mnras/stz3212},
archivePrefix = {arXiv},
       eprint = {1910.05282},
 primaryClass = {astro-ph.SR},
       adsurl = {https://ui.adsabs.harvard.edu/abs/2020MNRAS.491.1548G}
}

@ARTICLE{Queloz2001coralie,
       author = {{Queloz}, D. and {Mayor}, M. and {Udry}, S. and {Burnet}, M. and {Carrier}, F. and {Eggenberger}, A. and {Naef}, D. and {Santos}, N. and {Pepe}, F. and {Rupprecht}, G. and {Avila}, G. and {Baeza}, F. and {Benz}, W. and {Bertaux}, J. -L. and {Bouchy}, F. and {Cavadore}, C. and {Delabre}, B. and {Eckert}, W. and {Fischer}, J. and {Fleury}, M. and {Gilliotte}, A. and {Goyak}, D. and {Guzman}, J.~C. and {Kohler}, D. and {Lacroix}, D. and {Lizon}, J. -L. and {Megevand}, D. and {Sivan}, J. -P. and {Sosnowska}, D. and {Weilenmann}, U.},
        title = "{From CORALIE to HARPS. The way towards 1 m s$^{-1}$ precision Doppler measurements}",
      journal = {The Messenger},
         year = 2001,
        month = sep,
       volume = {105},
        pages = {1-7},
       adsurl = {https://ui.adsabs.harvard.edu/abs/2001Msngr.105....1Q}
}

@ARTICLE{Mayor2003harps,
       author = {{Mayor}, M. and {Pepe}, F. and {Queloz}, D. and {Bouchy}, F. and {Rupprecht}, G. and {Lo Curto}, G. and {Avila}, G. and {Benz}, W. and {Bertaux}, J.-L. and {Bonfils}, X. and et al.},
        title = "{Setting New Standards with HARPS}",
      journal = {The Messenger},
         year = 2003,
        month = dec,
       volume = {114},
        pages = {20-24},
       adsurl = {https://ui.adsabs.harvard.edu/abs/2003Msngr.114...20M}
}

@ARTICLE{allesfitter-paper,
       author = {{G{\"u}nther}, Maximilian N. and {Daylan}, Tansu},
        title = "{Allesfitter: Flexible Star and Exoplanet Inference from Photometry and Radial Velocity}",
      journal = {\apjs},
         year = 2021,
        month = may,
       volume = {254},
       number = {1},
          eid = {13},
        pages = {13},
          doi = {10.3847/1538-4365/abe70e},
archivePrefix = {arXiv},
       eprint = {2003.14371},
 primaryClass = {astro-ph.EP},
       adsurl = {https://ui.adsabs.harvard.edu/abs/2021ApJS..254...13G}
}

@MISC{allesfitter-code,
 author = {{G{\"u}nther}, Maximilian~N. and {Daylan}, Tansu},
 title = "{Allesfitter: Flexible Star and Exoplanet Inference From Photometry and Radial Velocity}",
 howpublished = {Astrophysics Source Code Library},
 year = 2019,
 month = mar,
 archivePrefix = "ascl",
 eprint = {1903.003},
 adsurl = {http://adsabs.harvard.edu/abs/2019ascl.soft03003G}
}

@ARTICLE{Maxted2016ellc,
       author = {{Maxted}, P.~F.~L.},
        title = "{ellc: A fast, flexible light curve model for detached eclipsing binary stars and transiting exoplanets}",
      journal = {\aap},
         year = 2016,
        month = jun,
       volume = {591},
          eid = {A111},
        pages = {A111},
          doi = {10.1051/0004-6361/201628579},
archivePrefix = {arXiv},
       eprint = {1603.08484},
 primaryClass = {astro-ph.IM},
       adsurl = {https://ui.adsabs.harvard.edu/abs/2016A&A...591A.111M}
}

@INPROCEEDINGS{Skilling2004,
       author = {{Skilling}, John},
        title = "{Nested Sampling}",
    booktitle = {Bayesian Inference and Maximum Entropy Methods in Science and Engineering: 24th International Workshop on Bayesian Inference and Maximum Entropy Methods in Science and Engineering},
         year = 2004,
       editor = {{Fischer}, Rainer and {Preuss}, Roland and {Toussaint}, Udo Von},
       series = {American Institute of Physics Conference Series},
       volume = {735},
        month = nov,
    publisher = {AIP},
        pages = {395-405},
          doi = {10.1063/1.1835238},
       adsurl = {https://ui.adsabs.harvard.edu/abs/2004AIPC..735..395S}
}

@article{Skilling2006nestedsampling,
       author = "Skilling, John",
          doi = "10.1214/06-BA127",
     fjournal = "Bayesian Analysis",
      journal = "Bayesian Anal.",
        month = "12",
       number = "4",
        pages = "833--859",
    publisher = "International Society for Bayesian Analysis",
        title = "Nested sampling for general Bayesian computation",
          url = "https://doi.org/10.1214/06-BA127",
       volume = "1",
         year = "2006"
}

@ARTICLE{Speagle2020,
       author = {{Speagle}, Joshua S.},
        title = "{DYNESTY: a dynamic nested sampling package for estimating Bayesian posteriors and evidences}",
      journal = {\mnras},
         year = 2020,
        month = apr,
       volume = {493},
       number = {3},
        pages = {3132-3158},
          doi = {10.1093/mnras/staa278},
archivePrefix = {arXiv},
       eprint = {1904.02180},
 primaryClass = {astro-ph.IM},
       adsurl = {https://ui.adsabs.harvard.edu/abs/2020MNRAS.493.3132S}
}

@ARTICLE{Kipping2013,
       author = {{Kipping}, David M.},
        title = "{Efficient, uninformative sampling of limb darkening coefficients for two-parameter laws}",
      journal = {\mnras},
         year = 2013,
        month = nov,
       volume = {435},
       number = {3},
        pages = {2152-2160},
          doi = {10.1093/mnras/stt1435},
archivePrefix = {arXiv},
       eprint = {1308.0009},
 primaryClass = {astro-ph.SR},
       adsurl = {https://ui.adsabs.harvard.edu/abs/2013MNRAS.435.2152K}
}

@article{Parviainen2015,
  author = {Parviainen, Hannu and Aigrain, Suzanne},
  doi = {10.1093/mnras/stv1857},
  journal = {MNRAS},
  month = nov,
  number = {4},
  pages = {3821--3826},
  title = {{ldtk: Limb Darkening Toolkit}},
  url = {http://mnras.oxfordjournals.org/lookup/doi/10.1093/mnras/stv1857},
  volume = {453},
  year = {2015}
}

@ARTICLE{Gill2020ngts11,
       author = {{Gill}, Samuel and {Wheatley}, Peter J. and {Cooke}, Benjamin F. and {Jord{\'a}n}, Andr{\'e}s and {Nielsen}, Louise D. and {Bayliss}, Daniel and {Anderson}, David R. and {Vines}, Jose I. and {Lendl}, Monika and {Acton}, Jack S. and {Armstrong}, David J. and {Bouchy}, Fran{\c{c}}ois and {Brahm}, Rafael and {Bryant}, Edward M. and {Burleigh}, Matthew R. and {Casewell}, Sarah L. and {Eigm{\"u}ller}, Philipp and {Espinoza}, N{\'e}stor and {Gillen}, Edward and {Goad}, Michael R. and {Grieves}, Nolan and {G{\"u}nther}, Maximilian N. and {Henning}, Thomas and {Hobson}, Melissa J. and {Hogan}, Aleisha and {Jenkins}, James S. and {McCormac}, James and {Moyano}, Maximiliano and {Osborn}, Hugh P. and {Pollacco}, Don and {Queloz}, Didier and {Rauer}, Heike and {Raynard}, Liam and {Rojas}, Felipe and {Sarkis}, Paula and {Smith}, Alexis M.~S. and {Tala Pinto}, Marcelo and {Tilbrook}, Rosanna H. and {Udry}, St{\'e}phane and {Watson}, Christopher A. and {West}, Richard G.},
        title = "{NGTS-11 b (TOI-1847 b): A Transiting Warm Saturn Recovered from a TESS Single-transit Event}",
      journal = {\apjl},
         year = 2020,
        month = jul,
       volume = {898},
       number = {1},
          eid = {L11},
        pages = {L11},
          doi = {10.3847/2041-8213/ab9eb9},
archivePrefix = {arXiv},
       eprint = {2005.00006},
 primaryClass = {astro-ph.EP},
       adsurl = {https://ui.adsabs.harvard.edu/abs/2020ApJ...898L..11G}
}

@ARTICLE{Akeson2013,
       author = {{Akeson}, R.~L. and {Chen}, X. and {Ciardi}, D. and {Crane}, M. and {Good}, J. and {Harbut}, M. and {Jackson}, E. and {Kane}, S.~R. and {Laity}, A.~C. and {Leifer}, S. and {Lynn}, M. and {McElroy}, D.~L. and {Papin}, M. and {Plavchan}, P. and {Ram{\'\i}rez}, S.~V. and {Rey}, R. and {von Braun}, K. and {Wittman}, M. and {Abajian}, M. and {Ali}, B. and {Beichman}, C. and {Beekley}, A. and {Berriman}, G.~B. and {Berukoff}, S. and {Bryden}, G. and {Chan}, B. and {Groom}, S. and {Lau}, C. and {Payne}, A.~N. and {Regelson}, M. and {Saucedo}, M. and {Schmitz}, M. and {Stauffer}, J. and {Wyatt}, P. and {Zhang}, A.},
        title = "{The NASA Exoplanet Archive: Data and Tools for Exoplanet Research}",
      journal = {\pasp},
         year = 2013,
        month = aug,
       volume = {125},
       number = {930},
        pages = {989},
          doi = {10.1086/672273},
archivePrefix = {arXiv},
       eprint = {1307.2944},
 primaryClass = {astro-ph.IM},
       adsurl = {https://ui.adsabs.harvard.edu/abs/2013PASP..125..989A}
}

@misc{anderson2025ngts11ctransitingneptunemass,
      title={NGTS-11 c: a transiting Neptune-mass planet interior to the warm Saturn NGTS-11 b}, 
      author={David R. Anderson and Jose I. Vines and Katharine Hesse and Louise Dyregaard Nielsen and Rafael Brahm and Maximiliano Moyano and Peter J. Wheatley and Khalid Barkaoui and Allyson Bieryla and Matthew R. Burleigh and Ryan Cloutier and Karen A. Collins and Phil Evans and Steve B. Howell and John Kielkopf and Pablo Lewin and Richard P. Schwarz and Avi Shporer and Thiam-Guan Tan and Mathilde Timmermans and Amaury H. M. J. Triaud and Carl Ziegler and Ioannis Apergis and David J. Armstrong and Douglas R. Alves and Daniel Bayliss and Francois Bouchy and Sarah L. Casewell and Alexander Chaushev and Benjamin D. R. Davies and Tansu Daylan and Elsa Ducrot and Mourad Ghachoui and Samuel Gill and Edward Gillen and Michael Gillon and Maximilian N. Gunther and Thomas Henning and Melissa Hobson and Keith Horne and Emmanuel Jehin and James S. Jenkins and Andres Jordan and Michelle Kunimoto and Regis Lachaume and Monika Lendl and James McCormac and Felipe Murgas and Catriona Murray and Ares Osborn and Francisco J. Pozuelos and Didier Queloz and Suman Saha and Daniel Sebastian and Alexis M. S. Smith and Stephane Udry and Solène Ulmer-Moll and Andrew Vanderburg and Richard G. West},
      year={2025},
      eprint={2510.14083},
      archivePrefix={arXiv},
      primaryClass={astro-ph.EP},
      url={https://arxiv.org/abs/2510.14083}, 
}

@ARTICLE{Hippke2019tls,
       author = {{Hippke}, Michael and {Heller}, Ren{\'e}},
        title = "{Optimized transit detection algorithm to search for periodic transits of small planets}",
      journal = {\aap},
         year = 2019,
        month = mar,
       volume = {623},
          eid = {A39},
        pages = {A39},
          doi = {10.1051/0004-6361/201834672},
archivePrefix = {arXiv},
       eprint = {1901.02015},
 primaryClass = {astro-ph.EP},
       adsurl = {https://ui.adsabs.harvard.edu/abs/2019A&A...623A..39H}
}

@ARTICLE{Hippke2019wotan,
       author = {{Hippke}, Michael and {David}, Trevor J. and {Mulders}, Gijs D. and {Heller}, Ren{\'e}},
        title = "{W{\={o}}tan: Comprehensive Time-series Detrending in Python}",
      journal = {\aj},
         year = 2019,
        month = oct,
       volume = {158},
       number = {4},
          eid = {143},
        pages = {143},
          doi = {10.3847/1538-3881/ab3984},
archivePrefix = {arXiv},
       eprint = {1906.00966},
 primaryClass = {astro-ph.EP},
       adsurl = {https://ui.adsabs.harvard.edu/abs/2019AJ....158..143H}
}

@ARTICLE{Rodel2024,
       author = {{Rodel}, Toby and {Bayliss}, Daniel and {Gill}, Samuel and {Hawthorn}, Faith},
        title = "{TIARA TESS 1: estimating exoplanet yields from Years 1 and 3 SPOC light curves}",
      journal = {\mnras},
         year = 2024,
        month = mar,
       volume = {529},
       number = {1},
        pages = {715-731},
          doi = {10.1093/mnras/stae474},
archivePrefix = {arXiv},
       eprint = {2402.07800},
 primaryClass = {astro-ph.EP},
       adsurl = {https://ui.adsabs.harvard.edu/abs/2024MNRAS.529..715R}
}

@ARTICLE{2012Albrecht,
       author = {{Albrecht}, Simon and {Winn}, Joshua N. and {Johnson}, John A. and {Howard}, Andrew W. and {Marcy}, Geoffrey W. and {Butler}, R. Paul and {Arriagada}, Pamela and {Crane}, Jeffrey D. and {Shectman}, Stephen A. and {Thompson}, Ian B. and {Hirano}, Teruyuki and {Bakos}, Gaspar and {Hartman}, Joel D.},
        title = "{Obliquities of Hot Jupiter Host Stars: Evidence for Tidal Interactions and Primordial Misalignments}",
      journal = {\apj},
         year = 2012,
        month = sep,
       volume = {757},
       number = {1},
          eid = {18},
        pages = {18},
          doi = {10.1088/0004-637X/757/1/18},
archivePrefix = {arXiv},
       eprint = {1206.6105},
 primaryClass = {astro-ph.SR},
       adsurl = {https://ui.adsabs.harvard.edu/abs/2012ApJ...757...18A}
}

@ARTICLE{Rossiter1924,
       author = {{Rossiter}, R.~A.},
        title = "{On the detection of an effect of rotation during eclipse in the velocity of the brighter component of beta Lyrae, and on the constancy of velocity of this system.}",
      journal = {\apj},
         year = 1924,
        month = jul,
       volume = {60},
        pages = {15-21},
          doi = {10.1086/142825},
       adsurl = {https://ui.adsabs.harvard.edu/abs/1924ApJ....60...15R}
}

@ARTICLE{McLaughlin1924,
       author = {{McLaughlin}, D.~B.},
        title = "{Some results of a spectrographic study of the Algol system.}",
      journal = {\apj},
         year = 1924,
        month = jul,
       volume = {60},
        pages = {22-31},
          doi = {10.1086/142826},
       adsurl = {https://ui.adsabs.harvard.edu/abs/1924ApJ....60...22M}
}

@ARTICLE{Dawson2018,
       author = {{Dawson}, Rebekah I. and {Johnson}, John Asher},
        title = "{Origins of Hot Jupiters}",
      journal = {\araa},
         year = 2018,
        month = sep,
       volume = {56},
        pages = {175-221},
          doi = {10.1146/annurev-astro-081817-051853},
archivePrefix = {arXiv},
       eprint = {1801.06117},
 primaryClass = {astro-ph.EP},
       adsurl = {https://ui.adsabs.harvard.edu/abs/2018ARA&A..56..175D}
}

@ARTICLE{Bryant2020,
       author = {{Bryant}, Edward M. and {Bayliss}, Daniel and {McCormac}, James and {Wheatley}, Peter J. and {Acton}, Jack S. and {Anderson}, David R. and {Armstrong}, David J. and {Bouchy}, Fran{\c{c}}ois and {Belardi}, Claudia and {Burleigh}, Matthew R. and {Tilbrook}, Rosie H. and {Casewell}, Sarah L. and {Cooke}, Benjamin F. and {Gill}, Samuel and {Goad}, Michael R. and {Jenkins}, James S. and {Lendl}, Monika and {Pollacco}, Don and {Queloz}, Didier and {Raynard}, Liam and {Smith}, Alexis M.~S. and {Vines}, Jose I. and {West}, Richard G. and {Udry}, Stephane},
        title = "{Simultaneous TESS and NGTS transit observations of WASP-166 b}",
      journal = {\mnras},
         year = 2020,
        month = jun,
       volume = {494},
       number = {4},
        pages = {5872-5881},
          doi = {10.1093/mnras/staa1075},
archivePrefix = {arXiv},
       eprint = {2004.07589},
 primaryClass = {astro-ph.EP},
       adsurl = {https://ui.adsabs.harvard.edu/abs/2020MNRAS.494.5872B}
}

@INPROCEEDINGS{Bayliss2022,
       author = {{Bayliss}, Daniel and {O'Brien}, Sean M. and {Bryant}, Edward and {Wheatley}, Peter and {West}, Richard and {McCormac}, James and {Chote}, Paul and {Gill}, Sam and {Anderson}, David R. and {Wise}, Adam and {Juvan-Beaulieu}, Ines and {Coates}, Colin and {Gillen}, Edward and {Smith}, Alexis M.~S. and {Jenkins}, James S. and {Moyano}, Maximiliano and {Alves}, Douglas R. and {Burleigh}, Matthew R. and {Goad}, Michael R. and {Casewell}, Sarah L. and {Acton}, Jack and {Tilbrook}, Rosanna L. and {Henderson}, Beth A. and {Kendall}, Alicia},
        title = "{High precision ground-based CCD photometry from the Next Generation Transit Survey}",
    booktitle = {X-Ray, Optical, and Infrared Detectors for Astronomy X},
         year = 2022,
       editor = {{Holland}, Andrew D. and {Beletic}, James},
       series = {Society of Photo-Optical Instrumentation Engineers (SPIE) Conference Series},
       volume = {12191},
        month = aug,
          eid = {121911A},
        pages = {121911A},
          doi = {10.1117/12.2628966},
       adsurl = {https://ui.adsabs.harvard.edu/abs/2022SPIE12191E..1AB}
}

@ARTICLE{Bayliss2020,
       author = {{Bayliss}, D. and {Wheatley}, P. and {West}, R. and {Pollacco}, D. and {Anderson}, D.~R. and {Armstrong}, D. and {Bryant}, E. and {Cegla}, H. and {Cooke}, B. and {G{\"a}nsicke}, B. and {Gill}, S. and {Jackman}, J. and {Loudon}, T. and {McCormac}, J. and {Acton}, J. and {Burleigh}, M.~R. and {Casewell}, S. and {Goad}, M. and {Henderson}, B. and {Hogan}, A. and {Raynard}, L. and {Tilbrook}, R.~H. and {Briegal}, J. and {Gillen}, E. and {Queloz}, D. and {Smith}, G. and {Eigm{\"u}ller}, P. and {Smith}, A.~M.~S. and {Watson}, C. and {Bouchy}, F. and {Lendl}, M. and {Nielsen}, L.~D. and {Udry}, S. and {Jenkins}, J. and {Vines}, J. and {Jord{\'a}n}, A. and {Moyano}, M. and {G{\"u}nther}, M.~N.},
        title = "{NGTS - Uncovering New Worlds with Ultra-Precise Photometry}",
      journal = {The Messenger},
         year = 2020,
        month = sep,
       volume = {181},
        pages = {28-32},
          doi = {10.18727/0722-6691/5208},
       adsurl = {https://ui.adsabs.harvard.edu/abs/2020Msngr.181...28B}
}

@ARTICLE{Doyle2024,
       author = {{Doyle}, Lauren and {Armstrong}, David J. and {Bayliss}, Daniel and {Rodel}, Toby and {Kunovac}, Vedad},
        title = "{The TESS-SPOC FFI target sample explored with Gaia}",
      journal = {\mnras},
         year = 2024,
        month = apr,
       volume = {529},
       number = {2},
        pages = {1802-1813},
          doi = {10.1093/mnras/stae616},
archivePrefix = {arXiv},
       eprint = {2403.02407},
 primaryClass = {astro-ph.SR},
       adsurl = {https://ui.adsabs.harvard.edu/abs/2024MNRAS.529.1802D}
}

@ARTICLE{Eschen2024,
       author = {{Eschen}, Yoshi Nike Emilia and {Bayliss}, Daniel and {Wilson}, Thomas G. and {Kunimoto}, Michelle and {Pelisoli}, Ingrid and {Rodel}, Toby},
        title = "{Viewing the PLATO LOPS2 field through the lenses of TESS}",
      journal = {\mnras},
         year = 2024,
        month = dec,
       volume = {535},
       number = {2},
        pages = {1778-1795},
          doi = {10.1093/mnras/stae2427},
archivePrefix = {arXiv},
       eprint = {2409.13039},
 primaryClass = {astro-ph.EP},
       adsurl = {https://ui.adsabs.harvard.edu/abs/2024MNRAS.535.1778E}
}

@ARTICLE{Johnson1987kinematics,
       author = {{Johnson}, Dean R.~H. and {Soderblom}, David R.},
        title = "{Calculating Galactic Space Velocities and Their Uncertainties, with an Application to the Ursa Major Group}",
      journal = {\aj},
         year = 1987,
        month = apr,
       volume = {93},
        pages = {864},
          doi = {10.1086/114370},
       adsurl = {https://ui.adsabs.harvard.edu/abs/1987AJ.....93..864J}
}

@ARTICLE{Bensby2003thinthickdisk,
       author = {{Bensby}, T. and {Feltzing}, S. and {Lundstr{\"o}m}, I.},
        title = "{Elemental abundance trends in the Galactic thin and thick disks as traced by nearby F and G dwarf stars}",
      journal = {\aap},
         year = 2003,
        month = nov,
       volume = {410},
        pages = {527-551},
          doi = {10.1051/0004-6361:20031213},
       adsurl = {https://ui.adsabs.harvard.edu/abs/2003A&A...410..527B}
}

@ARTICLE{Bensby2014thinthickdisk,
       author = {{Bensby}, T. and {Feltzing}, S. and {Oey}, M.~S.},
        title = "{Exploring the Milky Way stellar disk. A detailed elemental abundance study of 714 F and G dwarf stars in the solar neighbourhood}",
      journal = {\aap},
         year = 2014,
        month = feb,
       volume = {562},
          eid = {A71},
        pages = {A71},
          doi = {10.1051/0004-6361/201322631},
archivePrefix = {arXiv},
       eprint = {1309.2631},
 primaryClass = {astro-ph.GA},
       adsurl = {https://ui.adsabs.harvard.edu/abs/2014A&A...562A..71B}
}

@ARTICLE{Reddy2006thinthickdisk,
       author = {{Reddy}, Bacham E. and {Lambert}, David L. and {Allende Prieto}, Carlos},
        title = "{Elemental abundance survey of the Galactic thick disc}",
      journal = {\mnras},
         year = 2006,
        month = apr,
       volume = {367},
       number = {4},
        pages = {1329-1366},
          doi = {10.1111/j.1365-2966.2006.10148.x},
archivePrefix = {arXiv},
       eprint = {astro-ph/0512505},
 primaryClass = {astro-ph},
       adsurl = {https://ui.adsabs.harvard.edu/abs/2006MNRAS.367.1329R}
}

@ARTICLE{Chen2021thinthickdisk,
       author = {{Chen}, Di-Chang and {Xie}, Ji-Wei and {Zhou}, Ji-Lin and {Dong}, Subo and {Liu}, Chao and {Wang}, Hai-Feng and {Xiang}, Mao-Sheng and {Huang}, Yang and {Luo}, Ali and {Zheng}, Zheng},
        title = "{Planets Across Space and Time (PAST). I. Characterizing the Memberships of Galactic Components and Stellar Ages: Revisiting the Kinematic Methods and Applying to Planet Host Stars}",
      journal = {\apj},
         year = 2021,
        month = mar,
       volume = {909},
       number = {2},
          eid = {115},
        pages = {115},
          doi = {10.3847/1538-4357/abd5be},
archivePrefix = {arXiv},
       eprint = {2102.09424},
 primaryClass = {astro-ph.EP},
       adsurl = {https://ui.adsabs.harvard.edu/abs/2021ApJ...909..115C}
}

@ARTICLE{Bobylev_LSR,
       author = {{Bobylev}, V.~V. and {Bajkova}, A.~T.},
        title = "{The local standard of rest from data on young objects with account for the Galactic spiral density wave}",
      journal = {\mnras},
         year = 2014,
        month = jun,
       volume = {441},
       number = {1},
        pages = {142-149},
          doi = {10.1093/mnras/stu563},
archivePrefix = {arXiv},
       eprint = {1404.6987},
 primaryClass = {astro-ph.GA},
       adsurl = {https://ui.adsabs.harvard.edu/abs/2014MNRAS.441..142B}
}

@ARTICLE{Schoenrich_LSR,
       author = {{Sch{\"o}nrich}, Ralph and {Binney}, James and {Dehnen}, Walter},
        title = "{Local kinematics and the local standard of rest}",
      journal = {\mnras},
         year = 2010,
        month = apr,
       volume = {403},
       number = {4},
        pages = {1829-1833},
          doi = {10.1111/j.1365-2966.2010.16253.x},
archivePrefix = {arXiv},
       eprint = {0912.3693},
 primaryClass = {astro-ph.GA},
       adsurl = {https://ui.adsabs.harvard.edu/abs/2010MNRAS.403.1829S}
}

@ARTICLE{Koval_LSR,
       author = {{Koval'}, V.~V. and {Marsakov}, V.~A. and {Borkova}, T.~V.},
        title = "{Evolution of the velocity ellipsoids in the thin disk of the Galaxy and the radial migration of stars}",
      journal = {Astronomy Reports},
         year = 2009,
        month = dec,
       volume = {53},
       number = {12},
        pages = {1117-1126},
          doi = {10.1134/S106377290912004X},
archivePrefix = {arXiv},
       eprint = {1104.1562},
 primaryClass = {astro-ph.GA},
       adsurl = {https://ui.adsabs.harvard.edu/abs/2009ARep...53.1117K}
}

@ARTICLE{Coskunoglu_LSR,
       author = {{Co{\c{s}}kuno{\v{g}}lu}, B. and {Ak}, S. and {Bilir}, S. and {Karaali}, S. and {Yaz}, E. and {Gilmore}, G. and {Seabroke}, G.~M. and {Bienaym{\'e}}, O. and {Bland-Hawthorn}, J. and {Campbell}, R. and {Freeman}, K.~C. and {Gibson}, B. and {Grebel}, E.~K. and {Munari}, U. and {Navarro}, J.~F. and {Parker}, Q.~A. and {Siebert}, A. and {Siviero}, A. and {Steinmetz}, M. and {Watson}, F.~G. and {Wyse}, R.~F.~G. and {Zwitter}, T.},
        title = "{Local stellar kinematics from RAVE data - I. Local standard of rest}",
      journal = {\mnras},
         year = 2011,
        month = apr,
       volume = {412},
       number = {2},
        pages = {1237-1245},
          doi = {10.1111/j.1365-2966.2010.17983.x},
archivePrefix = {arXiv},
       eprint = {1011.1188},
 primaryClass = {astro-ph.GA},
       adsurl = {https://ui.adsabs.harvard.edu/abs/2011MNRAS.412.1237C}
}

@ARTICLE{Francis_LSR,
       author = {{Francis}, Charles and {Anderson}, Erik},
        title = "{The local standard of rest and the well in the velocity distribution}",
      journal = {Celestial Mechanics and Dynamical Astronomy},
         year = 2014,
        month = apr,
       volume = {118},
       number = {4},
        pages = {399-413},
          doi = {10.1007/s10569-014-9541-z},
archivePrefix = {arXiv},
       eprint = {1311.2069},
 primaryClass = {astro-ph.GA},
       adsurl = {https://ui.adsabs.harvard.edu/abs/2014CeMDA.118..399F}
}

@ARTICLE{Tian_LSR,
       author = {{Tian}, Hai-Jun and {Liu}, Chao and {Carlin}, Jeffrey L. and {Zhao}, Yong-Heng and {Chen}, Xue-Lei and {Wu}, Yue and {Li}, Guang-Wei and {Hou}, Yong-Hui and {Zhang}, Yong},
        title = "{The Stellar Kinematics in the Solar Neighborhood from LAMOST Data}",
      journal = {\apj},
         year = 2015,
        month = aug,
       volume = {809},
       number = {2},
          eid = {145},
        pages = {145},
          doi = {10.1088/0004-637X/809/2/145},
archivePrefix = {arXiv},
       eprint = {1507.05624},
 primaryClass = {astro-ph.GA},
       adsurl = {https://ui.adsabs.harvard.edu/abs/2015ApJ...809..145T}
}

@ARTICLE{Almeida_Fernandes_LSR,
       author = {{Almeida-Fernandes}, F. and {Rocha-Pinto}, H.~J.},
        title = "{A method to estimate stellar ages from kinematical data}",
      journal = {\mnras},
         year = 2018,
        month = may,
       volume = {476},
       number = {1},
        pages = {184-197},
          doi = {10.1093/mnras/sty119},
archivePrefix = {arXiv},
       eprint = {1801.04046},
 primaryClass = {astro-ph.SR},
       adsurl = {https://ui.adsabs.harvard.edu/abs/2018MNRAS.476..184A}
}

@ARTICLE{Kempton2018,
       author = {{Kempton}, Eliza M.-R. and {Bean}, Jacob L. and {Louie}, Dana R. and {Deming}, Drake and {Koll}, Daniel D.~B. and {Mansfield}, Megan and {Christiansen}, Jessie L. and {L{\'o}pez-Morales}, Mercedes and {Swain}, Mark R. and {Zellem}, Robert T. and {Ballard}, Sarah and {Barclay}, Thomas and {Barstow}, Joanna K. and {Batalha}, Natasha E. and {Beatty}, Thomas G. and {Berta-Thompson}, Zach and {Birkby}, Jayne and {Buchhave}, Lars A. and {Charbonneau}, David and {Cowan}, Nicolas B. and {Crossfield}, Ian and {de Val-Borro}, Miguel and {Doyon}, Ren{\'e} and {Dragomir}, Diana and {Gaidos}, Eric and {Heng}, Kevin and {Hu}, Renyu and {Kane}, Stephen R. and {Kreidberg}, Laura and {Mallonn}, Matthias and {Morley}, Caroline V. and {Narita}, Norio and {Nascimbeni}, Valerio and {Pall{\'e}}, Enric and {Quintana}, Elisa V. and {Rauscher}, Emily and {Seager}, Sara and {Shkolnik}, Evgenya L. and {Sing}, David K. and {Sozzetti}, Alessandro and {Stassun}, Keivan G. and {Valenti}, Jeff A. and {von Essen}, Carolina},
        title = "{A Framework for Prioritizing the TESS Planetary Candidates Most Amenable to Atmospheric Characterization}",
      journal = {\pasp},
         year = 2018,
        month = nov,
       volume = {130},
       number = {993},
        pages = {114401},
          doi = {10.1088/1538-3873/aadf6f},
archivePrefix = {arXiv},
       eprint = {1805.03671},
 primaryClass = {astro-ph.EP},
       adsurl = {https://ui.adsabs.harvard.edu/abs/2018PASP..130k4401K}
}

@ARTICLE{Fortney2020,
       author = {{Fortney}, Jonathan J. and {Visscher}, Channon and {Marley}, Mark S. and {Hood}, Callie E. and {Line}, Michael R. and {Thorngren}, Daniel P. and {Freedman}, Richard S. and {Lupu}, Roxana},
        title = "{Beyond Equilibrium Temperature: How the Atmosphere/Interior Connection Affects the Onset of Methane, Ammonia, and Clouds in Warm Transiting Giant Planets}",
      journal = {\aj},
         year = 2020,
        month = dec,
       volume = {160},
       number = {6},
          eid = {288},
        pages = {288},
          doi = {10.3847/1538-3881/abc5bd},
archivePrefix = {arXiv},
       eprint = {2010.00146},
 primaryClass = {astro-ph.EP},
       adsurl = {https://ui.adsabs.harvard.edu/abs/2020AJ....160..288F}
}

@ARTICLE{Onken2024,
       author = {{Onken}, Christopher A. and {Wolf}, Christian and {Bessell}, Michael S. and {Chang}, Seo-Won and {Luvaul}, Lance C. and {Tonry}, John L. and {White}, Marc C. and {Da Costa}, Gary S.},
        title = "{SkyMapper Southern Survey: Data release 4}",
      journal = {\pasa},
         year = 2024,
        month = oct,
       volume = {41},
          eid = {e061},
        pages = {e061},
          doi = {10.1017/pasa.2024.53},
archivePrefix = {arXiv},
       eprint = {2402.02015},
 primaryClass = {astro-ph.CO},
       adsurl = {https://ui.adsabs.harvard.edu/abs/2024PASA...41...61O}
}

@ARTICLE{Hog2000,
       author = {{H{\o}g}, E. and {Fabricius}, C. and {Makarov}, V.~V. and {Urban}, S. and {Corbin}, T. and {Wycoff}, G. and {Bastian}, U. and {Schwekendiek}, P. and {Wicenec}, A.},
        title = "{The Tycho-2 catalogue of the 2.5 million brightest stars}",
      journal = {\aap},
         year = 2000,
        month = mar,
       volume = {355},
        pages = {L27-L30},
       adsurl = {https://ui.adsabs.harvard.edu/abs/2000A&A...355L..27H}
}

@ARTICLE{Rauer2025,
       author = {{Rauer}, Heike and {Aerts}, Conny and {Cabrera}, Juan and {Deleuil}, Magali and {Erikson}, Anders and {Gizon}, Laurent and {Goupil}, Mariejo and {Heras}, Ana and {Walloschek}, Thomas and {Lorenzo-Alvarez}, Jose and {Marliani}, Filippo and {Martin-Garcia}, C{\'e}sar and {Mas-Hesse}, J. Miguel and {O'Rourke}, Laurence and {Osborn}, Hugh and {Pagano}, Isabella and {Piotto}, Giampaolo and {Pollacco}, Don and {Ragazzoni}, Roberto and {Ramsay}, Gavin and {Udry}, St{\'e}phane and {Appourchaux}, Thierry and {Benz}, Willy and {Brandeker}, Alexis and {G{\"u}del}, Manuel and {Janot-Pacheco}, Eduardo and {Kabath}, Petr and {Kjeldsen}, Hans and {Min}, Michiel and {Santos}, Nuno and {Smith}, Alan and {Suarez}, Juan-Carlos and {Werner}, Stephanie C. and {Aboudan}, Alessio and {Abreu}, Manuel and {Acu{\~n}a}, Lorena and {Adams}, Moritz and {Adibekyan}, Vardan and {Affer}, Laura and {Agneray}, Fran{\c{c}}ois and {Agnor}, Craig and {Aguirre B{\o}rsen-Koch}, Victor and {Ahmed}, Saad and {Aigrain}, Suzanne and {Al-Bahlawan}, Ashraf and {Alcacera Gil}, Ma de los Angeles and {Alei}, Eleonora and {Alencar}, Silvia and {Alexander}, Richard and {Alfonso-Garz{\'o}n}, Julia and {Alibert}, Yann and {Allende Prieto}, Carlos and {Almeida}, Leonardo and {Alonso Sobrino}, Roi and {Altavilla}, Giuseppe and {Althaus}, Christian and {Alvarez Trujillo}, Luis Alonso and {Amarsi}, Anish and {Ammler-von Eiff}, Matthias and {Am{\^o}res}, Eduardo and {Andrade}, Laerte and {Antoniadis-Karnavas}, Alexandros and {Ant{\'o}nio}, Carlos and {Aparicio del Moral}, Beatriz and {Appolloni}, Matteo and {Arena}, Claudio and {Armstrong}, David and {Aroca Aliaga}, Jose and {Asplund}, Martin and {Audenaert}, Jeroen and {Auricchio}, Natalia and {Avelino}, Pedro and {Baeke}, Ann and {Bailli{\'e}}, Kevin and {Balado}, Ana and {Ballber Balaguer{\'o}}, Pau and {Balestra}, Andrea and {Ball}, Warrick and {Ballans}, Herve and {Ballot}, Jerome and {Barban}, Caroline and {Barbary}, Ga{\"e}le and {Barbieri}, Mauro and {Barcel{\'o} Forteza}, Sebasti{\`a} and {Barker}, Adrian and {Barklem}, Paul and {Barnes}, Sydney and {Barrado Navascues}, David and {Barragan}, Oscar and {Baruteau}, Cl{\'e}ment and {Basu}, Sarbani and {Baudin}, Frederic and {Baumeister}, Philipp and {Bayliss}, Daniel and {Bazot}, Michael and {Beck}, Paul G. and {Belkacem}, Kevin and {Bellinger}, Earl and {Benatti}, Serena and {Benomar}, Othman and {B{\'e}rard}, Diane and {Bergemann}, Maria and {Bergomi}, Maria and {Bernardo}, Pierre and {Biazzo}, Katia and {Bignamini}, Andrea and {Bigot}, Lionel and {Billot}, Nicolas and {Binet}, Martin and {Biondi}, David and {Biondi}, Federico and {Birch}, Aaron C. and {Bitsch}, Bertram and {Bluhm Ceballos}, Paz Victoria and {B{\'o}di}, Attila and {Bogn{\'a}r}, Zs{\'o}fia and {Boisse}, Isabelle and {Bolmont}, Emeline and {Bonanno}, Alfio and {Bonavita}, Mariangela and {Bonfanti}, Andrea and {Bonfils}, Xavier and {Bonito}, Rosaria and {Bonomo}, Aldo Stefano and {B{\"o}rner}, Anko and {Boro Saikia}, Sudeshna and {Borreguero Mart{\'\i}n}, Elisa and {Borsa}, Francesco and {Borsato}, Luca and {Bossini}, Diego and {Bouchy}, Francois and {Bou{\'e}}, Gwena{\"e}l and {Boufleur}, Rodrigo and {Boumier}, Patrick and {Bourrier}, Vincent and {Bowman}, Dominic M. and {Bozzo}, Enrico and {Bradley}, Louisa and {Bray}, John and {Bressan}, Alessandro and {Breton}, Sylvain and {Brienza}, Daniele and {Brito}, Ana and {Brogi}, Matteo and {Brown}, Beverly and {Brown}, David J.~A. and {Brun}, Allan Sacha and {Bruno}, Giovanni and {Bruns}, Michael and {Buchhave}, Lars A. and {Bugnet}, Lisa and {Buldgen}, Ga{\"e}l and {Burgess}, Patrick and {Busatta}, Andrea and {Busso}, Giorgia and {Buzasi}, Derek and {Caballero}, Jos{\'e} A. and {Cabral}, Alexandre and {Cabrero Gomez}, Juan-Francisco and {Calderone}, Flavia and {Cameron}, Robert and {Cameron}, Andrew and {Campante}, Tiago and {Campos Gestal}, N{\'e}stor and {Canto Martins}, Bruno Leonardo and {Cara}, Christophe and {Carone}, Ludmila and {Carrasco}, Josep Manel and {Casagrande}, Luca and {Casewell}, Sarah L. and {Cassisi}, Santi and {Castellani}, Marco and {Castro}, Matthieu and {Catala}, Claude and {Catal{\'a}n Fern{\'a}ndez}, Irene and {Catelan}, M{\'a}rcio and {Cegla}, Heather and {Cerruti}, Chiara and {Cessa}, Virginie and {Chadid}, Merieme and {Chaplin}, William and {Charpinet}, Stephane and {Chiappini}, Cristina and {Chiarucci}, Simone and {Chiavassa}, Andrea and {Chinellato}, Simonetta and {Chirulli}, Giovanni and {Christensen-Dalsgaard}, J{\o}rgen and {Church}, Ross and {Claret}, Antonio and {Clarke}, Cathie and {Claudi}, Riccardo and {Clermont}, Lionel and {Coelho}, Hugo and {Coelho}, Joao and {Cogato}, Fabrizio and {Colom{\'e}}, Josep and {Condamin}, Mathieu and {Conde Garc{\'\i}a}, Fernando and {Conseil}, Simon},
        title = "{The PLATO mission}",
      journal = {Experimental Astronomy},
         year = 2025,
        month = jun,
       volume = {59},
       number = {3},
          eid = {26},
        pages = {26},
          doi = {10.1007/s10686-025-09985-9},
archivePrefix = {arXiv},
       eprint = {2406.05447},
 primaryClass = {astro-ph.IM},
       adsurl = {https://ui.adsabs.harvard.edu/abs/2025ExA....59...26R}
}

@ARTICLE{Nascimbeni2025,
       author = {{Nascimbeni}, V. and {Piotto}, G. and {Cabrera}, J. and {Montalto}, M. and {Marinoni}, S. and {Marrese}, P.~M. and {Aerts}, C. and {Altavilla}, G. and {Benatti}, S. and {B{\"o}rner}, A. and {Deleuil}, M. and {Desidera}, S. and {Gizon}, L. and {Goupil}, M.~J. and {Granata}, V. and {Heras}, A.~M. and {Magrin}, D. and {Malavolta}, L. and {Mas-Hesse}, J.~M. and {Osborn}, H.~P. and {Pagano}, I. and {Paproth}, C. and {Pollacco}, D. and {Prisinzano}, L. and {Ragazzoni}, R. and {Ramsay}, G. and {Rauer}, H. and {Tkachenko}, A. and {Udry}, S.},
        title = "{The PLATO field selection process: II. Characterization of LOPS2, the first long-pointing field}",
      journal = {\aap},
         year = 2025,
        month = feb,
       volume = {694},
          eid = {A313},
        pages = {A313},
          doi = {10.1051/0004-6361/202452325},
archivePrefix = {arXiv},
       eprint = {2501.07687},
 primaryClass = {astro-ph.EP},
       adsurl = {https://ui.adsabs.harvard.edu/abs/2025A&A...694A.313N}
}

@ARTICLE{Montalto2021,
       author = {{Montalto}, M. and {Piotto}, G. and {Marrese}, P.~M. and {Nascimbeni}, V. and {Prisinzano}, L. and {Granata}, V. and {Marinoni}, S. and {Desidera}, S. and {Ortolani}, S. and {Aerts}, C. and {Alei}, E. and {Altavilla}, G. and {Benatti}, S. and {B{\"o}rner}, A. and {Cabrera}, J. and {Claudi}, R. and {Deleuil}, M. and {Fabrizio}, M. and {Gizon}, L. and {Goupil}, M.~J. and {Heras}, A.~M. and {Magrin}, D. and {Malavolta}, L. and {Mas-Hesse}, J.~M. and {Pagano}, I. and {Paproth}, C. and {Pertenais}, M. and {Pollacco}, D. and {Ragazzoni}, R. and {Ramsay}, G. and {Rauer}, H. and {Udry}, S.},
        title = "{The all-sky PLATO input catalogue}",
      journal = {\aap},
         year = 2021,
        month = sep,
       volume = {653},
          eid = {A98},
        pages = {A98},
          doi = {10.1051/0004-6361/202140717},
archivePrefix = {arXiv},
       eprint = {2108.13712},
 primaryClass = {astro-ph.EP},
       adsurl = {https://ui.adsabs.harvard.edu/abs/2021A&A...653A..98M}
}

@ARTICLE{Börner2024,
       author = {{B{\"o}rner}, Anko and {Paproth}, Carsten and {Cabrera}, Juan and {Pertenais}, Martin and {Rauer}, Heike and {Mas-Hesse}, J. Miguel and {Pagano}, Isabella and {Alvarez}, Jose Lorenzo and {Erikson}, Anders and {Grie{\ss}bach}, Denis and {Levillain}, Yves and {Magrin}, Demetrio and {Mogulsky}, Valery and {Niemi}, Sami-Matias and {Prod'homme}, Thibaut and {Regibo}, Sara and {De Ridder}, Joris and {Rockstein}, Steve and {Samadi}, Reza and {Serrano-Velarde}, Dimitri and {Smith}, Alan and {Verhoeve}, Peter and {Walton}, Dave},
        title = "{PLATO's signal and noise budget}",
      journal = {Experimental Astronomy},
         year = 2024,
        month = aug,
       volume = {58},
       number = {1},
          eid = {1},
        pages = {1},
          doi = {10.1007/s10686-024-09948-6},
archivePrefix = {arXiv},
       eprint = {2406.11556},
 primaryClass = {astro-ph.EP},
       adsurl = {https://ui.adsabs.harvard.edu/abs/2024ExA....58....1B}
}

@ARTICLE{Freckelton2024,
       author = {{Freckelton}, Alix V. and {Sebastian}, Daniel and {Mortier}, Annelies and {Triaud}, Amaury H.~M.~J. and {Maxted}, Pierre F.~L. and {Acu{\~n}a}, Lorena and {Armstrong}, David J. and {Battley}, Matthew P. and {Baycroft}, Thomas A. and {Boisse}, Isabelle and {Bourrier}, Vincent and {Carmona}, Andres and {Coleman}, Gavin A.~L. and {Cameron}, Andrew Collier and {Cort{\'e}s-Zuleta}, P{\'\i}a and {Delfosse}, Xavier and {Dransfield}, Georgina and {Duck}, Alison and {Forveille}, Thierry and {French}, Jenni R. and {Hara}, Nathan and {Heidari}, Neda and {Hellier}, Coel and {Kunovac}, Vedad and {Martin}, David V. and {Martioli}, Eder and {McCormac}, James J. and {Nelson}, Richard P. and {Sairam}, Lalitha and {Sousa}, S{\'e}rgio G. and {Standing}, Matthew R. and {Willett}, Emma},
        title = "{BEBOP V. Homogeneous stellar analysis of potential circumbinary planet hosts}",
      journal = {\mnras},
         year = 2024,
        month = jul,
       volume = {531},
       number = {4},
        pages = {4085-4098},
          doi = {10.1093/mnras/stae1405},
archivePrefix = {arXiv},
       eprint = {2406.03094},
 primaryClass = {astro-ph.SR},
       adsurl = {https://ui.adsabs.harvard.edu/abs/2024MNRAS.531.4085F}
}

@ARTICLE{Freckelton2025,
       author = {{Freckelton}, Alix Violet and {Mortier}, Annelies and {Bedell}, Megan and {Morrell}, Sam and {Naylor}, Tim and {Buchhave}, Lars A. and {Davies}, Guy R. and {Gonz{\'a}lez Hern{\'a}ndez}, J.~I. and {Klein}, Baptiste and {de Mooij}, Ernst J.~W. and {Passegger}, Vera Maria and {Quirrenbach}, Andreas and {Roy}, Arpita and {Santos}, Nuno C. and {Sousa}, S{\'e}rgio G. and {Su{\'a}rez Mascare{\~n}o}, A. and {Tsantaki}, Maria and {Zhao}, Lily L.},
        title = "{gr8stars {\textendash} I. A homogeneous spectroscopic study of bright FGKM dwarfs and a public library of their high-resolution spectra}",
      journal = {\mnras},
         year = 2025,
        month = jun,
       volume = {540},
       number = {2},
        pages = {1786-1799},
          doi = {10.1093/mnras/staf825},
archivePrefix = {arXiv},
       eprint = {2505.12945},
 primaryClass = {astro-ph.SR},
       adsurl = {https://ui.adsabs.harvard.edu/abs/2025MNRAS.540.1786F}
}

@ARTICLE{Dotter2016,
       author = {{Dotter}, Aaron},
        title = "{MESA Isochrones and Stellar Tracks (MIST) 0: Methods for the Construction of Stellar Isochrones}",
      journal = {\apjs},
         year = 2016,
        month = jan,
       volume = {222},
       number = {1},
          eid = {8},
        pages = {8},
          doi = {10.3847/0067-0049/222/1/8},
archivePrefix = {arXiv},
       eprint = {1601.05144},
 primaryClass = {astro-ph.SR},
       adsurl = {https://ui.adsabs.harvard.edu/abs/2016ApJS..222....8D}
}

@ARTICLE{Choi2016,
       author = {{Choi}, Jieun and {Dotter}, Aaron and {Conroy}, Charlie and {Cantiello}, Matteo and {Paxton}, Bill and {Johnson}, Benjamin D.},
        title = "{Mesa Isochrones and Stellar Tracks (MIST). I. Solar-scaled Models}",
      journal = {\apj},
         year = 2016,
        month = jun,
       volume = {823},
       number = {2},
          eid = {102},
        pages = {102},
          doi = {10.3847/0004-637X/823/2/102},
archivePrefix = {arXiv},
       eprint = {1604.08592},
 primaryClass = {astro-ph.SR},
       adsurl = {https://ui.adsabs.harvard.edu/abs/2016ApJ...823..102C}
}

@misc{Morton2015,
       author = {{Morton}, Timothy D.},
        title = "{isochrones: Stellar model grid package}",
 howpublished = {Astrophysics Source Code Library, record ascl:1503.010},
         year = 2015,
        month = mar,
          eid = {ascl:1503.010},
archivePrefix = {ascl},
       eprint = {1503.010},
       adsurl = {https://ui.adsabs.harvard.edu/abs/2015ascl.soft03010M}
}

@ARTICLE{Lindegren2020,
       author = {{Lindegren}, Lennart},
        title = "{The Gaia reference frame for bright sources examined using VLBI observations of radio stars}",
      journal = {\aap},
         year = 2020,
        month = jan,
       volume = {633},
          eid = {A1},
        pages = {A1},
          doi = {10.1051/0004-6361/201936161},
archivePrefix = {arXiv},
       eprint = {1906.09827},
 primaryClass = {astro-ph.IM},
       adsurl = {https://ui.adsabs.harvard.edu/abs/2020A&A...633A...1L}
}

@ARTICLE{Tayar2022,
       author = {{Tayar}, Jamie and {Claytor}, Zachary R. and {Huber}, Daniel and {van Saders}, Jennifer},
        title = "{A Guide to Realistic Uncertainties on the Fundamental Properties of Solar-type Exoplanet Host Stars}",
      journal = {\apj},
         year = 2022,
        month = mar,
       volume = {927},
       number = {1},
          eid = {31},
        pages = {31},
          doi = {10.3847/1538-4357/ac4bbc},
archivePrefix = {arXiv},
       eprint = {2012.07957},
 primaryClass = {astro-ph.EP},
       adsurl = {https://ui.adsabs.harvard.edu/abs/2022ApJ...927...31T}
}

@ARTICLE{Kraft1967,
       author = {{Kraft}, Robert P.},
        title = "{Studies of Stellar Rotation. V. The Dependence of Rotation on Age among Solar-Type Stars}",
      journal = {\apj},
         year = 1967,
        month = nov,
       volume = {150},
        pages = {551},
          doi = {10.1086/149359},
       adsurl = {https://ui.adsabs.harvard.edu/abs/1967ApJ...150..551K}
}

@ARTICLE{Blanco-Cuaresma2014,
       author = {{Blanco-Cuaresma}, S. and {Soubiran}, C. and {Heiter}, U. and {Jofr{\'e}}, P.},
        title = "{Determining stellar atmospheric parameters and chemical abundances of FGK stars with iSpec}",
      journal = {\aap},
         year = 2014,
        month = sep,
       volume = {569},
          eid = {A111},
        pages = {A111},
          doi = {10.1051/0004-6361/201423945},
archivePrefix = {arXiv},
       eprint = {1407.2608},
 primaryClass = {astro-ph.IM},
       adsurl = {https://ui.adsabs.harvard.edu/abs/2014A&A...569A.111B}
}

@ARTICLE{Blanco-Cuaresma2019,
       author = {{Blanco-Cuaresma}, Sergi},
        title = "{Modern stellar spectroscopy caveats}",
      journal = {\mnras},
         year = 2019,
        month = jun,
       volume = {486},
       number = {2},
        pages = {2075-2101},
          doi = {10.1093/mnras/stz549},
archivePrefix = {arXiv},
       eprint = {1902.09558},
 primaryClass = {astro-ph.SR},
       adsurl = {https://ui.adsabs.harvard.edu/abs/2019MNRAS.486.2075B}
}

@ARTICLE{Watson2010,
       author = {{Watson}, C.~A. and {Littlefair}, S.~P. and {Collier Cameron}, A. and {Dhillon}, V.~S. and {Simpson}, E.~K.},
        title = "{Estimating the masses of extra-solar planets}",
      journal = {\mnras},
         year = 2010,
        month = nov,
       volume = {408},
       number = {3},
        pages = {1606-1622},
          doi = {10.1111/j.1365-2966.2010.17233.x},
archivePrefix = {arXiv},
       eprint = {1006.2069},
 primaryClass = {astro-ph.EP},
       adsurl = {https://ui.adsabs.harvard.edu/abs/2010MNRAS.408.1606W}
}

@ARTICLE{Pecaut2013,
       author = {{Pecaut}, Mark J. and {Mamajek}, Eric E.},
        title = "{Intrinsic Colors, Temperatures, and Bolometric Corrections of Pre-main-sequence Stars}",
      journal = {\apjs},
         year = 2013,
        month = sep,
       volume = {208},
       number = {1},
          eid = {9},
        pages = {9},
          doi = {10.1088/0067-0049/208/1/9},
archivePrefix = {arXiv},
       eprint = {1307.2657},
 primaryClass = {astro-ph.SR},
       adsurl = {https://ui.adsabs.harvard.edu/abs/2013ApJS..208....9P}
}

@INCOLLECTION{Winn2014,
       author = {{Winn}, J.~N.},
        title = "{Exoplanet Transits and Occultations}",
    booktitle = {Exoplanets},
         year = 2014,
       editor = {{Seager}, S.},
        pages = {55-77},
          doi = {10.48550/arXiv.1001.2010},
       adsurl = {https://ui.adsabs.harvard.edu/abs/2010exop.book...55W},
    publisher = {University of Arizona Press},
         isbn = {9780816529452},
         lccn = {2010042822},
}

@ARTICLE{Barnes2002,
       author = {{Barnes}, Jason W. and {O'Brien}, D.~P.},
        title = "{Stability of Satellites around Close-in Extrasolar Giant Planets}",
      journal = {\apj},
         year = 2002,
        month = aug,
       volume = {575},
       number = {2},
        pages = {1087-1093},
          doi = {10.1086/341477},
archivePrefix = {arXiv},
       eprint = {astro-ph/0205035},
 primaryClass = {astro-ph},
       adsurl = {https://ui.adsabs.harvard.edu/abs/2002ApJ...575.1087B}
}

@ARTICLE{Rosario-Franco2020,
       author = {{Rosario-Franco}, Marialis and {Quarles}, Billy and {Musielak}, Zdzislaw E. and {Cuntz}, Manfred},
        title = "{Orbital Stability of Exomoons and Submoons with Applications to Kepler 1625b-I}",
      journal = {\aj},
         year = 2020,
        month = jun,
       volume = {159},
       number = {6},
          eid = {260},
        pages = {260},
          doi = {10.3847/1538-3881/ab89a7},
archivePrefix = {arXiv},
       eprint = {2005.06521},
 primaryClass = {astro-ph.EP},
       adsurl = {https://ui.adsabs.harvard.edu/abs/2020AJ....159..260R}
}

@ARTICLE{Kipping2009,
       author = {{Kipping}, David M.},
        title = "{Transit timing effects due to an exomoon}",
      journal = {\mnras},
         year = 2009,
        month = jan,
       volume = {392},
       number = {1},
        pages = {181-189},
          doi = {10.1111/j.1365-2966.2008.13999.x},
archivePrefix = {arXiv},
       eprint = {0810.2243},
 primaryClass = {astro-ph},
       adsurl = {https://ui.adsabs.harvard.edu/abs/2009MNRAS.392..181K}
}

@INCOLLECTION{Triaud2018,
       author = {{Triaud}, Amaury H.~M.~J.},
        title = "{The Rossiter-McLaughlin Effect in Exoplanet Research}",
    booktitle = {Handbook of Exoplanets},
    publisher = "{Springer Cham}",
         year = 2018,
       editor = {{Deeg}, Hans J. and {Belmonte}, Juan Antonio},
          eid = {2},
        pages = {2},
          doi = {10.1007/978-3-319-55333-7_2},
       adsurl = {https://ui.adsabs.harvard.edu/abs/2018haex.bookE...2T}
}

@ARTICLE{Grouffal2025,
       author = {{Grouffal}, S. and {Santerne}, A. and {Bourrier}, V. and {Kunovac}, V. and {Dressing}, C. and {Akinsanmi}, B. and {Armstrong}, C. and {Baliwal}, S. and {Balsalobre-Ruza}, O. and {Barros}, S.~C.~C. and {Bayliss}, D. and {Crossfield}, I.~J.~M. and {Demangeon}, O. and {Dumusque}, X. and {Giacalone}, S. and {Harada}, C.~K. and {Isaacson}, H. and {Kellermann}, H. and {Lillo-Box}, J. and {Llama}, J. and {Mortier}, A. and {Palle}, E. and {Rajpurohit}, A.~S. and {Rice}, M. and {Santos}, N.~C. and {Seidel}, J.~V. and {Sharma}, R. and {Sousa}, S.~G. and {Thomas}, L. and {Turtelboom}, E.~V. and {Udry}, S. and {Wheatley}, P.~J.},
        title = "{The star HIP 41378 potentially misaligned with its cohort of long-period planets}",
      journal = {\aap},
         year = 2025,
        month = sep,
       volume = {701},
          eid = {A173},
        pages = {A173},
          doi = {10.1051/0004-6361/202555487},
archivePrefix = {arXiv},
       eprint = {2507.01807},
 primaryClass = {astro-ph.EP},
       adsurl = {https://ui.adsabs.harvard.edu/abs/2025A&A...701A.173G}
}

@article{Barnes2004,
      doi = {10.1086/425067},
      url = {https://doi.org/10.1086/425067},
      year = {2004},
      month = {dec},
      publisher = {},
      volume = {616},
      number = {2},
      pages = {1193},
      author = {Barnes, Jason W. and Fortney, Jonathan J.},
      title = {Transit Detectability of Ring Systems around Extrasolar Giant Planets},
      journal = {The Astrophysical Journal}
}

@ARTICLE{Kenworthy2015,
       author = {{Kenworthy}, M.~A. and {Mamajek}, E.~E.},
        title = "{Modeling Giant Extrasolar Ring Systems in Eclipse and the Case of J1407b: Sculpting by Exomoons?}",
      journal = {\apj},
         year = 2015,
        month = feb,
       volume = {800},
       number = {2},
          eid = {126},
        pages = {126},
          doi = {10.1088/0004-637X/800/2/126},
archivePrefix = {arXiv},
       eprint = {1501.05652},
 primaryClass = {astro-ph.SR},
       adsurl = {https://ui.adsabs.harvard.edu/abs/2015ApJ...800..126K}
}

@ARTICLE{Aizawa2017,
       author = {{Aizawa}, Masataka and {Uehara}, Sho and {Masuda}, Kento and {Kawahara}, Hajime and {Suto}, Yasushi},
        title = "{Toward Detection of Exoplanetary Rings via Transit Photometry: Methodology and a Possible Candidate}",
      journal = {\aj},
         year = 2017,
        month = apr,
       volume = {153},
       number = {4},
          eid = {193},
        pages = {193},
          doi = {10.3847/1538-3881/aa6336},
archivePrefix = {arXiv},
       eprint = {1702.08252},
 primaryClass = {astro-ph.EP},
       adsurl = {https://ui.adsabs.harvard.edu/abs/2017AJ....153..193A}
}

@INCOLLECTION{Heller2018,
       author = {{Heller}, Ren{\'e}},
        title = "{Detecting and Characterizing Exomoons and Exorings}",
    booktitle = {Handbook of Exoplanets},
         year = 2018,
    publisher = "{Springer Cham}",
       editor = {{Deeg}, Hans J. and {Belmonte}, Juan Antonio},
          eid = {35},
        pages = {35},
          doi = {10.1007/978-3-319-55333-7_35},
       adsurl = {https://ui.adsabs.harvard.edu/abs/2018haex.bookE..35H}
}

@ARTICLE{deMooij2017,
       author = {{de Mooij}, Ernst J.~W. and {Watson}, Christopher A. and {Kenworthy}, Matthew A.},
        title = "{Characterizing exo-ring systems around fast-rotating stars using the Rossiter-McLaughlin effect}",
      journal = {\mnras},
         year = 2017,
        month = dec,
       volume = {472},
       number = {3},
        pages = {2713-2721},
          doi = {10.1093/mnras/stx2142},
archivePrefix = {arXiv},
       eprint = {1709.00680},
 primaryClass = {astro-ph.EP},
       adsurl = {https://ui.adsabs.harvard.edu/abs/2017MNRAS.472.2713D}
}

@ARTICLE{Mayor1995,
       author = {{Mayor}, Michel and {Queloz}, Didier},
        title = "{A Jupiter-mass companion to a solar-type star}",
      journal = {\nat},
         year = 1995,
        month = nov,
       volume = {378},
       number = {6555},
        pages = {355-359},
          doi = {10.1038/378355a0},
       adsurl = {https://ui.adsabs.harvard.edu/abs/1995Natur.378..355M}
}

@ARTICLE{Bodenheimer2000,
       author = {{Bodenheimer}, Peter and {Hubickyj}, Olenka and {Lissauer}, Jack J.},
        title = "{Models of the in Situ Formation of Detected Extrasolar Giant Planets}",
      journal = {\icarus},
         year = 2000,
        month = jan,
       volume = {143},
       number = {1},
        pages = {2-14},
          doi = {10.1006/icar.1999.6246},
archivePrefix = {arXiv},
       eprint = {2111.08776},
 primaryClass = {astro-ph.EP},
       adsurl = {https://ui.adsabs.harvard.edu/abs/2000Icar..143....2B}
}

@ARTICLE{Rasio1996,
       author = {{Rasio}, Frederic A. and {Ford}, Eric B.},
        title = "{Dynamical instabilities and the formation of extrasolar planetary systems}",
      journal = {Science},
         year = 1996,
        month = nov,
       volume = {274},
        pages = {954-956},
          doi = {10.1126/science.274.5289.954},
       adsurl = {https://ui.adsabs.harvard.edu/abs/1996Sci...274..954R}
}

@ARTICLE{Goldreich1980,
       author = {{Goldreich}, P. and {Tremaine}, S.},
        title = "{Disk-satellite interactions.}",
      journal = {\apj},
         year = 1980,
        month = oct,
       volume = {241},
        pages = {425-441},
          doi = {10.1086/158356},
       adsurl = {https://ui.adsabs.harvard.edu/abs/1980ApJ...241..425G}
}

@ARTICLE{Albrecht2022,
       author = {{Albrecht}, Simon H. and {Dawson}, Rebekah I. and {Winn}, Joshua N.},
        title = "{Stellar Obliquities in Exoplanetary Systems}",
      journal = {\pasp},
         year = 2022,
        month = aug,
       volume = {134},
       number = {1038},
          eid = {082001},
        pages = {082001},
          doi = {10.1088/1538-3873/ac6c09},
archivePrefix = {arXiv},
       eprint = {2203.05460},
 primaryClass = {astro-ph.EP},
       adsurl = {https://ui.adsabs.harvard.edu/abs/2022PASP..134h2001A}
}

@ARTICLE{Rice2022,
       author = {{Rice}, Malena and {Wang}, Songhu and {Wang}, Xian-Yu and {Stef{\'a}nsson}, Gu{\dj}mundur and {Isaacson}, Howard and {Howard}, Andrew W. and {Logsdon}, Sarah E. and {Schweiker}, Heidi and {Dai}, Fei and {Brinkman}, Casey and {Giacalone}, Steven and {Holcomb}, Rae},
        title = "{A Tendency Toward Alignment in Single-star Warm-Jupiter Systems}",
      journal = {\aj},
         year = 2022,
        month = sep,
       volume = {164},
       number = {3},
          eid = {104},
        pages = {104},
          doi = {10.3847/1538-3881/ac8153},
archivePrefix = {arXiv},
       eprint = {2207.06511},
 primaryClass = {astro-ph.EP},
       adsurl = {https://ui.adsabs.harvard.edu/abs/2022AJ....164..104R}
}

@ARTICLE{Wang2024,
       author = {{Wang}, Xian-Yu and {Rice}, Malena and {Wang}, Songhu and {Kanodia}, Shubham and {Dai}, Fei and {Logsdon}, Sarah E. and {Schweiker}, Heidi and {Teske}, Johanna K. and {Butler}, R. Paul and {Crane}, Jeffrey D. and {Shectman}, Stephen and {Quinn}, Samuel N. and {Kostov}, Veselin and {Osborn}, Hugh P. and {Goeke}, Robert F. and {Eastman}, Jason D. and {Shporer}, Avi and {Rapetti}, David and {Collins}, Karen A. and {Watkins}, Cristilyn N. and {Relles}, Howard M. and {Ricker}, George R. and {Seager}, Sara and {Winn}, Joshua N. and {Jenkins}, Jon M.},
        title = "{Single-star Warm-Jupiter Systems Tend to Be Aligned, Even around Hot Stellar Hosts: No T $_{eff}${\textendash}{\ensuremath{\lambda}} Dependency}",
      journal = {\apjl},
         year = 2024,
        month = sep,
       volume = {973},
       number = {1},
          eid = {L21},
        pages = {L21},
          doi = {10.3847/2041-8213/ad7469},
archivePrefix = {arXiv},
       eprint = {2408.10038},
 primaryClass = {astro-ph.EP},
       adsurl = {https://ui.adsabs.harvard.edu/abs/2024ApJ...973L..21W}
}

@ARTICLE{Sanz-Forcada2011,
       author = {{Sanz-Forcada}, J. and {Micela}, G. and {Ribas}, I. and {Pollock}, A.~M.~T. and {Eiroa}, C. and {Velasco}, A. and {Solano}, E. and {Garc{\'\i}a-{\'A}lvarez}, D.},
        title = "{Estimation of the XUV radiation onto close planets and their evaporation}",
      journal = {\aap},
         year = 2011,
        month = aug,
       volume = {532},
          eid = {A6},
        pages = {A6},
          doi = {10.1051/0004-6361/201116594},
archivePrefix = {arXiv},
       eprint = {1105.0550},
 primaryClass = {astro-ph.EP},
       adsurl = {https://ui.adsabs.harvard.edu/abs/2011A&A...532A...6S}
}

@ARTICLE{Charbonneau2002,
       author = {{Charbonneau}, David and {Brown}, Timothy M. and {Noyes}, Robert W. and {Gilliland}, Ronald L.},
        title = "{Detection of an Extrasolar Planet Atmosphere}",
      journal = {\apj},
         year = 2002,
        month = mar,
       volume = {568},
       number = {1},
        pages = {377-384},
          doi = {10.1086/338770},
archivePrefix = {arXiv},
       eprint = {astro-ph/0111544},
 primaryClass = {astro-ph},
       adsurl = {https://ui.adsabs.harvard.edu/abs/2002ApJ...568..377C}
}

@ARTICLE{Oberg2011,
       author = {{{\"O}berg}, Karin I. and {Murray-Clay}, Ruth and {Bergin}, Edwin A.},
        title = "{The Effects of Snowlines on C/O in Planetary Atmospheres}",
      journal = {\apjl},
         year = 2011,
        month = dec,
       volume = {743},
       number = {1},
          eid = {L16},
        pages = {L16},
          doi = {10.1088/2041-8205/743/1/L16},
archivePrefix = {arXiv},
       eprint = {1110.5567},
 primaryClass = {astro-ph.GA},
       adsurl = {https://ui.adsabs.harvard.edu/abs/2011ApJ...743L..16O}
}

@ARTICLE{Cooke2018,
       author = {{Cooke}, Benjamin F. and {Pollacco}, Don and {West}, Richard and {McCormac}, James and {Wheatley}, Peter J.},
        title = "{Single site observations of TESS single transit detections}",
      journal = {\aap},
         year = 2018,
        month = nov,
       volume = {619},
          eid = {A175},
        pages = {A175},
          doi = {10.1051/0004-6361/201834014},
archivePrefix = {arXiv},
       eprint = {1809.10687},
 primaryClass = {astro-ph.EP},
       adsurl = {https://ui.adsabs.harvard.edu/abs/2018A&A...619A.175C}
}

@ARTICLE{Cooke2019,
       author = {{Cooke}, Benjamin F. and {Pollacco}, Don and {Bayliss}, Daniel},
        title = "{An examination of the effect of the TESS extended mission on southern hemisphere monotransits}",
      journal = {\aap},
         year = 2019,
        month = nov,
       volume = {631},
          eid = {A83},
        pages = {A83},
          doi = {10.1051/0004-6361/201936703},
archivePrefix = {arXiv},
       eprint = {1909.13546},
 primaryClass = {astro-ph.EP},
       adsurl = {https://ui.adsabs.harvard.edu/abs/2019A&A...631A..83C}
}

@ARTICLE{Villanueva2019,
       author = {{Villanueva}, Jr., Steven and {Dragomir}, Diana and {Gaudi}, B. Scott},
        title = "{An Estimate of the Yield of Single-transit Planetary Events from the Transiting Exoplanet Survey Satellite}",
      journal = {\aj},
         year = 2019,
        month = feb,
       volume = {157},
       number = {2},
          eid = {84},
        pages = {84},
          doi = {10.3847/1538-3881/aaf85e},
archivePrefix = {arXiv},
       eprint = {1805.00956},
 primaryClass = {astro-ph.EP},
       adsurl = {https://ui.adsabs.harvard.edu/abs/2019AJ....157...84V}
}

@ARTICLE{Battley2024,
       author = {{Battley}, M.~P. and {Collins}, K.~A. and {Ulmer-Moll}, S. and {Quinn}, S.~N. and {Lendl}, M. and {Gill}, S. and {Brahm}, R. and {Hobson}, M.~J. and {Osborn}, H.~P. and {Deline}, A. and {Faria}, J.~P. and {Claringbold}, A.~B. and {Chakraborty}, H. and {Stassun}, K.~G. and {Hellier}, C. and {Alves}, D.~R. and {Ziegler}, C. and {Anderson}, D.~R. and {Apergis}, I. and {Armstrong}, D.~J. and {Bayliss}, D. and {Beletsky}, Y. and {Bieryla}, A. and {Bouchy}, F. and {Burleigh}, M.~R. and {Butler}, R.~P. and {Casewell}, S.~L. and {Christiansen}, J.~L. and {Crane}, J.~D. and {Dalba}, P.~A. and {Daylan}, T. and {Figueira}, P. and {Gillen}, E. and {Goad}, M.~R. and {G{\"u}nther}, M.~N. and {Henderson}, B.~A. and {Henning}, T. and {Jenkins}, J.~S. and {Jord{\'a}n}, A. and {Kanodia}, S. and {Kendall}, A. and {Kunimoto}, M. and {Latham}, D.~W. and {Levine}, A.~M. and {McCormac}, J. and {Moyano}, M. and {Osborn}, A. and {Osip}, D. and {Pritchard}, T.~A. and {Psaridi}, A. and {Rice}, M. and {Rodriguez}, J.~E. and {Saha}, S. and {Seager}, S. and {Shectman}, S.~A. and {Smith}, A.~M.~S. and {Teske}, J.~K. and {Ting}, E.~B. and {Udry}, S. and {Vines}, J.~I. and {Watson}, C.~A. and {West}, R.~G. and {Wheatley}, P.~J. and {Winn}, J.~N. and {Yee}, S.~W. and {Zhao}, Y.},
        title = "{NGTS-30b/TOI-4862b: An  1 Gyr old 98-day transiting warm Jupiter}",
      journal = {\aap},
         year = 2024,
        month = jun,
       volume = {686},
          eid = {A230},
        pages = {A230},
          doi = {10.1051/0004-6361/202449307},
archivePrefix = {arXiv},
       eprint = {2404.02974},
 primaryClass = {astro-ph.EP},
       adsurl = {https://ui.adsabs.harvard.edu/abs/2024A&A...686A.230B}
}

@ARTICLE{Gill2024,
       author = {{Gill}, Samuel and {Bayliss}, Daniel and {Ulmer-Moll}, Sol{\`e}ne and {Wheatley}, Peter J. and {Brahm}, Rafael and {Anderson}, David R. and {Armstrong}, David and {Apergis}, Ioannis and {Alves}, Douglas R. and {Burleigh}, Matthew R. and {Butler}, R.~P. and {Bouchy}, Fran{\c{c}}ois and {Battley}, Matthew P. and {Bryant}, Edward M. and {Bieryla}, Allyson and {Crane}, Jeffrey D. and {Collins}, Karen A. and {Casewell}, Sarah L. and {Carleo}, Ilaria and {Claringbold}, Alastair B. and {Dalba}, Paul A. and {Dragomir}, Diana and {Eigm{\"u}ller}, Philipp and {Eberhardt}, Jan and {Fausnaugh}, Michael and {G{\"u}nther}, Maximilian N. and {Grieves}, Nolan and {Goad}, Michael R. and {Gillen}, Edward and {Hagelberg}, Janis and {Hobson}, Melissa and {Hedges}, Christina and {Henderson}, Beth A. and {Hawthorn}, Faith and {Henning}, Thomas and {Jones}, Mat{\'\i}as I. and {Jord{\'a}n}, Andr{\'e}s and {Jenkins}, James S. and {Kunimoto}, Michelle and {Krenn}, Andreas F. and {Kendall}, Alicia and {Lendl}, Monika and {McCormac}, James and {Moyano}, Maximiliano and {Torres-Miranda}, Pascal and {Nielsen}, Louise D. and {Osborn}, Ares and {Otegi}, Jon and {Osborn}, Hugh and {Quinn}, Samuel N. and {Rodriguez}, Joseph E. and {Ramsay}, Gavin and {Schlecker}, Martin and {Shectman}, Stephen A. and {Seager}, Sara and {Tilbrook}, Rosanna H. and {Trifonov}, Trifon and {Teske}, Johanna K. and {Udry}, Stephane and {Vines}, Jose I. and {West}, Richard R. and {Wohler}, Bill and {Winn}, Joshua N. and {Wang}, Sharon X. and {Zhou}, George and {Zivave}, Tafadzwa},
        title = "{TOI-2447 b / NGTS-29 b: a 69-day Saturn around a Solar analogue}",
      journal = {\mnras},
         year = 2024,
        month = aug,
       volume = {532},
       number = {2},
        pages = {1444-1458},
          doi = {10.1093/mnras/stae1256},
       adsurl = {https://ui.adsabs.harvard.edu/abs/2024MNRAS.532.1444G}
}

@ARTICLE{Ulmer-Moll2022,
       author = {{Ulmer-Moll}, S. and {Lendl}, M. and {Gill}, S. and {Villanueva}, S. and {Hobson}, M.~J. and {Bouchy}, F. and {Brahm}, R. and {Dragomir}, D. and {Grieves}, N. and {Mordasini}, C. and {Anderson}, D.~R. and {Acton}, J.~S. and {Bayliss}, D. and {Bieryla}, A. and {Burleigh}, M.~R. and {Casewell}, S.~L. and {Chaverot}, G. and {Eigm{\"u}ller}, P. and {Feliz}, D. and {Gaudi}, B.~S. and {Gillen}, E. and {Goad}, M.~R. and {Gupta}, A.~F. and {G{\"u}nther}, M.~N. and {Henderson}, B.~A. and {Henning}, T. and {Jenkins}, J.~S. and {Jones}, M. and {Jord{\'a}n}, A. and {Kendall}, A. and {Latham}, D.~W. and {Mireles}, I. and {Moyano}, M. and {Nadol}, J. and {Osborn}, H.~P. and {Pepper}, J. and {Tala Pinto}, M. and {Psaridi}, A. and {Queloz}, D. and {Quinn}, S. and {Rojas}, F. and {Sarkis}, P. and {Schlecker}, M. and {Tilbrook}, R.~H. and {Torres}, P. and {Trifonov}, T. and {Udry}, S. and {Vines}, J.~I. and {West}, R. and {Wheatley}, P. and {Yao}, X. and {Zhao}, Y. and {Zhou}, G.},
        title = "{Two long-period transiting exoplanets on eccentric orbits: NGTS-20 b (TOI-5152 b) and TOI-5153 b}",
      journal = {\aap},
         year = 2022,
        month = oct,
       volume = {666},
          eid = {A46},
        pages = {A46},
          doi = {10.1051/0004-6361/202243583},
archivePrefix = {arXiv},
       eprint = {2207.03911},
 primaryClass = {astro-ph.EP},
       adsurl = {https://ui.adsabs.harvard.edu/abs/2022A&A...666A..46U}
}

@ARTICLE{Ulmer-Moll2025,
       author = {{Ulmer-Moll}, S. and {Gill}, S. and {Brahm}, R. and {Claringbold}, A. and {Lendl}, M. and {Al Moulla}, K. and {Anderson}, D. and {Battley}, M. and {Bayliss}, D. and {Bonfanti}, A. and {Bouchy}, F. and {Brice{\~n}o}, C. and {Bryant}, E.~M. and {Burleigh}, M.~R. and {Collins}, K.~A. and {Deline}, A. and {Dumusque}, X. and {Eberhardt}, J. and {Espinoza}, N. and {Falk}, B. and {Faria}, J.~P. and {Fern{\'a}ndez Fern{\'a}ndez}, J. and {Figueira}, P. and {Fridlund}, M. and {Furlan}, E. and {Goad}, M.~R. and {Goeke}, R.~F. and {Hagelberg}, J. and {Hawthorn}, F. and {Helled}, R. and {Henning}, Th. and {Hobson}, M. and {Howell}, S.~B. and {Jafariyazani}, M. and {Jenkins}, J.~M. and {Jenkins}, J.~S. and {Jones}, M.~I. and {Jord{\'a}n}, A. and {Kendall}, A. and {Law}, N. and {Littlefield}, C. and {Mann}, A.~W. and {McCormac}, J. and {Mordasini}, C. and {Moyano}, M. and {Osborn}, H. and {Pezzotti}, C. and {Psaridi}, A. and {Quinn}, S.~N. and {Rodel}, T. and {Rodriguez}, J.~E. and {Rojas}, F. and {Saha}, S. and {Schlecker}, M. and {Seager}, S. and {Sousa}, S.~G. and {Tala Pinto}, M. and {Trifonov}, T. and {Udry}, S. and {Vines}, J.~I. and {Viviani}, G. and {Watson}, C.~A. and {Wheatley}, P.~J. and {Wilson}, T.~G. and {Winn}, J.~N. and {Zhou}, G. and {Ziegler}, C.},
        title = "{Detection and characterisation of a 106-day transiting Jupiter: TOI-2449 b/NGTS-36 b}",
      journal = {\aap},
         year = 2025,
        month = nov,
       volume = {703},
          eid = {A258},
        pages = {A258},
          doi = {10.1051/0004-6361/202555168},
archivePrefix = {arXiv},
       eprint = {2509.15424},
 primaryClass = {astro-ph.EP},
       adsurl = {https://ui.adsabs.harvard.edu/abs/2025A&A...703A.258U}
}

@ARTICLE{Benz2021,
       author = {{Benz}, W. and {Broeg}, C. and {Fortier}, A. and {Rando}, N. and {Beck}, T. and {Beck}, M. and {Queloz}, D. and {Ehrenreich}, D. and {Maxted}, P.~F.~L. and {Isaak}, K.~G. and {Billot}, N. and {Alibert}, Y. and {Alonso}, R. and {Ant{\'o}nio}, C. and {Asquier}, J. and {Bandy}, T. and {B{\'a}rczy}, T. and {Barrado}, D. and {Barros}, S.~C.~C. and {Baumjohann}, W. and {Bekkelien}, A. and {Bergomi}, M. and {Biondi}, F. and {Bonfils}, X. and {Borsato}, L. and {Brandeker}, A. and {Busch}, M.-D. and {Cabrera}, J. and {Cessa}, V. and {Charnoz}, S. and {Chazelas}, B. and {Collier Cameron}, A. and {Corral Van Damme}, C. and {Cortes}, D. and {Davies}, M.~B. and {Deleuil}, M. and {Deline}, A. and {Delrez}, L. and {Demangeon}, O. and {Demory}, B.~O. and {Erikson}, A. and {Farinato}, J. and {Fossati}, L. and {Fridlund}, M. and {Futyan}, D. and {Gandolfi}, D. and {Garcia Munoz}, A. and {Gillon}, M. and {Guterman}, P. and {Gutierrez}, A. and {Hasiba}, J. and {Heng}, K. and {Hernandez}, E. and {Hoyer}, S. and {Kiss}, L.~L. and {Kovacs}, Z. and {Kuntzer}, T. and {Laskar}, J. and {Lecavelier des Etangs}, A. and {Lendl}, M. and {L{\'o}pez}, A. and {Lora}, I. and {Lovis}, C. and {L{\"u}ftinger}, T. and {Magrin}, D. and {Malvasio}, L. and {Marafatto}, L. and {Michaelis}, H. and {de Miguel}, D. and {Modrego}, D. and {Munari}, M. and {Nascimbeni}, V. and {Olofsson}, G. and {Ottacher}, H. and {Ottensamer}, R. and {Pagano}, I. and {Palacios}, R. and {Pall{\'e}}, E. and {Peter}, G. and {Piazza}, D. and {Piotto}, G. and {Pizarro}, A. and {Pollaco}, D. and {Ragazzoni}, R. and {Ratti}, F. and {Rauer}, H. and {Ribas}, I. and {Rieder}, M. and {Rohlfs}, R. and {Safa}, F. and {Salatti}, M. and {Santos}, N.~C. and {Scandariato}, G. and {S{\'e}gransan}, D. and {Simon}, A.~E. and {Smith}, A.~M.~S. and {Sordet}, M. and {Sousa}, S.~G. and {Steller}, M. and {Szab{\'o}}, G.~M. and {Szoke}, J. and {Thomas}, N. and {Tschentscher}, M. and {Udry}, S. and {Van Grootel}, V. and {Viotto}, V. and {Walter}, I. and {Walton}, N.~A. and {Wildi}, F. and {Wolter}, D.},
        title = "{The CHEOPS mission}",
      journal = {Experimental Astronomy},
         year = 2021,
        month = feb,
       volume = {51},
       number = {1},
        pages = {109-151},
          doi = {10.1007/s10686-020-09679-4},
archivePrefix = {arXiv},
       eprint = {2009.11633},
 primaryClass = {astro-ph.IM},
       adsurl = {https://ui.adsabs.harvard.edu/abs/2021ExA....51..109B}
}

@ARTICLE{Tuson2023,
       author = {{Tuson}, A. and {Queloz}, D. and {Osborn}, H.~P. and {Wilson}, T.~G. and {Hooton}, M.~J. and {Beck}, M. and {Lendl}, M. and {Olofsson}, G. and {Fortier}, A. and {Bonfanti}, A. and {Brandeker}, A. and {Buchhave}, L.~A. and {Collier Cameron}, A. and {Ciardi}, D.~R. and {Collins}, K.~A. and {Gandolfi}, D. and {Garai}, Z. and {Giacalone}, S. and {Gomes da Silva}, J. and {Howell}, S.~B. and {Patel}, J.~A. and {Persson}, C.~M. and {Serrano}, L.~M. and {Sousa}, S.~G. and {Ulmer-Moll}, S. and {Vanderburg}, A. and {Ziegler}, C. and {Alibert}, Y. and {Alonso}, R. and {Anglada}, G. and {B{\'a}rczy}, T. and {Barrado Navascues}, D. and {Barros}, S.~C.~C. and {Baumjohann}, W. and {Beck}, T. and {Benz}, W. and {Billot}, N. and {Bonfils}, X. and {Borsato}, L. and {Broeg}, C. and {Cabrera}, J. and {Charnoz}, S. and {Conti}, D.~M. and {Csizmadia}, Sz and {Cubillos}, P.~E. and {Davies}, M.~B. and {Deleuil}, M. and {Delrez}, L. and {Demangeon}, O.~D.~S. and {Demory}, B.-O. and {Dragomir}, D. and {Dressing}, C.~D. and {Ehrenreich}, D. and {Erikson}, A. and {Essack}, Z. and {Farinato}, J. and {Fossati}, L. and {Fridlund}, M. and {Furlan}, E. and {Gill}, H. and {Gillon}, M. and {Gnilka}, C.~L. and {Gonzales}, E. and {G{\"u}del}, M. and {G{\"u}nther}, M.~N. and {Hoyer}, S. and {Isaak}, K.~G. and {Jenkins}, J.~M. and {Kiss}, L.~L. and {Laskar}, J. and {Latham}, D.~W. and {Law}, N. and {Lecavelier des Etangs}, A. and {Curto}, G. Lo and {Lovis}, C. and {Luque}, R. and {Magrin}, D. and {Mann}, A.~W. and {Maxted}, P.~F.~L. and {Mayor}, M. and {McDermott}, S. and {Mecina}, M. and {Mordasini}, C. and {Mortier}, A. and {Nascimbeni}, V. and {Ottensamer}, R. and {Pagano}, I. and {Pall{\'e}}, E. and {Peter}, G. and {Piotto}, G. and {Pollacco}, D. and {Pritchard}, T. and {Ragazzoni}, R. and {Rando}, N. and {Ratti}, F. and {Rauer}, H. and {Ribas}, I. and {Ricker}, G.~R. and {Rieder}, M. and {Santos}, N.~C. and {Savel}, A.~B. and {Scandariato}, G. and {Schwarz}, R.~P. and {Seager}, S. and {S{\'e}gransan}, D. and {Shporer}, A. and {Simon}, A.~E. and {Smith}, A.~M.~S. and {Steller}, M. and {Stockdale}, C. and {Szab{\'o}}, Gy M. and {Thomas}, N. and {Torres}, G. and {Tronsgaard}, R. and {Udry}, S. and {Ulmer}, B. and {Van Grootel}, V. and {Vanderspek}, R. and {Venturini}, J. and {Walton}, N.~A. and {Winn}, J.~N. and {Wohler}, B.},
        title = "{TESS and CHEOPS discover two warm sub-Neptunes transiting the bright K-dwarf HD 15906}",
      journal = {\mnras},
         year = 2023,
        month = aug,
       volume = {523},
       number = {2},
        pages = {3090-3118},
          doi = {10.1093/mnras/stad1369},
archivePrefix = {arXiv},
       eprint = {2306.04511},
 primaryClass = {astro-ph.EP},
       adsurl = {https://ui.adsabs.harvard.edu/abs/2023MNRAS.523.3090T}
}

@ARTICLE{Hawthorn2024,
       author = {{Hawthorn}, Faith and {Gill}, Sam and {Bayliss}, Daniel and {Osborn}, Hugh P. and {Pelisoli}, Ingrid and {Rodel}, Toby and {Smith Darnbrook}, Kaylen and {Wheatley}, Peter J. and {Anderson}, David R. and {Apergis}, Ioannis and {Battley}, Matthew P. and {Burleigh}, Matthew R. and {Casewell}, Sarah L. and {Eigm{\"u}ller}, Philipp and {G{\"u}nther}, Maximilian N. and {Jenkins}, James S. and {Lendl}, Monika and {Moyano}, Maximiliano and {Osborn}, Ares and {Ramsay}, Gavin and {Ulmer-Moll}, Sol{\`e}ne and {Vines}, Jose I. and {West}, Richard},
        title = "{TESS duotransit candidates from the Southern Ecliptic Hemisphere}",
      journal = {\mnras},
         year = 2024,
        month = feb,
       volume = {528},
       number = {2},
        pages = {1841-1862},
          doi = {10.1093/mnras/stad3783},
archivePrefix = {arXiv},
       eprint = {2310.17268},
 primaryClass = {astro-ph.EP},
       adsurl = {https://ui.adsabs.harvard.edu/abs/2024MNRAS.528.1841H}
}

@ARTICLE{Osborn2023,
       author = {{Osborn}, H.~P. and {Nowak}, G. and {H{\'e}brard}, G. and {Masseron}, T. and {Lillo-Box}, J. and {Pall{\'e}}, E. and {Bekkelien}, A. and {Flor{\'e}n}, H.-G. and {Guterman}, P. and {Simon}, A.~E. and {Adibekyan}, V. and {Bieryla}, A. and {Borsato}, L. and {Brandeker}, A. and {Ciardi}, D.~R. and {Collier Cameron}, A. and {Collins}, K.~A. and {Egger}, J.~A. and {Gandolfi}, D. and {Hooton}, M.~J. and {Latham}, D.~W. and {Lendl}, M. and {Matthews}, E.~C. and {Tuson}, A. and {Ulmer-Moll}, S. and {Vanderburg}, A. and {Wilson}, T.~G. and {Ziegler}, C. and {Alibert}, Y. and {Alonso}, R. and {Anglada}, G. and {Arnold}, L. and {Asquier}, J. and {Barrado y Navascues}, D. and {Baumjohann}, W. and {Beck}, T. and {Belinski}, A.~A. and {Benz}, W. and {Biondi}, F. and {Boisse}, I. and {Bonfils}, X. and {Broeg}, C. and {Buchhave}, L.~A. and {B{\'a}rczy}, T. and {Barros}, S.~C.~C. and {Cabrera}, J. and {Cardona Guillen}, C. and {Carleo}, I. and {Castro-Gonz{\'a}lez}, A. and {Charnoz}, S. and {Christiansen}, J. and {Cortes-Zuleta}, P. and {Csizmadia}, S. and {Dalal}, S. and {Davies}, M.~B. and {Deleuil}, M. and {Delfosse}, X. and {Delrez}, L. and {Demory}, B.-O. and {Dunlavey}, A.~B. and {Ehrenreich}, D. and {Erikson}, A. and {Fernandes}, R.~B. and {Fortier}, A. and {Forveille}, T. and {Fossati}, L. and {Fridlund}, M. and {Gillon}, M. and {Goeke}, R.~F. and {Goliguzova}, M.~V. and {Gonzales}, E.~J. and {G{\"u}nther}, M.~N. and {G{\"u}del}, M. and {Heidari}, N. and {Henze}, C.~E. and {Howell}, S. and {Hoyer}, S. and {Frey}, J.~I. and {Isaak}, K.~G. and {Jenkins}, J.~M. and {Kiefer}, F. and {Kiss}, L. and {Korth}, J. and {Maxted}, P.~F.~L. and {Laskar}, J. and {Lecavelier des Etangs}, A. and {Lovis}, C. and {Lund}, M.~B. and {Luque}, R. and {Magrin}, D. and {Almenara}, J.~M. and {Martioli}, E. and {Mecina}, M. and {Medina}, J.~V. and {Moldovan}, D. and {Morales-Calder{\'o}n}, M. and {Morello}, G. and {Moutou}, C. and {Murgas}, F. and {Jensen}, E.~L.~N. and {Nascimbeni}, V. and {Olofsson}, G. and {Ottensamer}, R. and {Pagano}, I. and {Peter}, G. and {Piotto}, G. and {Pollacco}, D. and {Queloz}, D. and {Ragazzoni}, R. and {Rando}, N. and {Rauer}, H. and {Ribas}, I. and {Ricker}, G. and {Demangeon}, O.~D.~S. and {Smith}, A.~M.~S. and {Santos}, N. and {Scandariato}, G. and {Seager}, S. and {Sousa}, S.~G. and {Steller}, M. and {Szab{\'o}}, G.~M. and {S{\'e}gransan}, D. and {Thomas}, N. and {Udry}, S. and {Ulmer}, B. and {Van Grootel}, V. and {Vanderspek}, R. and {Walton}, N. and {Winn}, J.~N.},
        title = "{Two warm Neptunes transiting HIP 9618 revealed by TESS and Cheops}",
      journal = {\mnras},
         year = 2023,
        month = aug,
       volume = {523},
       number = {2},
        pages = {3069-3089},
          doi = {10.1093/mnras/stad1319},
archivePrefix = {arXiv},
       eprint = {2306.04450},
 primaryClass = {astro-ph.EP},
       adsurl = {https://ui.adsabs.harvard.edu/abs/2023MNRAS.523.3069O}
}

@ARTICLE{Ulmer-Moll2023,
       author = {{Ulmer-Moll}, S. and {Osborn}, H.~P. and {Tuson}, A. and {Egger}, J.~A. and {Lendl}, M. and {Maxted}, P. and {Bekkelien}, A. and {Simon}, A.~E. and {Olofsson}, G. and {Adibekyan}, V. and {Alibert}, Y. and {Bonfanti}, A. and {Bouchy}, F. and {Brandeker}, A. and {Fridlund}, M. and {Gandolfi}, D. and {Mordasini}, C. and {Persson}, C.~M. and {Salmon}, S. and {Serrano}, L.~M. and {Sousa}, S.~G. and {Wilson}, T.~G. and {Rieder}, M. and {Hasiba}, J. and {Asquier}, J. and {Sicilia}, D. and {Walter}, I. and {Alonso}, R. and {Anglada}, G. and {Barrado y Navascues}, D. and {Barros}, S.~C.~C. and {Baumjohann}, W. and {Beck}, M. and {Beck}, T. and {Benz}, W. and {Billot}, N. and {Bonfils}, X. and {Borsato}, L. and {Broeg}, C. and {B{\'a}rczy}, T. and {Cabrera}, J. and {Charnoz}, S. and {Cointepas}, M. and {Cameron}, A. Collier and {Csizmadia}, Sz. and {Cubillos}, P.~E. and {Davies}, M.~B. and {Deleuil}, M. and {Deline}, A. and {Delrez}, L. and {Demangeon}, O.~D.~S. and {Demory}, B.-O. and {Dumusque}, X. and {Ehrenreich}, D. and {Eisner}, N.~L. and {Erikson}, A. and {Fortier}, A. and {Fossati}, L. and {Gillon}, M. and {Grieves}, N. and {G{\"u}del}, M. and {Hagelberg}, J. and {Helled}, R. and {Hoyer}, S. and {Isaak}, K.~G. and {Kiss}, L.~L. and {Laskar}, J. and {des Etangs}, A. Lecavelier and {Lovis}, C. and {Magrin}, D. and {Nascimbeni}, V. and {Otegi}, J. and {Ottensammer}, R. and {Pagano}, I. and {Pall{\'e}}, E. and {Peter}, G. and {Piotto}, G. and {Pollacco}, D. and {Psaridi}, A. and {Queloz}, D. and {Ragazzoni}, R. and {Rando}, N. and {Rauer}, H. and {Ribas}, I. and {Santos}, N.~C. and {Scandariato}, G. and {Smith}, A.~M.~S. and {Steller}, M. and {Szab{\'o}}, G.~M. and {S{\'e}gransan}, D. and {Thomas}, N. and {Udry}, S. and {Van Grootel}, V. and {Venturini}, J. and {Walton}, N.~A.},
        title = "{TOI-5678b: A 48-day transiting Neptune-mass planet characterized with CHEOPS and HARPS★}",
      journal = {\aap},
         year = 2023,
        month = jun,
       volume = {674},
          eid = {A43},
        pages = {A43},
          doi = {10.1051/0004-6361/202245478},
archivePrefix = {arXiv},
       eprint = {2306.04295},
 primaryClass = {astro-ph.EP},
       adsurl = {https://ui.adsabs.harvard.edu/abs/2023A&A...674A..43U}
}

@ARTICLE{Schlecker2020,
       author = {{Schlecker}, Martin and {Kossakowski}, Diana and {Brahm}, Rafael and {Espinoza}, N{\'e}stor and {Henning}, Thomas and {Carone}, Ludmila and {Molaverdikhani}, Karan and {Trifonov}, Trifon and {Molli{\`e}re}, Paul and {Hobson}, Melissa J. and {Jord{\'a}n}, Andr{\'e}s and {Rojas}, Felipe I. and {Klahr}, Hubert and {Sarkis}, Paula and {Bakos}, G{\'a}sp{\'a}r {\'A}. and {Bhatti}, Waqas and {Osip}, David and {Suc}, Vincent and {Ricker}, George and {Vanderspek}, Roland and {Latham}, David W. and {Seager}, Sara and {Winn}, Joshua N. and {Jenkins}, Jon M. and {Vezie}, Michael and {Villase{\~n}or}, Jesus Noel and {Rose}, Mark E. and {Rodriguez}, David R. and {Rodriguez}, Joseph E. and {Quinn}, Samuel N. and {Shporer}, Avi},
        title = "{A Highly Eccentric Warm Jupiter Orbiting TIC 237913194}",
      journal = {\aj},
         year = 2020,
        month = dec,
       volume = {160},
       number = {6},
          eid = {275},
        pages = {275},
          doi = {10.3847/1538-3881/abbe03},
archivePrefix = {arXiv},
       eprint = {2010.03570},
 primaryClass = {astro-ph.EP},
       adsurl = {https://ui.adsabs.harvard.edu/abs/2020AJ....160..275S}
}

@ARTICLE{Hobson2021,
       author = {{Hobson}, Melissa J. and {Brahm}, Rafael and {Jord{\'a}n}, Andr{\'e}s and {Espinoza}, Nestor and {Kossakowski}, Diana and {Henning}, Thomas and {Rojas}, Felipe and {Schlecker}, Martin and {Sarkis}, Paula and {Trifonov}, Trifon and {Thorngren}, Daniel and {Binnenfeld}, Avraham and {Shahaf}, Sahar and {Zucker}, Shay and {Ricker}, George R. and {Latham}, David W. and {Seager}, S. and {Winn}, Joshua N. and {Jenkins}, Jon M. and {Addison}, Brett and {Bouchy}, Fran{\c{c}}ois and {Bowler}, Brendan P. and {Briegal}, Joshua T. and {Bryant}, Edward M. and {Collins}, Karen A. and {Daylan}, Tansu and {Grieves}, Nolan and {Horner}, Jonathan and {Huang}, Chelsea and {Kane}, Stephen R. and {Kielkopf}, John and {McLean}, Brian and {Mengel}, Matthew W. and {Nielsen}, Louise D. and {Okumura}, Jack and {Jones}, Matias and {Plavchan}, Peter and {Shporer}, Avi and {Smith}, Alexis M.~S. and {Tilbrook}, Rosanna and {Tinney}, C.~G. and {Twicken}, Joseph D. and {Udry}, St{\'e}phane and {Unger}, Nicolas and {West}, Richard and {Wittenmyer}, Robert A. and {Wohler}, Bill and {Torres}, Pascal and {Wright}, Duncan J.},
        title = "{A Transiting Warm Giant Planet around the Young Active Star TOI-201}",
      journal = {\aj},
         year = 2021,
        month = may,
       volume = {161},
       number = {5},
          eid = {235},
        pages = {235},
          doi = {10.3847/1538-3881/abeaa1},
archivePrefix = {arXiv},
       eprint = {2103.02685},
 primaryClass = {astro-ph.EP},
       adsurl = {https://ui.adsabs.harvard.edu/abs/2021AJ....161..235H}
}

@ARTICLE{Eberhardt2023,
       author = {{Eberhardt}, Jan and {Hobson}, Melissa J. and {Henning}, Thomas and {Trifonov}, Trifon and {Brahm}, Rafael and {Espinoza}, Nestor and {Jord{\'a}n}, Andr{\'e}s and {Thorngren}, Daniel and {Burn}, Remo and {Rojas}, Felipe I. and {Sarkis}, Paula and {Schlecker}, Martin and {Tala Pinto}, Marcelo and {Barkaoui}, Khalid and {Schwarz}, Richard P. and {Suarez}, Olga and {Guillot}, Tristan and {Triaud}, Amaury H.~M.~J. and {G{\"u}nther}, Maximilian N. and {Abe}, Lyu and {Boyle}, Gavin and {Leiva}, Rodrigo and {Suc}, Vincent and {Evans}, Phil and {Dunckel}, Nick and {Ziegler}, Carl and {Falk}, Ben and {Fong}, William and {Rudat}, Alexander and {Shporer}, Avi and {Striegel}, Stephanie and {Watanabe}, David and {Jenkins}, Jon M. and {Seager}, Sara and {Winn}, Joshua N.},
        title = "{Three Warm Jupiters around Solar-analog Stars Detected with TESS}",
      journal = {\aj},
         year = 2023,
        month = dec,
       volume = {166},
       number = {6},
          eid = {271},
        pages = {271},
          doi = {10.3847/1538-3881/ad06bc},
archivePrefix = {arXiv},
       eprint = {2402.17592},
 primaryClass = {astro-ph.EP},
       adsurl = {https://ui.adsabs.harvard.edu/abs/2023AJ....166..271E}
}

@ARTICLE{Brahm2023,
       author = {{Brahm}, Rafael and {Ulmer-Moll}, Sol{\`e}ne and {Hobson}, Melissa J. and {Jord{\'a}n}, Andr{\'e}s and {Henning}, Thomas and {Trifonov}, Trifon and {Jones}, Mat{\'\i}as I. and {Schlecker}, Martin and {Espinoza}, Nestor and {Rojas}, Felipe I. and {Torres}, Pascal and {Sarkis}, Paula and {Tala}, Marcelo and {Eberhardt}, Jan and {Kossakowski}, Diana and {Mu{\~n}oz}, Diego J. and {Hartman}, Joel D. and {Boyle}, Gavin and {Suc}, Vincent and {Bouchy}, Fran{\c{c}}ois and {Deline}, Adrien and {Chaverot}, Guillaume and {Grieves}, Nolan and {Lendl}, Monika and {Suarez}, Olga and {Guillot}, Tristan and {Triaud}, Amaury H.~M.~J. and {Crouzet}, Nicolas and {Dransfield}, Georgina and {Cloutier}, Ryan and {Barkaoui}, Khalid and {Schwarz}, Rick P. and {Stockdale}, Chris and {Harris}, Mallory and {Mireles}, Ismael and {Evans}, Phil and {Mann}, Andrew W. and {Ziegler}, Carl and {Dragomir}, Diana and {Villanueva}, Steven and {Mordasini}, Christoph and {Ricker}, George and {Vanderspek}, Roland and {Latham}, David W. and {Seager}, Sara and {Winn}, Joshua N. and {Jenkins}, Jon M. and {Vezie}, Michael and {Youngblood}, Allison and {Daylan}, Tansu and {Collins}, Karen A. and {Caldwell}, Douglas A. and {Ciardi}, David R. and {Palle}, Enric and {Murgas}, Felipe},
        title = "{Three Long-period Transiting Giant Planets from TESS}",
      journal = {\aj},
         year = 2023,
        month = jun,
       volume = {165},
       number = {6},
          eid = {227},
        pages = {227},
          doi = {10.3847/1538-3881/accadd},
archivePrefix = {arXiv},
       eprint = {2304.02139},
 primaryClass = {astro-ph.EP},
       adsurl = {https://ui.adsabs.harvard.edu/abs/2023AJ....165..227B}
}

@INPROCEEDINGS{Pepe2000,
       author = {{Pepe}, Francesco and {Mayor}, Michel and {Delabre}, Bernard and {Kohler}, Dominique and {Lacroix}, Daniel and {Queloz}, Didier and {Udry}, Stephane and {Benz}, Willy and {Bertaux}, Jean-Loup and {Sivan}, Jean-Pierre},
        title = "{HARPS: a new high-resolution spectrograph for the search of extrasolar planets}",
    booktitle = {Optical and IR Telescope Instrumentation and Detectors},
         year = 2000,
       editor = {{Iye}, Masanori and {Moorwood}, Alan F.},
       series = {Society of Photo-Optical Instrumentation Engineers (SPIE) Conference Series},
       volume = {4008},
        month = aug,
        pages = {582-592},
          doi = {10.1117/12.395516},
       adsurl = {https://ui.adsabs.harvard.edu/abs/2000SPIE.4008..582P}
}

@ARTICLE{Kochanek2017,
       author = {{Kochanek}, C.~S. and {Shappee}, B.~J. and {Stanek}, K.~Z. and {Holoien}, T.~W.-S. and {Thompson}, Todd A. and {Prieto}, J.~L. and {Dong}, Subo and {Shields}, J.~V. and {Will}, D. and {Britt}, C. and {Perzanowski}, D. and {Pojma{\'n}ski}, G.},
        title = "{The All-Sky Automated Survey for Supernovae (ASAS-SN) Light Curve Server v1.0}",
      journal = {\pasp},
         year = 2017,
        month = oct,
       volume = {129},
       number = {980},
        pages = {104502},
          doi = {10.1088/1538-3873/aa80d9},
archivePrefix = {arXiv},
       eprint = {1706.07060},
 primaryClass = {astro-ph.SR},
       adsurl = {https://ui.adsabs.harvard.edu/abs/2017PASP..129j4502K}
}

@ARTICLE{Claytor2020,
       author = {{Claytor}, Zachary R. and {van Saders}, Jennifer L. and {Santos}, {\^A}ngela R.~G. and {Garc{\'\i}a}, Rafael A. and {Mathur}, Savita and {Tayar}, Jamie and {Pinsonneault}, Marc H. and {Shetrone}, Matthew},
        title = "{Chemical Evolution in the Milky Way: Rotation-based Ages for APOGEE-Kepler Cool Dwarf Stars}",
      journal = {\apj},
         year = 2020,
        month = jan,
       volume = {888},
       number = {1},
          eid = {43},
        pages = {43},
          doi = {10.3847/1538-4357/ab5c24},
archivePrefix = {arXiv},
       eprint = {1911.04518},
 primaryClass = {astro-ph.SR},
       adsurl = {https://ui.adsabs.harvard.edu/abs/2020ApJ...888...43C}
}

@ARTICLE{Torres2012,
       author = {{Torres}, Guillermo and {Fischer}, Debra A. and {Sozzetti}, Alessandro and {Buchhave}, Lars A. and {Winn}, Joshua N. and {Holman}, Matthew J. and {Carter}, Joshua A.},
        title = "{Improved Spectroscopic Parameters for Transiting Planet Hosts}",
      journal = {\apj},
         year = 2012,
        month = oct,
       volume = {757},
       number = {2},
          eid = {161},
        pages = {161},
          doi = {10.1088/0004-637X/757/2/161},
archivePrefix = {arXiv},
       eprint = {1208.1268},
 primaryClass = {astro-ph.SR},
       adsurl = {https://ui.adsabs.harvard.edu/abs/2012ApJ...757..161T}
}

@ARTICLE{Mortier2014,
       author = {{Mortier}, A. and {Sousa}, S.~G. and {Adibekyan}, V. Zh. and {Brand{\~a}o}, I.~M. and {Santos}, N.~C.},
        title = "{Correcting the spectroscopic surface gravity using transits and asteroseismology. No significant effect on temperatures or metallicities with ARES and MOOG in local thermodynamic equilibrium}",
      journal = {\aap},
         year = 2014,
        month = dec,
       volume = {572},
          eid = {A95},
        pages = {A95},
          doi = {10.1051/0004-6361/201424537},
archivePrefix = {arXiv},
       eprint = {1410.1310},
 primaryClass = {astro-ph.SR},
       adsurl = {https://ui.adsabs.harvard.edu/abs/2014A&A...572A..95M}
}

@ARTICLE{Mordasini2012,
       author = {{Mordasini}, C. and {Alibert}, Y. and {Klahr}, H. and {Henning}, T.},
        title = "{Characterization of exoplanets from their formation. I. Models of combined planet formation and evolution}",
      journal = {\aap},
         year = 2012,
        month = nov,
       volume = {547},
          eid = {A111},
        pages = {A111},
          doi = {10.1051/0004-6361/201118457},
archivePrefix = {arXiv},
       eprint = {1206.6103},
 primaryClass = {astro-ph.EP},
       adsurl = {https://ui.adsabs.harvard.edu/abs/2012A&A...547A.111M}
}

@misc{Paxton2010,
       author = {{Paxton}, Bill and {Bildsten}, Lars and {Dotter}, Aaron and {Herwig}, Falk and {Lesaffre}, Pierre and {Timmes}, Frank},
        title = "{MESA: Modules for Experiments in Stellar Astrophysics}",
 howpublished = {Astrophysics Source Code Library, record ascl:1010.083},
         year = 2010,
        month = oct,
          eid = {ascl:1010.083},
archivePrefix = {ascl},
       eprint = {1010.083},
       adsurl = {https://ui.adsabs.harvard.edu/abs/2010ascl.soft10083P}
}

@ARTICLE{Dotter2008,
       author = {{Dotter}, Aaron and {Chaboyer}, Brian and {Jevremovi{\'c}}, Darko and {Kostov}, Veselin and {Baron}, E. and {Ferguson}, Jason W.},
        title = "{The Dartmouth Stellar Evolution Database}",
      journal = {\apjs},
         year = 2008,
        month = sep,
       volume = {178},
       number = {1},
        pages = {89-101},
          doi = {10.1086/589654},
archivePrefix = {arXiv},
       eprint = {0804.4473},
 primaryClass = {astro-ph},
       adsurl = {https://ui.adsabs.harvard.edu/abs/2008ApJS..178...89D}
}

@ARTICLE{Demarque2008,
       author = {{Demarque}, P. and {Guenther}, D.~B. and {Li}, L.~H. and {Mazumdar}, A. and {Straka}, C.~W.},
        title = "{YREC: the Yale rotating stellar evolution code. Non-rotating version, seismology applications}",
      journal = {\apss},
         year = 2008,
        month = aug,
       volume = {316},
       number = {1-4},
        pages = {31-41},
          doi = {10.1007/s10509-007-9698-y},
archivePrefix = {arXiv},
       eprint = {0710.4003},
 primaryClass = {astro-ph},
       adsurl = {https://ui.adsabs.harvard.edu/abs/2008Ap&SS.316...31D}
}

@ARTICLE{Weiss2008,
       author = {{Weiss}, Achim and {Schlattl}, Helmut},
        title = "{GARSTEC{\textemdash}the Garching Stellar Evolution Code. The direct descendant of the legendary Kippenhahn code}",
      journal = {\apss},
         year = 2008,
        month = aug,
       volume = {316},
       number = {1-4},
        pages = {99-106},
          doi = {10.1007/s10509-007-9606-5},
       adsurl = {https://ui.adsabs.harvard.edu/abs/2008Ap&SS.316...99W}
}

@ARTICLE{Pierrehumbert2019,
    author = "Pierrehumbert, Raymond T. and Hammond, Mark",
    title = "Atmospheric Circulation of Tide-Locked Exoplanets", 
    journal= "Annual Review of Fluid Mechanics",
    year = "2019",
    volume = "51",
    number = "Volume 51, 2019",
    pages = "275-303",
    doi = "https://doi.org/10.1146/annurev-fluid-010518-040516",
    url = "https://www.annualreviews.org/content/journals/10.1146/annurev-fluid-010518-040516",
    publisher = "Annual Reviews",
    issn = "1545-4479",
    ype = "Journal Article",
}

@misc{Kenworthy_software,
       author = {{Kenworthy}, Matthew A. and {Mamajek}, Eric E.},
        title = "{Exorings: Exoring modelling software}",
 howpublished = {Astrophysics Source Code Library, record ascl:1501.012},
         year = 2015,
        month = jan,
          eid = {ascl:1501.012},
       adsurl = {https://ui.adsabs.harvard.edu/abs/2015ascl.soft01012K}
}

@ARTICLE{Eschen2025,
       author = {{Eschen}, Yoshi Nike Emilia and {Wilson}, Thomas G. and {Bonfanti}, Andrea and {Persson}, Carina M. and {Sousa}, S{\'e}rgio G. and {Lendl}, Monika and {Heitzmann}, Alexis and {Simon}, Attila E. and {Olofsson}, G{\"o}ran and {Castro-Gonz{\'a}lez}, Amadeo and {Egger}, Jo Ann and {Fossati}, Luca and {Mustill}, Alexander James and {Osborn}, Hugh P. and {Vivien}, Hugo G. and {Alibert}, Yann and {Alonso}, Roi and {B{\'a}rczy}, Tamas and {Barrado}, David and {Barros}, Susana C.~C. and {Baumjohann}, Wolfgang and {Benz}, Willy and {Billot}, Nicolas and {Borsato}, Luca and {Brandeker}, Alexis and {Broeg}, Christopher and {Buder}, Maximilian and {Caldwell}, Douglas A. and {Cameron}, Andrew Collier and {Correia}, Alexandre C.~M. and {Csizmadia}, Szilard and {Cubillos}, Patricio E. and {Davies}, Melvyn B. and {Deleuil}, Magali and {Deline}, Adrien and {Demangeon}, Olivier D.~S. and {Demory}, Brice-Olivier and {Derekas}, Aliz and {Edwards}, Billy and {Ehrenreich}, David and {Erikson}, Anders and {Farinato}, Jacopo and {Fortier}, Andrea and {Fridlund}, Malcolm and {Gandolfi}, Davide and {Gazeas}, Kosmas and {Gillon}, Micha{\"e}l and {Goeke}, Robert and {G{\"u}del}, Manuel and {G{\"u}nther}, Maximilian N. and {Hasiba}, Johann and {Helling}, Ch and {Isaak}, Kate G. and {Jenkins}, Jon M. and {Keller}, Tatiana and {Kiss}, Laszlo L. and {Kitzmann}, Daniel and {Korth}, Judith and {Lam}, Kristine W.~F. and {Laskar}, Jacques and {Etangs}, Alain Lecavelier des and {Leleu}, Adrien and {Magrin}, Demetrio and {Maxted}, Pierre F.~L. and {Mer{\'\i}n}, Bruno and {Mordasini}, Christoph and {Nascimbeni}, Valerio and {Ottensamer}, Roland and {Pagano}, Isabella and {Pall{\'e}}, Enric and {Peter}, Gisbert and {Piazza}, Daniele and {Piotto}, Giampaolo and {Pollacco}, Don and {Queloz}, Didier and {Ragazzoni}, Roberto and {Rando}, Nicola and {Ratti}, Francesco and {Rauer}, Heike and {Ribas}, Ignasi and {Santos}, Nuno C. and {Scandariato}, Gaetano and {S{\'e}gransan}, Damien and {Shporer}, Avi and {Smith}, Alexis M.~S. and {Stalport}, Manu and {Sulis}, Sophia and {Szab{\'o}}, Gyula M. and {Udry}, St{\'e}phane and {Ulmer-Moll}, Sol{\`e}ne and {Van Grootel}, Val{\'e}rie and {Venturini}, Julia and {Villaver}, Eva and {Walton}, Nicholas A. and {Watanabe}, David and {Wolf}, Sebastian and {Ziegler}, Carl},
        title = "{An ultra-short period super-Earth and sub-Neptune spanning the Radius Valley orbiting the kinematic thick disc star TOI-2345}",
      journal = {\mnras},
         year = 2025,
        month = dec,
       volume = {544},
       number = {2},
        pages = {2614-2636},
          doi = {10.1093/mnras/staf1806},
archivePrefix = {arXiv},
       eprint = {2510.12783},
 primaryClass = {astro-ph.EP},
       adsurl = {https://ui.adsabs.harvard.edu/abs/2025MNRAS.544.2614E}
}

@ARTICLE{Pollacco2006,
       author = {{Pollacco}, D.~L. and {Skillen}, I. and {Collier Cameron}, A. and {Christian}, D.~J. and {Hellier}, C. and {Irwin}, J. and {Lister}, T.~A. and {Street}, R.~A. and {West}, R.~G. and {Anderson}, D.~R. and {Clarkson}, W.~I. and {Deeg}, H. and {Enoch}, B. and {Evans}, A. and {Fitzsimmons}, A. and {Haswell}, C.~A. and {Hodgkin}, S. and {Horne}, K. and {Kane}, S.~R. and {Keenan}, F.~P. and {Maxted}, P.~F.~L. and {Norton}, A.~J. and {Osborne}, J. and {Parley}, N.~R. and {Ryans}, R.~S.~I. and {Smalley}, B. and {Wheatley}, P.~J. and {Wilson}, D.~M.},
        title = "{The WASP Project and the SuperWASP Cameras}",
      journal = {\pasp},
         year = 2006,
        month = oct,
       volume = {118},
       number = {848},
        pages = {1407-1418},
          doi = {10.1086/508556},
archivePrefix = {arXiv},
       eprint = {astro-ph/0608454},
 primaryClass = {astro-ph},
       adsurl = {https://ui.adsabs.harvard.edu/abs/2006PASP..118.1407P}
}

@ARTICLE{Brown2013,
       author = {{Brown}, T.~M. and {Baliber}, N. and {Bianco}, F.~B. and {Bowman}, M. and {Burleson}, B. and {Conway}, P. and {Crellin}, M. and {Depagne}, {\'E}. and {De Vera}, J. and {Dilday}, B. and {Dragomir}, D. and {Dubberley}, M. and {Eastman}, J.~D. and {Elphick}, M. and {Falarski}, M. and {Foale}, S. and {Ford}, M. and {Fulton}, B.~J. and {Garza}, J. and {Gomez}, E.~L. and {Graham}, M. and {Greene}, R. and {Haldeman}, B. and {Hawkins}, E. and {Haworth}, B. and {Haynes}, R. and {Hidas}, M. and {Hjelstrom}, A.~E. and {Howell}, D.~A. and {Hygelund}, J. and {Lister}, T.~A. and {Lobdill}, R. and {Martinez}, J. and {Mullins}, D.~S. and {Norbury}, M. and {Parrent}, J. and {Paulson}, R. and {Petry}, D.~L. and {Pickles}, A. and {Posner}, V. and {Rosing}, W.~E. and {Ross}, R. and {Sand}, D.~J. and {Saunders}, E.~S. and {Shobbrook}, J. and {Shporer}, A. and {Street}, R.~A. and {Thomas}, D. and {Tsapras}, Y. and {Tufts}, J.~R. and {Valenti}, S. and {Vander Horst}, K. and {Walker}, Z. and {White}, G. and {Willis}, M.},
        title = "{Las Cumbres Observatory Global Telescope Network}",
      journal = {\pasp},
         year = 2013,
        month = sep,
       volume = {125},
       number = {931},
        pages = {1031},
          doi = {10.1086/673168},
archivePrefix = {arXiv},
       eprint = {1305.2437},
 primaryClass = {astro-ph.IM},
       adsurl = {https://ui.adsabs.harvard.edu/abs/2013PASP..125.1031B}
}

@ARTICLE{Caldwell2020,
       author = {{Caldwell}, Douglas A. and {Tenenbaum}, Peter and {Twicken}, Joseph D. and {Jenkins}, Jon M. and {Ting}, Eric and {Smith}, Jeffrey C. and {Hedges}, Christina and {Fausnaugh}, Michael M. and {Rose}, Mark and {Burke}, Christopher},
        title = "{TESS Science Processing Operations Center FFI Target List Products}",
      journal = {Research Notes of the American Astronomical Society},
         year = 2020,
        month = nov,
       volume = {4},
       number = {11},
          eid = {201},
        pages = {201},
          doi = {10.3847/2515-5172/abc9b3},
archivePrefix = {arXiv},
       eprint = {2011.05495},
 primaryClass = {astro-ph.EP},
       adsurl = {https://ui.adsabs.harvard.edu/abs/2020RNAAS...4..201C}
}

@ARTICLE{Jackson2008,
	doi = {10.1086/529187},
	url = {https://doi.org/10.1086/529187},
	year = {2008},
	month = {may},
	publisher = {},
	volume = {678},
	number = {2},
	pages = {1396},
	author = {Jackson, Brian and Greenberg, Richard and Barnes, Rory},
	title = {Tidal Evolution of Close-in Extrasolar Planets},
	journal = {The Astrophysical Journal},
}

@ARTICLE{Showman2020,
       author = {{Showman}, Adam P. and {Tan}, Xianyu and {Parmentier}, Vivien},
        title = "{Atmospheric Dynamics of Hot Giant Planets and Brown Dwarfs}",
      journal = {\ssr},
         year = 2020,
        month = dec,
       volume = {216},
       number = {8},
          eid = {139},
        pages = {139},
          doi = {10.1007/s11214-020-00758-8},
archivePrefix = {arXiv},
       eprint = {2007.15363},
 primaryClass = {astro-ph.EP},
       adsurl = {https://ui.adsabs.harvard.edu/abs/2020SSRv..216..139S}
}

@ARTICLE{Guillot1996,
       author = {{Guillot}, T. and {Burrows}, A. and {Hubbard}, W.~B. and {Lunine}, J.~I. and {Saumon}, D.},
        title = "{Giant Planets at Small Orbital Distances}",
      journal = {\apjl},
         year = 1996,
        month = mar,
       volume = {459},
        pages = {L35},
          doi = {10.1086/309935},
archivePrefix = {arXiv},
       eprint = {astro-ph/9511109},
 primaryClass = {astro-ph},
       adsurl = {https://ui.adsabs.harvard.edu/abs/1996ApJ...459L..35G}
}

@ARTICLE{Grouffal2022,
       author = {{Grouffal}, S. and {Santerne}, A. and {Bourrier}, V. and {Dumusque}, X. and {Triaud}, A.~H.~M.~J. and {Malavolta}, L. and {Kunovac}, V. and {Armstrong}, D.~J. and {Attia}, M. and {Barros}, S.~C.~C. and {Boisse}, I. and {Deleuil}, M. and {Demangeon}, O.~D.~S. and {Dressing}, C.~D. and {Figueira}, P. and {Lillo-Box}, J. and {Mortier}, A. and {Nardiello}, D. and {Santos}, N.~C. and {Sousa}, S.~G.},
        title = "{Rossiter-McLaughlin detection of the 9-month period transiting exoplanet HIP41378 d}",
      journal = {\aap},
         year = 2022,
        month = dec,
       volume = {668},
          eid = {A172},
        pages = {A172},
          doi = {10.1051/0004-6361/202244182},
archivePrefix = {arXiv},
       eprint = {2210.14125},
 primaryClass = {astro-ph.EP},
       adsurl = {https://ui.adsabs.harvard.edu/abs/2022A&A...668A.172G}
}

@ARTICLE{Winn2010,
       author = {{Winn}, Joshua N. and {Fabrycky}, Daniel and {Albrecht}, Simon and {Johnson}, John Asher},
        title = "{Hot Stars with Hot Jupiters Have High Obliquities}",
      journal = {\apjl},
         year = 2010,
        month = aug,
       volume = {718},
       number = {2},
        pages = {L145-L149},
          doi = {10.1088/2041-8205/718/2/L145},
archivePrefix = {arXiv},
       eprint = {1006.4161},
 primaryClass = {astro-ph.EP},
       adsurl = {https://ui.adsabs.harvard.edu/abs/2010ApJ...718L.145W}
}

@ARTICLE{Albrecht2012,
       author = {{Albrecht}, Simon and {Winn}, Joshua N. and {Johnson}, John A. and {Howard}, Andrew W. and {Marcy}, Geoffrey W. and {Butler}, R. Paul and {Arriagada}, Pamela and {Crane}, Jeffrey D. and {Shectman}, Stephen A. and {Thompson}, Ian B. and {Hirano}, Teruyuki and {Bakos}, Gaspar and {Hartman}, Joel D.},
        title = "{Obliquities of Hot Jupiter Host Stars: Evidence for Tidal Interactions and Primordial Misalignments}",
      journal = {\apj},
         year = 2012,
        month = sep,
       volume = {757},
       number = {1},
          eid = {18},
        pages = {18},
          doi = {10.1088/0004-637X/757/1/18},
archivePrefix = {arXiv},
       eprint = {1206.6105},
 primaryClass = {astro-ph.SR},
       adsurl = {https://ui.adsabs.harvard.edu/abs/2012ApJ...757...18A}
}

@ARTICLE{Winn2015,
       author = {{Winn}, Joshua N. and {Fabrycky}, Daniel C.},
        title = "{The Occurrence and Architecture of Exoplanetary Systems}",
      journal = {\araa},
         year = 2015,
        month = aug,
       volume = {53},
        pages = {409-447},
          doi = {10.1146/annurev-astro-082214-122246},
archivePrefix = {arXiv},
       eprint = {1410.4199},
 primaryClass = {astro-ph.EP},
       adsurl = {https://ui.adsabs.harvard.edu/abs/2015ARA&A..53..409W}
}

@ARTICLE{Knudstrup2024,
       author = {{Knudstrup}, E. and {Albrecht}, S.~H. and {Winn}, J.~N. and {Gandolfi}, D. and {Zanazzi}, J.~J. and {Persson}, C.~M. and {Fridlund}, M. and {Marcussen}, M.~L. and {Chontos}, A. and {Keniger}, M.~A.~F. and {Eisner}, N.~L. and {Bieryla}, A. and {Isaacson}, H. and {Howard}, A.~W. and {Hirsch}, L.~A. and {Murgas}, F. and {Narita}, N. and {Palle}, E. and {Kawai}, Y. and {Baker}, D.},
        title = "{Obliquities of exoplanet host stars: Nineteen new and updated measurements, and trends in the sample of 205 measurements}",
      journal = {\aap},
         year = 2024,
        month = oct,
       volume = {690},
          eid = {A379},
        pages = {A379},
          doi = {10.1051/0004-6361/202450627},
archivePrefix = {arXiv},
       eprint = {2408.09793},
 primaryClass = {astro-ph.EP},
       adsurl = {https://ui.adsabs.harvard.edu/abs/2024A&A...690A.379K}
}

@ARTICLE{Hertzsprung1913,
       author = {{Hertzsprung}, Ejnar},
        title = "{{\"U}ber die r{\"a}umliche Verteilung der Ver{\"a}nderlichen vom {\ensuremath{\delta}} Cephei-Typus}",
      journal = {Astronomische Nachrichten},
         year = 1913,
        month = nov,
       volume = {196},
        pages = {201},
       adsurl = {https://ui.adsabs.harvard.edu/abs/1913AN....196..201H}
}

@ARTICLE{Russell1913,
       author = {{Russell}, H.~N.},
        title = "{``Giant'' and ``dwarf'' stars}",
      journal = {The Observatory},
         year = 1913,
        month = aug,
       volume = {36},
        pages = {324-329},
       adsurl = {https://ui.adsabs.harvard.edu/abs/1913Obs....36..324R}
}

@ARTICLE{Grieves2022,
       author = {{Grieves}, Nolan and {Bouchy}, Fran{\c{c}}ois and {Ulmer-Moll}, Sol{\`e}ne and {Gill}, Samuel and {Anderson}, David R. and {Psaridi}, Angelica and {Lendl}, Monika and {Stassun}, Keivan G. and {Jenkins}, Jon M. and {Burleigh}, Matthew R. and {Acton}, Jack S. and {Boyd}, Patricia T. and {Casewell}, Sarah L. and {Eigm{\"u}ller}, Philipp and {Goad}, Michael R. and {Goeke}, Robert F. and {G{\"u}nther}, Maximilian N. and {Hawthorn}, Faith and {Henderson}, Beth A. and {Henze}, Christopher E. and {Jord{\'a}n}, Andr{\'e}s and {Kendall}, Alicia and {Mishra}, Lokesh and {Moyano}, Maximiliano and {Osborn}, Hugh and {Revol}, Alexandre and {Sefako}, Ramotholo R. and {Tilbrook}, Rosanna H. and {Udry}, St{\'e}phane and {Unger}, Nicolas and {Vines}, Jose I. and {West}, Richard G. and {Worters}, Hannah L.},
        title = "{An old warm Jupiter orbiting the metal-poor G-dwarf TOI-5542}",
      journal = {\aap},
         year = 2022,
        month = dec,
       volume = {668},
          eid = {A29},
        pages = {A29},
          doi = {10.1051/0004-6361/202244077},
archivePrefix = {arXiv},
       eprint = {2209.14830},
 primaryClass = {astro-ph.EP},
       adsurl = {https://ui.adsabs.harvard.edu/abs/2022A&A...668A..29G}
}

@ARTICLE{Lendl2020,
       author = {{Lendl}, Monika and {Bouchy}, Fran{\c{c}}ois and {Gill}, Samuel and {Nielsen}, Louise D. and {Turner}, Oliver and {Stassun}, Keivan and {Acton}, Jack S. and {Anderson}, David R. and {Armstrong}, David J. and {Bayliss}, Daniel and {Belardi}, Claudia and {Bryant}, Edward M. and {Burleigh}, Matthew R. and {Chaushev}, Alexander and {Casewell}, Sarah L. and {Cooke}, Benjamin F. and {Eigm{\"u}ller}, Philipp and {Gillen}, Edward and {Goad}, Michael R. and {G{\"u}nther}, Maximilian N. and {Hagelberg}, Janis and {Jenkins}, James S. and {Louden}, Tom and {Marmier}, Maxime and {McCormac}, James and {Moyano}, Maximiliano and {Pollacco}, Don and {Raynard}, Liam and {Tilbrook}, Rosanna H. and {Udry}, St{\'e}phane and {Vines}, Jose I. and {West}, Richard G. and {Wheatley}, Peter J. and {Ricker}, George and {Vanderspek}, Roland and {Latham}, David W. and {Seager}, Sara and {Winn}, Josh and {Jenkins}, Jon M. and {Addison}, Brett and {Brice{\~n}o}, C{\'e}sar and {Brahm}, Rafael and {Caldwell}, Douglas A. and {Doty}, John and {Espinoza}, N{\'e}stor and {Goeke}, Bob and {Henning}, Thomas and {Jord{\'a}n}, Andr{\'e}s and {Krishnamurthy}, Akshata and {Law}, Nicholas and {Morris}, Robert and {Okumura}, Jack and {Mann}, Andrew W. and {Rodriguez}, Joseph E. and {Sarkis}, Paula and {Schlieder}, Joshua and {Twicken}, Joseph D. and {Villanueva}, Steven and {Wittenmyer}, Robert A. and {Wright}, Duncan J. and {Ziegler}, Carl},
        title = "{TOI-222: a single-transit TESS candidate revealed to be a 34-d eclipsing binary with CORALIE, EulerCam, and NGTS}",
      journal = {\mnras},
         year = 2020,
        month = feb,
       volume = {492},
       number = {2},
        pages = {1761-1769},
          doi = {10.1093/mnras/stz3545},
archivePrefix = {arXiv},
       eprint = {1910.05050},
 primaryClass = {astro-ph.EP},
       adsurl = {https://ui.adsabs.harvard.edu/abs/2020MNRAS.492.1761L}
}

@ARTICLE{Gill2020eblm,
       author = {{Gill}, Samuel and {Cooke}, Benjamin F. and {Bayliss}, Daniel and {Nielsen}, Louise D. and {Lendl}, Monika and {Wheatley}, Peter J. and {Anderson}, David R. and {Moyano}, Maximiliano and {Bryant}, Edward M. and {Acton}, Jack S. and {Belardi}, Claudia and {Bouchy}, Fran{\c{c}}ois and {Burleigh}, Matthew R. and {Casewell}, Sarah L. and {Chaushev}, Alexander and {Goad}, Michael R. and {Jackman}, James A.~G. and {Jenkins}, James S. and {McCormac}, James and {G{\"u}nther}, Maximilian N. and {Osborn}, Hugh P. and {Pollacco}, Don and {Raynard}, Liam and {Smith}, Alexis M.~S. and {Tilbrook}, Rosanna H. and {Turner}, Oliver and {Udry}, St{\'e}phane and {Vines}, Jose I. and {Watson}, Christopher A. and {West}, Richard G.},
        title = "{A long-period (P = 61.8 d) M5V dwarf eclipsing a Sun-like star from TESS and NGTS}",
      journal = {\mnras},
         year = 2020,
        month = jul,
       volume = {495},
       number = {3},
        pages = {2713-2719},
          doi = {10.1093/mnras/staa1248},
archivePrefix = {arXiv},
       eprint = {2002.09311},
 primaryClass = {astro-ph.SR},
       adsurl = {https://ui.adsabs.harvard.edu/abs/2020MNRAS.495.2713G}
}

@ARTICLE{Gill2022,
       author = {{Gill}, Samuel and {Ulmer-Moll}, Sol{\`e}ne and {Wheatley}, Peter J. and {Bayliss}, Daniel and {Burleigh}, Matthew R. and {Acton}, Jack S. and {Casewell}, Sarah L. and {Watson}, Christopher A. and {Lendl}, Monika and {Worters}, Hannah L. and {Sefako}, Ramotholo R. and {Anderson}, David R. and {Alves}, Douglas R. and {Bouchy}, Fran{\c{c}}ois and {Bryant}, Edward M. and {Eigm{\"u}ller}, Philipp and {Gillen}, Edward and {Goad}, Michael R. and {Grieves}, Nolan and {G{\"u}nther}, Maximilian N. and {Henderson}, Beth A. and {Jenkins}, James S. and {Mishra}, Lokesh and {Moyano}, Maximiliano and {Osborn}, Hugh P. and {Tilbrook}, Rosanna H. and {Udry}, St{\'e}phane and {Vines}, Jose I. and {West}, Richard G.},
        title = "{TIC-320687387 B: a long-period eclipsing M-dwarf close to the hydrogen burning limit}",
      journal = {\mnras},
         year = 2022,
        month = jun,
       volume = {513},
       number = {2},
        pages = {1785-1793},
          doi = {10.1093/mnras/stac798},
archivePrefix = {arXiv},
       eprint = {2201.01713},
 primaryClass = {astro-ph.SR},
       adsurl = {https://ui.adsabs.harvard.edu/abs/2022MNRAS.513.1785G}
}

@ARTICLE{Henderson2024,
       author = {{Henderson}, Beth A. and {Casewell}, Sarah L. and {Jord{\'a}n}, Andr{\'e}s and {Brahm}, Rafael and {Henning}, Thomas and {Gill}, Samuel and {Mayorga}, L.~C. and {Ziegler}, Carl and {Stassun}, Keivan G. and {Goad}, Michael R. and {Acton}, Jack and {Alves}, Douglas R. and {Anderson}, David R. and {Apergis}, Ioannis and {Armstrong}, David J. and {Bayliss}, Daniel and {Burleigh}, Matthew R. and {Dragomir}, Diana and {Gillen}, Edward and {G{\"u}nther}, Maximilian N. and {Hedges}, Christina and {Hesse}, Katharine M. and {Hobson}, Melissa J. and {Jenkins}, James S. and {Jenkins}, Jon M. and {Kendall}, Alicia and {Lendl}, Monika and {Lund}, Michael B. and {McCormac}, James and {Moyano}, Maximiliano and {Osborn}, Ares and {Pinto}, Marcelo Tala and {Ramsay}, Gavin and {Rapetti}, David and {Saha}, Suman and {Seager}, Sara and {Trifonov}, Trifon and {Udry}, St{\'e}phane and {Vines}, Jose I. and {West}, Richard G. and {Wheatley}, Peter J. and {Winn}, Joshua N. and {Zivave}, Tafadzwa},
        title = "{TOI-2490b - the most eccentric brown dwarf transiting in the brown dwarf desert}",
      journal = {\mnras},
         year = 2024,
        month = sep,
       volume = {533},
       number = {3},
        pages = {2823-2842},
          doi = {10.1093/mnras/stae1940},
archivePrefix = {arXiv},
       eprint = {2408.04475},
 primaryClass = {astro-ph.EP},
       adsurl = {https://ui.adsabs.harvard.edu/abs/2024MNRAS.533.2823H}
}

@ARTICLE{Rodel2025,
       author = {{Rodel}, Toby and {Watson}, Christopher A. and {Ulmer-Moll}, Sol{\`e}ne and {Gill}, Samuel and {Maxted}, Pierre F.~L. and {Casewell}, Sarah L. and {Brahm}, Rafael and {Wilson}, Thomas G. and {Costes}, Jean C. and {Eschen}, Yoshi Nike Emilia and {Doyle}, Lauren and {Freckelton}, Alix V. and {Alves}, Douglas R. and {Apergis}, Ioannis and {Bayliss}, Daniel and {Bouchy}, Francois and {Burleigh}, Matthew R. and {Dumusque}, Xavier and {Eberhardt}, Jan and {Fern{\'a}ndez Fern{\'a}ndez}, Jorge and {Gillen}, Edward and {Goad}, Michael R. and {Hawthorn}, Faith and {Helled}, Ravit and {Henning}, Thomas and {Hobbs}, Katlyn L. and {Jenkins}, James S. and {Jord{\'a}n}, Andr{\'e}s and {Kendall}, Alicia and {Lendl}, Monika and {McCormac}, James and {de Mooij}, Ernst J.~W. and {O'Brien}, Sean M. and {Saha}, Suman and {Pinto}, Marcelo Tala and {Trifonov}, Trifon and {Udry}, St{\'e}phane and {Wheatley}, Peter J.},
        title = "{NGTS-EB-7, an eccentric, long-period, low-mass eclipsing binary}",
      journal = {\mnras},
         year = 2025,
        month = feb,
       volume = {537},
       number = {1},
        pages = {35-55},
          doi = {10.1093/mnras/stae2799},
archivePrefix = {arXiv},
       eprint = {2501.04523},
 primaryClass = {astro-ph.SR},
       adsurl = {https://ui.adsabs.harvard.edu/abs/2025MNRAS.537...35R}
}

@ARTICLE{1962Lidov,
       author = {{Lidov}, M.~L.},
        title = "{The evolution of orbits of artificial satellites of planets under the action of gravitational perturbations of external bodies}",
      journal = {\planss},
         year = 1962,
        month = oct,
       volume = {9},
       number = {10},
        pages = {719-759},
          doi = {10.1016/0032-0633(62)90129-0},
       adsurl = {https://ui.adsabs.harvard.edu/abs/1962P&SS....9..719L}
}

@ARTICLE{1962Kozai,
       author = {{Kozai}, Yoshihide},
        title = "{Secular perturbations of asteroids with high inclination and eccentricity}",
      journal = {\aj},
         year = 1962,
        month = nov,
       volume = {67},
        pages = {591-598},
          doi = {10.1086/108790},
       adsurl = {https://ui.adsabs.harvard.edu/abs/1962AJ.....67..591K}
}

@ARTICLE{Kaufer1999,
       author = {{Kaufer}, A. and {Stahl}, O. and {Tubbesing}, S. and {N{\o}rregaard}, P. and {Avila}, G. and {Francois}, P. and {Pasquini}, L. and {Pizzella}, A.},
        title = "{Commissioning FEROS, the new high-resolution spectrograph at La-Silla.}",
      journal = {The Messenger},
         year = 1999,
        month = mar,
       volume = {95},
        pages = {8-12},
       adsurl = {https://ui.adsabs.harvard.edu/abs/1999Msngr..95....8K}
}

@ARTICLE{Cooke2021,
       author = {{Cooke}, Benjamin F. and {Pollacco}, Don and {Anderson}, David R. and {Bayliss}, Daniel and {Bouchy}, Fran{\c{c}}ois and {Gill}, Samuel and {Grieves}, Nolan and {Lendl}, Monika and {Nielsen}, Louise D. and {Udry}, St{\'e}phane and {Wheatley}, Peter J.},
        title = "{Resolving period aliases for TESS monotransits recovered during the extended mission}",
      journal = {\mnras},
         year = 2021,
        month = jan,
       volume = {500},
       number = {4},
        pages = {5088-5097},
          doi = {10.1093/mnras/staa3569},
archivePrefix = {arXiv},
       eprint = {2011.05832},
 primaryClass = {astro-ph.EP},
       adsurl = {https://ui.adsabs.harvard.edu/abs/2021MNRAS.500.5088C}
}

@ARTICLE{Kendall2025,
       author = {{Kendall}, Alicia and {Ulmer-Moll}, Sol{\`e}ne and {Gill}, Samuel and {Burleigh}, Matthew R. and {Goad}, Michael R. and {Anderson}, David R. and {Bryant}, Edward M. and {Lavie}, Baptiste and {Bugatti}, Maddalena and {Acevedo Barroso}, Javier A. and {Steiner}, Michal and {Dragomir}, Diana and {Villanueva}, Jr., Steven and {Stevens}, Daniel J. and {Gupta}, Arvind F. and {Gaudi}, Scott and {Sun}, Guoyou and {Claringbold}, Alastair and {Doyle}, Lauren and {Guillot}, Tristan and {Suarez}, Olga and {M{\'e}karnia}, Djamel and {Triaud}, Amaury H.~M.~J. and {Bendjoya}, Philippe and {Ziegler}, Carl and {Mann}, Andrew W. and {Howell}, Steve B. and {Fajardo-Acosta}, Sergio B. and {Littlefield}, Colin and {Caldwell}, Douglas A. and {Kunimoto}, Michelle and {Rowden}, Pamela and {Kostov}, Veselin and {Villase{\~n}or}, Jesus Noel and {Alves}, Douglas and {Apergis}, Ioannis and {Armstrong}, David J. and {Battley}, Matthew P. and {Bayliss}, Daniel and {Bouchy}, Fran{\c{c}}ois and {Casewell}, Sarah L. and {G{\"u}nther}, Maximilian N. and {Harvey}, George T. and {Hawthorn}, Faith and {Jenkins}, James S. and {Lendl}, Monika and {McCormac}, James and {Moyano}, Maximilano and {Nielsen}, Louise D. and {Osborn}, Ares and {Rodel}, Toby and {Saha}, Suman and {Udry}, Stephane and {Vines}, Jose I. and {Wheatley}, Peter J. and {Zivave}, Tafadzwa},
        title = "{A 43 d transiting Neptune and two 25 d Saturns from TESS, NGTS, and ASTEP}",
      journal = {\mnras},
         year = 2026,
        month = apr,
       volume = {547},
       number = {2},
          eid = {staf2189},
        pages = {staf2189},
          doi = {10.1093/mnras/staf2189},
archivePrefix = {arXiv},
       eprint = {2512.07716},
 primaryClass = {astro-ph.EP},
       adsurl = {https://ui.adsabs.harvard.edu/abs/2026MNRAS.547f2189K}
}

@ARTICLE{Aller2020,
       author = {{Aller}, A. and {Lillo-Box}, J. and {Jones}, D. and {Miranda}, L.~F. and {Barcel{\'o} Forteza}, S.},
        title = "{Planetary nebulae seen with TESS: Discovery of new binary central star candidates from Cycle 1}",
      journal = {\aap},
         year = 2020,
        month = mar,
       volume = {635},
          eid = {A128},
        pages = {A128},
          doi = {10.1051/0004-6361/201937118},
archivePrefix = {arXiv},
       eprint = {1911.09991},
 primaryClass = {astro-ph.SR},
       adsurl = {https://ui.adsabs.harvard.edu/abs/2020A&A...635A.128A}
}

@INPROCEEDINGS{McCully2018,
       author = {{McCully}, Curtis and {Volgenau}, Nikolaus H. and {Harbeck}, Daniel-Rolf and {Lister}, Tim A. and {Saunders}, Eric S. and {Turner}, Monica L. and {Siiverd}, Robert J. and {Bowman}, Mark},
        title = "{Real-time processing of the imaging data from the network of Las Cumbres Observatory Telescopes using BANZAI}",
    booktitle = {Software and Cyberinfrastructure for Astronomy V},
         year = 2018,
       editor = {{Guzman}, Juan C. and {Ibsen}, Jorge},
       series = {Society of Photo-Optical Instrumentation Engineers (SPIE) Conference Series},
       volume = {10707},
        month = jul,
          eid = {107070K},
        pages = {107070K},
          doi = {10.1117/12.2314340},
archivePrefix = {arXiv},
       eprint = {1811.04163},
 primaryClass = {astro-ph.IM},
       adsurl = {https://ui.adsabs.harvard.edu/abs/2018SPIE10707E..0KM}
}

@article{Collins2017,
	doi = {10.3847/1538-3881/153/2/77},
	url = {https://doi.org/10.3847/1538-3881/153/2/77},
	year = 2017,
	month = {jan},
	publisher = {American Astronomical Society},
	volume = {153},
	number = {2},
	pages = {77},
	author = {Karen A. Collins and John F. Kielkopf and Keivan G. Stassun and Frederic V. Hessman},
	title = {{ASTROIMAGEJ}: {IMAGE} {PROCESSING} {AND} {PHOTOMETRIC} {EXTRACTION} {FOR} {ULTRA}-{PRECISE} {ASTRONOMICAL} {LIGHT} {CURVES}},
	journal = {The Astronomical Journal}
}

@ARTICLE{Akinsanmi2018,
       author = {{Akinsanmi}, B. and {Oshagh}, M. and {Santos}, N.~C. and {Barros}, S.~C.~C.},
        title = "{Detecting transit signatures of exoplanetary rings using SOAP3.0}",
      journal = {\aap},
         year = 2018,
        month = jan,
       volume = {609},
          eid = {A21},
        pages = {A21},
          doi = {10.1051/0004-6361/201731215},
archivePrefix = {arXiv},
       eprint = {1709.06443},
 primaryClass = {astro-ph.EP},
       adsurl = {https://ui.adsabs.harvard.edu/abs/2018A&A...609A..21A}
}

@ARTICLE{Dalba2021,
       author = {{Dalba}, Paul A. and {Kane}, Stephen R. and {Li}, Zhexing and {MacDougall}, Mason G. and {Rosenthal}, Lee J. and {Cherubim}, Collin and {Isaacson}, Howard and {Thorngren}, Daniel P. and {Fulton}, Benjamin and {Howard}, Andrew W. and {Petigura}, Erik A. and {Schwieterman}, Edward W. and {Peluso}, Dan O. and {Esposito}, Thomas M. and {Marchis}, Franck and {Payne}, Matthew J.},
        title = "{Giant Outer Transiting Exoplanet Mass (GOT 'EM) Survey. II. Discovery of a Failed Hot Jupiter on a 2.7 Yr, Highly Eccentric Orbit}",
      journal = {\aj},
         year = 2021,
        month = oct,
       volume = {162},
       number = {4},
          eid = {154},
        pages = {154},
          doi = {10.3847/1538-3881/ac134b},
archivePrefix = {arXiv},
       eprint = {2107.06901},
 primaryClass = {astro-ph.EP},
       adsurl = {https://ui.adsabs.harvard.edu/abs/2021AJ....162..154D}
}



\clearpage
\onecolumn

\appendix

\section{\ngts\ photometric data}
\label{section:appendix:ngts_data}
In Table\,\ref{tab:ngts_data} we show the \ngts\ data for \name\ separated into the runs shown in Table\,\ref{tab:phot_summary}. The table is truncated in print to only the first five rows for each run, however, full machine readable versions will be made available online.

ActionID is an identification number given to each night of observations from each telescope. Camera is the ID of the telescope making the observations, in runs 1-3 this is only 809 while run 4 contains 809 and 810. Timestamps are listed in BJD. Relative Flux is the `raw' flux output of the \bsproc\ pipeline and is the target counts divided by the total master comparison counts. Normalised Flux is this relative flux median normalised for each run. The Sky background is given as the total number of electron counts for that exposure. The FWHM is measured (in pixels) by creating a super-sampled Point Spread Function (PSF) across the central third of the image.

\begin{table}
	\centering
	\caption{\ngts\ data.}
	\label{tab:ngts_data}
	\begin{tabular}{cccccccc}
		\toprule
		ActionID & Camera & Time & Relative Flux & Normalised Flux & Airmass & Sky Background & FWHM\\
		 & & BJD & & & & e$^-$ counts & Pixels\\
		\hline
		\multicolumn{8}{c}{\it Run 1}\\
		\hline
		354981&809&2460182.87854167&$0.02155\pm0.00026$&$1.002\pm0.012$&1.98486&30572.598&2.0288\\
		354981&809&2460182.87877315&$0.02177\pm0.00026$&$1.012\pm0.012$&1.9807&30488.725&1.9034\\
		354981&809&2460182.87892361&$0.02162\pm0.00026$&$1.005\pm0.012$&1.97804&30561.318&1.9016\\
		354981&809&2460182.87907407&$0.02140\pm0.00026$&$0.995\pm0.012$&1.97533&30451.943&1.9763\\
		354981&809&2460182.87922454&$0.02146\pm0.00026$&$0.998\pm0.012$&1.97268&30389.973&1.9915\\
		... & ... & ... & ... & ... & ... & ... & ... \\
		\hline
		\multicolumn{8}{c}{\it Run 2}\\
		\hline
		371970&809&2460417.47934028&$0.01771\pm0.00009$&$0.993\pm0.005$&1.04657&31509.871&1.7379\\
		371970&809&2460417.47966435&$0.01789\pm0.00009$&$1.002\pm0.005$&1.04714&30847.371&1.7083\\
		371970&809&2460417.47993056&$0.01784\pm0.00009$&$1.000\pm0.005$&1.0476&30342.18&1.7198\\
		371970&809&2460417.48008102&$0.01776\pm0.00009$&$0.995\pm0.005$&1.04786&30145.863&1.7168\\
		371970&809&2460417.48023148&$0.01781\pm0.00009$&$0.998\pm0.005$&1.04813&29873.92&1.7392\\
		... & ... & ... & ... & ... & ... & ... & ... \\
		\hline
		\multicolumn{8}{c}{\it Run 3}\\
		\hline
		380563&809&2460541.9097338&$0.02013\pm0.00026$&$1.018\pm0.013$&1.75903&67949.33&2.0486\\
		380563&809&2460541.90997685&$0.01997\pm0.00026$&$1.010\pm0.013$&1.75561&67824.34&2.0401\\
		380563&809&2460541.91012731&$0.01986\pm0.00025$&$1.004\pm0.013$&1.75361&67808.19&2.0626\\
		380563&809&2460541.91027778&$0.01980\pm0.00025$&$1.002\pm0.013$&1.7516&68065.3&1.9862\\
		380563&809&2460541.91042824&$0.01966\pm0.00025$&$0.994\pm0.013$&1.7496&67986.94&2.0328\\
		... & ... & ... & ... & ... & ... & ... & ... \\
		\hline
		\multicolumn{8}{c}{\it Run 4}\\
		\hline
		389750&809&2460637.63924769&$0.01995\pm0.00023$&$1.013\pm0.011$&1.9713&20280.932&1.9383\\
		389750&809&2460637.63939815&$0.01961\pm0.00022$&$0.995\pm0.011$&1.96865&20225.166&1.8716\\
		389750&809&2460637.63954861&$0.01984\pm0.00022$&$1.007\pm0.011$&1.96599&20302.121&1.9032\\
		389750&809&2460637.63969907&$0.01976\pm0.00022$&$1.003\pm0.011$&1.96334&20340.297&1.9533\\
		389750&809&2460637.63984954&$0.01978\pm0.00022$&$1.004\pm0.011$&1.96071&20321.646&1.9033\\
		... & ... & ... & ... & ... & ... & ... & ... \\
		\hline
		\multicolumn{8}{c}{\it Run 5}\\
		\hline
		422211&806&2461012.61271991&$0.01583\pm0.00021$&$1.017\pm0.013$&1.97804&56994.95&1.8726\\
		422211&806&2461012.61297454&$0.01530\pm0.00020$&$0.983\pm0.013$&1.9736&56874.996&1.8558\\
		422211&806&2461012.61311343&$0.01542\pm0.00020$&$0.991\pm0.013$&1.97096&56803.188&1.8159\\
		422211&806&2461012.61326389&$0.01541\pm0.00020$&$0.990\pm0.013$&1.96832&56776.184&1.853\\
		422211&806&2461012.61341435&$0.01558\pm0.00021$&$1.001\pm0.013$&1.96568&56766.62&1.8438\\
		... & ... & ... & ... & ... & ... & ... & ... \\
		\bottomrule
	\end{tabular}
\end{table}

\section{Radial Velocity data}
\label{section:appendix:RV_data}
Here we present the radial velocity timeseries data from both \coralie\,(Table\,\ref{tab:rvs_coralie}) and \harps\,(Table\,\ref{tab:rvs_harps}). These tables will be available online in a machine readable format.

For both datasets Time is the Time in Barycentric \tess\ Julian Date (BTJD = BJD-2457000) rounded to 3 decimal places and RV is the radial velocity in \kms. For \harps\ we list the exposure time T\textsubscript{exp} in minutes for each spectrum whilst we omit this for \coralie\ as every spectrum was taken with an exposure time of 20\,minutes. For \coralie\ SNR$_{62}$ is the signal to noise ratio of spectral order 62 and for \harps\ SNR$_{64}$ is the signal to noise ratio of spectral order 64. For both datasets, `Airmass' is the airmass at the midpoint of the exposure for each spectrum. FWHM and BIS are the Full Width Half Maximum and Bisector Inverse Slopes of the Cross Correlation Function used to determine the RVs of each spectrum in \kms\ and \ms\ respectively. For Both datasets we also include the Contrast as well as the Hydrogen alpha, Sodium and Calcium indexes as dimensionless quantities. We list the Barycentric Radial Velocity correction (Berv) in \ms\ for both datasets. For \harps\ we also include the \logrhk\ value calculated for each spectra.

\begin{table*}
	\centering
	\caption{\coralie\ data}
	\label{tab:rvs_coralie}
	\begin{tabular}{cccccccccccc}
		\toprule
		Time & RV & T$_\text{exp}$ & SNR$_{62}$ & AM & FWHM & BIS & Contrast & \halpha\,index & Na\,index & Ca\,index & Berv \\
		(BTJD) & \kms & (mins) & & & \kms & \ms & & & & & \ms \\
		\hline
		\multicolumn{12}{c}{\coralie\,14} \\
		\hline
		3050.560&$20.19\pm0.02$&20&23.3&1.29&$13.00\pm0.05$&$-18\pm45$&$30.61\pm0.11$&$0.189\pm0.001$&$0.413\pm0.001$&$0.103\pm0.005$&-18.87\\
		3289.701&$20.05\pm0.01$&20&31.8&1.11&$13.00\pm0.03$&$-38\pm28$&$30.72\pm0.07$&$0.187\pm0.001$&$0.423\pm0.001$&$0.115\pm0.002$&11.72\\
		3310.871&$20.02\pm0.01$&20&36.1&1.41&$13.04\pm0.03$&$-46\pm25$&$30.44\pm0.06$&$0.187\pm0.001$&$0.406\pm0.001$&$0.132\pm0.002$&5.05\\
		3367.678&$19.96\pm0.01$&20&34.5&1.20&$12.94\pm0.03$&$39\pm28$&$30.48\pm0.07$&$0.195\pm0.001$&$0.412\pm0.001$&$0.126\pm0.003$&-12.48\\
		3381.712&$20.02\pm0.01$&20&35.3&1.74&$13.00\pm0.03$&$45\pm28$&$30.50\pm0.07$&$0.189\pm0.001$&$0.408\pm0.001$&$0.124\pm0.004$&-15.63\\
		3394.673&$20.13\pm0.01$&20&37.9&1.76&$12.96\pm0.03$&$-5\pm27$&$30.55\pm0.06$&$0.182\pm0.001$&$0.399\pm0.001$&$0.123\pm0.003$&-17.60\\
		3397.534&$20.15\pm0.02$&20&31.6&1.03&$12.95\pm0.03$&$18\pm34$&$30.66\pm0.08$&$0.180\pm0.001$&$0.394\pm0.001$&$0.133\pm0.004$&-17.71\\
		3401.572&$20.19\pm0.01$&20&40.4&1.15&$12.95\pm0.03$&$-8\pm25$&$30.58\pm0.06$&$0.187\pm0.001$&$0.401\pm0.001$&$0.097\pm0.002$&-18.19\\
		3404.564&$20.18\pm0.01$&20&39.2&1.16&$12.97\pm0.03$&$-4\pm26$&$30.52\pm0.06$&$0.190\pm0.001$&$0.399\pm0.001$&$0.124\pm0.003$&-18.42\\
		3409.579&$20.22\pm0.02$&20&23.5&1.28&$12.96\pm0.05$&$-84\pm49$&$30.57\pm0.12$&$0.192\pm0.001$&$0.403\pm0.001$&$0.046\pm0.006$&-18.74\\
		3424.509&$20.19\pm0.02$&20&34.3&1.16&$12.98\pm0.03$&$-23\pm31$&$30.42\pm0.07$&$0.193\pm0.001$&$0.409\pm0.001$&$0.099\pm0.003$&-18.66\\
		3428.526&$20.20\pm0.02$&20&30.5&1.29&$12.93\pm0.04$&$-25\pm37$&$30.70\pm0.09$&$0.184\pm0.001$&$0.392\pm0.001$&$0.128\pm0.004$&-18.50\\
		3435.483&$20.19\pm0.01$&20&35.2&1.18&$12.94\pm0.03$&$-2\pm28$&$30.65\pm0.07$&$0.190\pm0.001$&$0.413\pm0.001$&$0.147\pm0.003$&-17.91\\
		\hline
		\multicolumn{12}{c}{\coralie\,24} \\
		\hline
		3605.816&$20.21\pm0.02$&20&24.9&1.17&$13.10\pm0.05$&$29\pm45$&$30.67\pm0.11$&$0.194\pm0.002$&$0.437\pm0.001$&$0.148\pm0.004$&18.93\\
		3614.760&$20.18\pm0.02$&20&19.6&1.32&$12.92\pm0.05$&$-108\pm48$&$30.69\pm0.11$&$0.200\pm0.002$&$0.427\pm0.001$&$0.136\pm0.005$&18.63\\
		3621.746&$20.15\pm0.01$&20&31.1&1.29&$12.83\pm0.03$&$-29\pm28$&$30.55\pm0.07$&$0.193\pm0.001$&$0.404\pm0.001$&$0.130\pm0.003$&18.03\\
		3630.688&$20.15\pm0.02$&20&25.9&1.55&$13.03\pm0.04$&$3\pm38$&$30.74\pm0.09$&$0.185\pm0.001$&$0.411\pm0.001$&$0.106\pm0.004$&16.93\\
		3637.714&$20.12\pm0.02$&20&31.3&1.23&$12.98\pm0.03$&$-40\pm30$&$30.57\pm0.07$&$0.198\pm0.001$&$0.414\pm0.001$&$0.130\pm0.003$&15.68\\
		3659.728&$20.05\pm0.01$&20&33.7&1.02&$13.01\pm0.03$&$-39\pm27$&$30.53\pm0.06$&$0.191\pm0.001$&$0.417\pm0.001$&$0.125\pm0.002$&10.35\\
		3666.642&$20.04\pm0.02$&20&27.9&1.20&$12.93\pm0.03$&$-30\pm34$&$30.68\pm0.08$&$0.191\pm0.001$&$0.402\pm0.001$&$0.128\pm0.003$&8.49\\
		3677.698&$20.03\pm0.02$&20&27.8&1.00&$12.92\pm0.03$&$-27\pm34$&$30.65\pm0.08$&$0.194\pm0.001$&$0.395\pm0.001$&$0.178\pm0.003$&4.88\\
		3684.604&$19.99\pm0.02$&20&26.3&1.16&$12.89\pm0.04$&$-59\pm36$&$30.58\pm0.09$&$0.188\pm0.001$&$0.386\pm0.001$&$0.124\pm0.004$&2.78\\
		3694.557&$20.01\pm0.02$&20&29.5&1.25&$13.03\pm0.03$&$14\pm33$&$30.65\pm0.08$&$0.176\pm0.001$&$0.398\pm0.001$&$0.119\pm0.003$&-0.47\\
		3703.621&$19.95\pm0.01$&20&34.2&1.01&$12.83\pm0.03$&$55\pm29$&$30.64\pm0.07$&$0.185\pm0.001$&$0.398\pm0.001$&$0.134\pm0.003$&-3.63\\
		3710.720&$19.94\pm0.02$&20&29.3&1.13&$12.95\pm0.03$&$42\pm34$&$30.70\pm0.08$&$0.182\pm0.001$&$0.397\pm0.001$&$0.125\pm0.004$&-6.16\\
		3715.699&$19.95\pm0.04$&25&15.3&1.11&$13.04\pm0.07$&$-16\pm73$&$30.72\pm0.17$&$0.188\pm0.003$&$0.404\pm0.002$&$0.138\pm0.010$&-7.67\\
		3722.635&$19.95\pm0.02$&20&28.3&1.02&$12.90\pm0.03$&$-58\pm32$&$30.46\pm0.08$&$0.183\pm0.001$&$0.395\pm0.001$&$0.146\pm0.003$&-9.60\\
		3727.648&$19.95\pm0.01$&20&31.0&1.06&$13.03\pm0.03$&$-35\pm30$&$30.53\pm0.07$&$0.181\pm0.001$&$0.396\pm0.001$&$0.121\pm0.003$&-11.03\\
		3734.659&$19.99\pm0.02$&20&27.3&1.15&$12.96\pm0.03$&$43\pm33$&$30.59\pm0.08$&$0.193\pm0.001$&$0.398\pm0.001$&$0.137\pm0.003$&-12.88\\
		3747.588&$20.06\pm0.01$&20&30.4&1.05&$12.96\pm0.03$&$72\pm30$&$30.46\pm0.07$&$0.187\pm0.001$&$0.407\pm0.001$&$0.148\pm0.003$&-15.54\\
		3754.634&$20.13\pm0.01$&20&34.9&1.28&$13.03\pm0.03$&$16\pm26$&$30.58\pm0.06$&$0.194\pm0.001$&$0.398\pm0.001$&$0.127\pm0.002$&-16.82\\
		3760.643&$20.20\pm0.02$&20&23.3&1.47&$12.96\pm0.04$&$40\pm41$&$30.57\pm0.10$&$0.190\pm0.001$&$0.409\pm0.001$&$0.120\pm0.005$&-17.66\\
		3765.529&$20.22\pm0.02$&20&22.5&1.04&$12.93\pm0.04$&$21\pm41$&$30.54\pm0.10$&$0.195\pm0.002$&$0.407\pm0.001$&$0.122\pm0.004$&-17.99\\
		3773.532&$20.23\pm0.01$&20&33.4&1.09&$12.93\pm0.03$&$6\pm27$&$30.64\pm0.06$&$0.195\pm0.001$&$0.422\pm0.001$&$0.108\pm0.003$&-18.58\\
		3777.506&$20.22\pm0.02$&20&28.4&1.05&$12.93\pm0.03$&$-16\pm33$&$30.63\pm0.08$&$0.210\pm0.001$&$0.416\pm0.001$&$0.151\pm0.003$&-18.69\\
		3782.500&$20.20\pm0.02$&20&32.9&1.07&$12.90\pm0.03$&$-5\pm31$&$30.35\pm0.07$&$0.189\pm0.001$&$0.404\pm0.001$&$0.097\pm0.003$&-18.76\\
		3784.553&$20.19\pm0.02$&20&22.8&1.31&$12.97\pm0.04$&$-58\pm44$&$30.56\pm0.10$&$0.192\pm0.002$&$0.415\pm0.001$&$0.097\pm0.005$&-18.85\\
		3789.498&$20.16\pm0.02$&20&26.6&1.12&$12.94\pm0.04$&$-40\pm36$&$30.61\pm0.09$&$0.201\pm0.001$&$0.402\pm0.001$&$0.092\pm0.004$&-18.66\\
		3795.509&$20.18\pm0.02$&20&29.1&1.23&$12.94\pm0.04$&$-45\pm36$&$30.62\pm0.08$&$0.189\pm0.001$&$0.406\pm0.001$&$0.094\pm0.004$&-18.37\\
		3801.497&$20.17\pm0.02$&20&28.5&1.25&$12.94\pm0.03$&$-7\pm34$&$30.60\pm0.08$&$0.195\pm0.001$&$0.420\pm0.001$&$0.127\pm0.004$&-17.85\\
		3808.494&$20.17\pm0.02$&20&28.6&1.35&$13.01\pm0.05$&$4\pm46$&$30.29\pm0.11$&$0.172\pm0.001$&$0.421\pm0.001$&$0.187\pm0.005$&-17.03\\
		3818.466&$20.13\pm0.01$&20&33.1&1.35&$12.93\pm0.03$&$-33\pm30$&$30.52\pm0.07$&$0.189\pm0.001$&$0.400\pm0.001$&$0.118\pm0.003$&-15.46\\
		3826.445&$20.09\pm0.02$&20&30.2&1.34&$12.86\pm0.03$&$6\pm34$&$30.32\pm0.08$&$0.196\pm0.001$&$0.404\pm0.001$&$0.133\pm0.004$&-13.88\\
		3834.458&$20.08\pm0.02$&20&27.9&1.66&$12.79\pm0.04$&$9\pm38$&$30.44\pm0.09$&$0.194\pm0.001$&$0.409\pm0.001$&$0.134\pm0.005$&-12.09\\
		3949.837&$20.18\pm0.02$&20&23.2&1.34&$12.96\pm0.04$&$-7\pm40$&$30.63\pm0.09$&$0.192\pm0.001$&$0.419\pm0.001$&$0.092\pm0.004$&18.06\\
		3957.846&$20.22\pm0.02$&20&29.0&1.17&$13.05\pm0.03$&$-35\pm31$&$30.47\pm0.07$&$0.199\pm0.001$&$0.426\pm0.001$&$0.089\pm0.003$&18.64\\
		3964.873&$20.21\pm0.02$&20&29.2&1.05&$12.87\pm0.03$&$6\pm31$&$30.53\pm0.07$&$0.196\pm0.001$&$0.416\pm0.001$&$0.134\pm0.003$&18.84\\
		\bottomrule
	\end{tabular}
\end{table*}

\begin{landscape}
    \begin{table*}
	\centering
	\caption{\harps\ data}
	\label{tab:rvs_harps}
	\begin{tabular}{cccccccccccccc}
		\toprule
		Time & RV & T\textsubscript{exp} & SNR$_{64}$ & AM & FWHM & BIS & Contrast & \halpha\,index & Na\,index & Ca\,index & Berv & \logrhk \\
		(BTJD) & \kms & (mins) & & & \kms & \ms & & & & & \ms & \\
		\hline
		3299.723&$20.038\pm0.003$&25&75.5&1.02&$12.22\pm0.01$&$16\pm7$&$31.95\pm0.02$&$0.1926\pm0.0002$&$0.3893\pm0.0002$&$0.1084\pm0.0004$&8.82&$-5.012\pm0.003$\\
		3325.788&$19.967\pm0.004$&25&63.6&1.19&$12.24\pm0.01$&$-4\pm8$&$31.92\pm0.02$&$0.1948\pm0.0003$&$0.3886\pm0.0002$&$0.1059\pm0.0006$&0.22&$-5.032\pm0.005$\\
		3328.702&$19.967\pm0.005$&25&56.8&1.01&$12.21\pm0.01$&$11\pm9$&$31.91\pm0.03$&$0.1971\pm0.0003$&$0.3871\pm0.0002$&$0.1077\pm0.0006$&-0.59&$-5.017\pm0.005$\\
		3331.723&$19.951\pm0.005$&25&51.4&1.05&$12.21\pm0.01$&$3\pm10$&$31.92\pm0.03$&$0.1973\pm0.0004$&$0.3862\pm0.0003$&$0.1072\pm0.0007$&-1.65&$-5.022\pm0.006$\\
		3342.747&$19.939\pm0.004$&25&75.9&1.21&$12.26\pm0.01$&$-0\pm7$&$31.93\pm0.02$&$0.1898\pm0.0002$&$0.3847\pm0.0002$&$0.1067\pm0.0005$&-5.33&$-5.026\pm0.004$\\
		3346.790&$19.929\pm0.005$&25&58.9&1.59&$12.22\pm0.01$&$0\pm10$&$31.97\pm0.03$&$0.1903\pm0.0003$&$0.3785\pm0.0002$&$0.0993\pm0.0010$&-6.68&$-5.091\pm0.009$\\
		3350.712&$19.922\pm0.003$&25&78.4&1.16&$12.23\pm0.01$&$16\pm7$&$31.89\pm0.02$&$0.1915\pm0.0002$&$0.3822\pm0.0002$&$0.1056\pm0.0005$&-7.78&$-5.035\pm0.004$\\
		3375.668&$19.956\pm0.004$&20&65.1&1.27&$12.22\pm0.01$&$20\pm8$&$31.95\pm0.02$&$0.1918\pm0.0002$&$0.3801\pm0.0002$&$0.1067\pm0.0006$&-14.34&$-5.026\pm0.005$\\
		3380.527&$19.988\pm0.005$&20&53.2&1.00&$12.22\pm0.01$&$11\pm9$&$31.97\pm0.03$&$0.1954\pm0.0003$&$0.3861\pm0.0003$&$0.1101\pm0.0005$&-15.07&$-4.999\pm0.004$\\
		3598.796&$20.195\pm0.006$&20&46.1&1.36&$12.20\pm0.01$&$24\pm11$&$31.99\pm0.03$&$0.2033\pm0.0004$&$0.3962\pm0.0003$&$0.0924\pm0.0008$&18.95&$-5.161\pm0.009$\\
		3601.803&$20.182\pm0.005$&20&44.9&1.26&$12.20\pm0.01$&$-1\pm11$&$32.00\pm0.03$&$0.2021\pm0.0004$&$0.3996\pm0.0003$&$0.0906\pm0.0008$&18.98&$-5.182\pm0.009$\\
		3614.774&$20.171\pm0.006$&20&42.7&1.24&$12.23\pm0.01$&$-2\pm12$&$31.99\pm0.04$&$0.2038\pm0.0004$&$0.3955\pm0.0004$&$0.0964\pm0.0010$&18.61&$-5.119\pm0.010$\\
		3660.672&$20.042\pm0.004$&20&55.0&1.15&$12.24\pm0.01$&$8\pm9$&$31.98\pm0.03$&$0.1986\pm0.0003$&$0.3919\pm0.0003$&$0.0937\pm0.0006$&10.19&$-5.147\pm0.007$\\
		3702.720&$19.954\pm0.005$&20&53.0&1.07&$12.24\pm0.01$&$16\pm9$&$31.98\pm0.03$&$0.1944\pm0.0003$&$0.3968\pm0.0002$&$0.1025\pm0.0008$&-3.55&$-5.061\pm0.007$\\
		3731.658&$19.949\pm0.005$&20&52.8&1.13&$12.23\pm0.01$&$26\pm10$&$31.98\pm0.03$&$0.1951\pm0.0003$&$0.3979\pm0.0003$&$0.0945\pm0.0007$&-12.12&$-5.138\pm0.008$\\
		3737.647&$19.966\pm0.005$&20&49.7&1.14&$12.22\pm0.01$&$29\pm11$&$31.93\pm0.03$&$0.1972\pm0.0003$&$0.3905\pm0.0003$&$0.0945\pm0.0009$&-13.57&$-5.139\pm0.009$\\
		3747.656&$20.048\pm0.005$&20&59.0&1.31&$12.24\pm0.01$&$6\pm9$&$31.86\pm0.03$&$0.1898\pm0.0003$&$0.3819\pm0.0002$&$0.0987\pm0.0008$&-15.67&$-5.097\pm0.007$\\
		3753.588&$20.106\pm0.005$&20&52.2&1.10&$12.23\pm0.01$&$-6\pm10$&$31.86\pm0.03$&$0.1968\pm0.0003$&$0.3973\pm0.0002$&$0.0833\pm0.0007$&-16.57&$-5.277\pm0.010$\\
		3765.630&$20.185\pm0.005$&20&49.9&1.49&$12.19\pm0.01$&$-6\pm11$&$31.92\pm0.03$&$0.1981\pm0.0004$&$0.3960\pm0.0003$&$0.0872\pm0.0010$&-18.17&$-5.223\pm0.013$\\
		3784.578&$20.188\pm0.008$&20&35.3&1.49&$12.23\pm0.02$&$19\pm16$&$32.04\pm0.05$&$0.2028\pm0.0006$&$0.3988\pm0.0004$&$0.0643\pm0.0018$&-18.89&$-5.716\pm0.071$\\
		3800.528&$20.152\pm0.004$&20&63.2&1.45&$12.24\pm0.01$&$14\pm8$&$31.88\pm0.02$&$0.1908\pm0.0002$&$0.4011\pm0.0002$&$0.0921\pm0.0007$&-17.99&$-5.165\pm0.008$\\
		3801.463&$20.158\pm0.004$&20&57.0&1.13&$12.22\pm0.01$&$2\pm9$&$31.93\pm0.03$&$0.1921\pm0.0003$&$0.4009\pm0.0002$&$0.0900\pm0.0006$&-17.79&$-5.189\pm0.007$\\
		4045.820&$19.987\pm0.005$&20&56.0&1.20&$12.25\pm0.01$&$14\pm10$&$32.04\pm0.03$&$0.2000\pm0.0003$&$0.3794\pm0.0003$&$0.0965\pm0.0008$&3.69&$-5.118\pm0.008$\\
		\bottomrule
	\end{tabular}
\end{table*}
\end{landscape}


\section{\alles\ results tables}
\label{section:appendix:alles}
Here we present the full set of fitted (Table\,\ref{tab:ns_table}) and derived (Table\,\ref{tab:ns_derived_table}) parameters of \name\,b produced by \alles\ (see Section\,\ref{section:orbit-fit}). Some of these parameters are repeated in Table\,\ref{tab:planet_properties}. These will be available online in a machine readable format.

\begin{table}
	\centering
	\caption{\alles\ global fitted values and priors.}
	\label{tab:ns_table}
	\begin{tabular}{c c c c c}
		\toprule
		Parameter & Initial guess & Prior & Fitted value & units\\
		\hline
        \multicolumn{5}{c}{\it Astrophysical parameters}\\
        \hline
		$R_b / R_\star$&0.06&$\mathcal{U}\left(0,0.1\right)$&\brr&\\
		$(R_\star + R_b) / a_b$&0.014&$\mathcal{U}\left(0,0.05\right)$&\brsuma&\\
		$\cos{i_b}$&0.009&$\mathcal{U}\left(0,1\right)$&\bcosi&\\
		$T_{0;b}$&2459209.228&$\mathcal{U}\left(2459208,2459210\right)$&\bepoch&BJD\\
		$P_b$&180.5&$\mathcal{U}\left(179,181\right)$&\bperiod&days\\
		$\sqrt{e_b} \cos{\omega_b}$&0.1&$\mathcal{U}\left(-1,1\right)$&\bfc&\\
		$\sqrt{e_b} \sin{\omega_b}$&-0.5&$\mathcal{U}\left(-1,1\right)$&\bfs&\\
		$K_b$&0.14&$\mathcal{U}\left(0,0.5\right)$&\bK&\kms\\
        \hline
        \multicolumn{5}{c}{\it Instrumental limb darkening parameters}\\
        \hline
		$q_{1; \mathrm{TESS}}$&0.278&$\mathcal{N}\left(0.278,0.002\right)$&\hostldcqoneTESS&\\
		$q_{2; \mathrm{TESS}}$&0.375&$\mathcal{N}\left(0.375,0.025\right)$&\hostldcqtwoTESS&\\
		$q_{1; \mathrm{NGTS}}$&0.337&$\mathcal{N}\left(0.337,0.003\right)$&\hostldcqoneNGTS&\\
		$q_{2; \mathrm{NGTS}}$&0.385&$\mathcal{N}\left(0.385,0.025\right)$&\hostldcqtwoNGTS&\\
		$q_{1; \mathrm{LCO}}$&0.283&$\mathcal{N}\left(0.283,0.003\right)$&\hostldcqoneLCO&\\
		$q_{2; \mathrm{LCO}}$&0.379&$\mathcal{N}\left(0.379,0.026\right)$&\hostldcqtwoLCO&\\
		$q_{1; \mathrm{ElSauce}}$&0.376&$\mathcal{N}\left(0.376,0.009\right)$&\hostldcqoneElSauce&\\
		$q_{2; \mathrm{ElSauce}}$&0.388&$\mathcal{N}\left(0.388,0.073\right)$&\hostldcqtwoElSauce&\\
		$q_{1; \mathrm{DSC0.4m}}$&0.337&$\mathcal{N}\left(0.337,0.009\right)$&\hostldcqoneVignes&\\
		$q_{2; \mathrm{DSC0.4m}}$&0.384&$\mathcal{N}\left(0.384,0.076\right)$&\hostldcqtwoVignes&\\
		$q_{1; \mathrm{OACC-CAO}}$&0.269&$\mathcal{N}\left(0.269,0.008\right)$&\hostldcqoneMallia&\\
		$q_{2; \mathrm{OACC-CAO}}$&0.376&$\mathcal{N}\left(0.376,0.079\right)$&\hostldcqtwoMallia&\\
        \hline
        \multicolumn{5}{c}{\it Instrumental white noise parameters}\\
        \hline
		$\ln{\sigma_\mathrm{TESS}}$&-6.0&$\mathcal{U}\left(-14.0,0.0\right)$&\lnerrfluxTESS&$\ln{ \mathrm{rel. flux.} }$\\
		$\ln{\sigma_\mathrm{NGTS}}$&-6.0&$\mathcal{U}\left(-14.0,0.0\right)$&\lnerrfluxNGTS&$\ln{ \mathrm{rel. flux.} }$\\
		$\ln{\sigma_\mathrm{LCO-CTIO (1m)}}$&-6.0&$\mathcal{U}\left(-14.0,0.0\right)$&\lnerrfluxLCOCTIOL&$\ln{ \mathrm{rel. flux.} }$\\
		$\ln{\sigma_\mathrm{LCO-CTIO (0.35m)}}$&-6.0&$\mathcal{U}\left(-14.0,0.0\right)$&\lnerrfluxLCOCTIOS&$\ln{ \mathrm{rel. flux.} }$\\
		$\ln{\sigma_\mathrm{LCO-SSO (1m)}}$&-6.0&$\mathcal{U}\left(-14.0,0.0\right)$&\lnerrfluxLCOSSOL&$\ln{ \mathrm{rel. flux.} }$\\
		$\ln{\sigma_\mathrm{LCO-SSO (0.35m)}}$&-6.0&$\mathcal{U}\left(-14.0,0.0\right)$&\lnerrfluxLCOSSOS&$\ln{ \mathrm{rel. flux.} }$\\
		$\ln{\sigma_\mathrm{ElSauce}}$&-6.0&$\mathcal{U}\left(-14.0,0.0\right)$&\lnerrfluxElSauce&$\ln{ \mathrm{rel. flux.} }$\\
		$\ln{\sigma_\mathrm{DSC0.4m}}$&-6.0&$\mathcal{U}\left(-14.0,0.0\right)$&\lnerrfluxVignes&$\ln{ \mathrm{rel. flux.} }$\\
        \smallskip
		$\ln{\sigma_\mathrm{OACC-CAO}}$&-6.0&$\mathcal{U}\left(-14.0,0.0\right)$&\lnerrfluxMallia&$\ln{ \mathrm{rel. flux.} }$\\
        \smallskip
		$\ln{\sigma_\mathrm{CORALIE14}}$&-5.0&$\mathcal{U}\left(-7.6,-4.6\right)$&\lnjitterrvCORALIEonefour&$\ln$\,\kms\\
        \smallskip
		$\ln{\sigma_\mathrm{CORALIE24}}$&-5.0&$\mathcal{U}\left(-7.6,-4.6\right)$&\lnjitterrvCORALIEtwofour&$\ln$\,\kms\\
		$\ln{\sigma_\mathrm{HARPS}}$&-5.0&$\mathcal{U}\left(-7.6,-4.6\right)$&\lnjitterrvHARPS&$\ln$\,\kms\\
        \hline
        \multicolumn{5}{c}{\it Instrumental red noise / offset parameters}\\
        \hline
		offset CORALIE14&20.03&$\mathcal{U}\left(19.9,20.1\right)$&\baselineoffsetrvCORALIEonefour&\kms\\
		offset CORALIE24&20.03&$\mathcal{U}\left(19.9,20.1\right)$&\baselineoffsetrvCORALIEtwofour&\kms\\
        \smallskip
		offset HARPS&19.93&$\mathcal{U}\left(19.9,20.1\right)$&\baselineoffsetrvHARPS&\kms\\
		offset NGTS&-0.0005&$\mathcal{U}\left(-0.001,0.001\right)$&\baselineoffsetfluxNGTS&\\
		offset LCO-CTIO (1m)&0.0&$\mathcal{U}\left(-0.001,0.001\right)$&\baselineoffsetfluxLCOCTIOL&\\
		offset LCO-CTIO (0.35m)&0.0&$\mathcal{U}\left(-0.001,0.001\right)$&\baselineoffsetfluxLCOCTIOS&\\
		offset LCO-SSO (1m)&0.0&$\mathcal{U}\left(-0.001,0.001\right)$&\baselineoffsetfluxLCOSSOL&\\
		offset LCO-SSO (0.35m)&0.0&$\mathcal{U}\left(-0.001,0.001\right)$&\baselineoffsetfluxLCOSSOS&\\
		offset ElSauce&0.0&$\mathcal{U}\left(-0.001,0.001\right)$&\baselineoffsetfluxElSauce&\\
		offset DSC0.4m&0.0&$\mathcal{U}\left(-0.001,0.001\right)$&\baselineoffsetfluxVignes&\\
		offset OACC-CAO&0.0&$\mathcal{U}\left(-0.001,0.001\right)$&\baselineoffsetfluxMallia&\\
		\bottomrule
	\end{tabular}
\end{table}

\begin{table}
    \centering
    \caption{\alles\ global model derived values.}
    \label{tab:ns_derived_table}
    \begin{tabular}{c c c}
        \toprule
        Parameter & Value & Unit\\ 
        \hline
        \multicolumn{3}{c}{\it Astrophysical parameters} \\
        \hline
        $R_\star/a_\mathrm{b}$ & \bRstarovera &  \\ 
        $a_\mathrm{b}/R_\star$ & \baoverRstar &  \\ 
        $R_\mathrm{b}/a_\mathrm{b}$ & \bRcompanionovera & \\ 
        $R_\mathrm{b}$ & \bRcompanionRearth & \rearth \\ 
        $R_\mathrm{b}$ & \bRcompanionRjup & \rjup \\ 
        $a_\mathrm{b}$ & \baRsun & \rsun \\ 
        $a_\mathrm{b}$ & \baAU & au \\ 
        $i_\mathrm{b}$ & \bi & deg \\ 
        $e_\mathrm{b}$ & \be & \\ 
        $\omega_\mathrm{b}$ & \bw & deg \\
        \smallskip
        $q_\mathrm{b}$ & \bq &  \\
        \smallskip
        $M_\mathrm{b}$ & \bMcompanionMearth & \mearth \\
        \smallskip
        $M_\mathrm{b}$ & \bMcompanionMjup & \mjup \\
        \smallskip
        $M_\mathrm{b}$ & \bMcompanionMsun & \msun  \\ 
        $b_\mathrm{tra;b}$ & \bbtra  \\ 
        $T_\mathrm{tot;b}$ & \bTtratot & hours \\ 
        \smallskip
        $T_\mathrm{full;b}$ & \bTtrafull & hours \\
        \smallskip
        $\rho_\mathrm{\star}$ & \combinedhostdensity & cgs  \\
        \smallskip
        $\rho_\mathrm{b}$ & \bdensity & cgs \\ 
        $g_\mathrm{b}$ & \bsurfacegravity & cgs \\ 
        $T_\mathrm{eq;b}$ & \bTeq & K \\
        \hline
        \multicolumn{3}{c}{\it Transit depth per Instrument} \\
        \hline
        \smallskip
        $\delta_\mathrm{tr; b; \tess}$ & \bdepthtrundilTESS & parts per thousand (ppt) \\
        \smallskip
        $\delta_\mathrm{tr; b; \ngts}$ & \bdepthtrundilNGTS & parts per thousand (ppt) \\
        \smallskip
        $\delta_\mathrm{tr; b; LCO-SSO 1m}$ & \bdepthtrundilLCOSSOL & parts per thousand (ppt) \\
        \smallskip
        $\delta_\mathrm{tr; b; LCO-SSO 0.35m}$ & \bdepthtrundilLCOSSOS & parts per thousand (ppt) \\
        \smallskip
        $\delta_\mathrm{tr; b; LCO-CTIO 1m}$ & \bdepthtrundilLCOCTIOL & parts per thousand (ppt) \\
        \smallskip
        $\delta_\mathrm{tr; b; LCO-CTIO 0.35m}$ & \bdepthtrundilLCOCTIOS & parts per thousand (ppt) \\
        \smallskip
        $\delta_\mathrm{tr; b; \elsauce}$ & \bdepthtrundilElSauce & parts per thousand (ppt) \\
        \smallskip
        $\delta_\mathrm{tr; b; DSC0.4m}$ & \bdepthtrundilVignes & parts per thousand (ppt) \\
        \smallskip
        $\delta_\mathrm{tr; b; OACC-CAO}$ & \bdepthtrundilMallia & parts per thousand (ppt) \\
        \hline
        \multicolumn{3}{c}{\it Classical limb darkening parameters per instrument} \\
        \hline
        $u_\mathrm{1; \tess}$ & \hostldcuoneTESS & \\ 
        $u_\mathrm{2; \tess}$ & \hostldcutwoTESS & \\ 
        $u_\mathrm{1; \ngts}$ & \hostldcuoneNGTS & \\ 
        $u_\mathrm{2; \ngts}$ & \hostldcutwoNGTS & \\
        $u_\mathrm{1; \lco}$ & \hostldcuoneLCO & \\ 
        $u_\mathrm{2; \lco}$ & \hostldcutwoLCO & \\
        $u_\mathrm{1; \elsauce}$ & \hostldcuoneElSauce & \\ 
        $u_\mathrm{2; \elsauce}$ & \hostldcutwoElSauce & \\
        $u_\mathrm{1; DSC0.4m}$ & \hostldcuoneVignes & \\ 
        $u_\mathrm{2; DSC0.4m}$ & \hostldcutwoVignes & \\
        $u_\mathrm{1; OACC-CAO}$ & \hostldcuoneMallia & \\ 
        $u_\mathrm{2; OACC-CAO}$ & \hostldcutwoMallia & \\
        \bottomrule
    \end{tabular}
\end{table}


\bsp	
\label{lastpage}
\end{document}